%% file: main.tex
\documentclass[10pt,journal,compsoc]{IEEEtran}
\usepackage{cite}
\usepackage{amsmath,amssymb,amsfonts}
\usepackage{graphicx}
\usepackage{textcomp}
\usepackage{xcolor}
\usepackage{caption}
\usepackage{tikz}
\usepackage{array}
\usepackage{hyperref,xcolor}
\usepackage{mdwmath}
\usepackage{mdwtab}
\usepackage{eqparbox}

\usepackage[tight,footnotesize]{subfigure}
\usepackage{amsfonts}
\usepackage{tikz}
\usepackage{textcomp}
\usepackage{wrapfig}
\usepackage{comment}
\usepackage{tcolorbox}

\usepackage{tabularx}
\usepackage{lscape}
\usepackage{mathtools}
\usepackage{booktabs}
\usepackage{algpseudocode}
\usepackage{filecontents}
\usepackage{algorithm}

\usepackage{times}

\usepackage{soul}
\usepackage{url}
\usepackage[utf8]{inputenc}
\usepackage{caption}
\usepackage{amsmath,amssymb,amsfonts}
\usepackage{booktabs}
\usepackage{tikz}
\usepackage{amsmath}

\usepackage{times}

\usepackage{soul}
\usepackage{url}

\usepackage[utf8]{inputenc}
\usepackage{graphicx}
\usepackage{booktabs}
\usepackage{booktabs}
\usepackage{algpseudocode}
\usepackage{algorithm}

\usepackage[none]{hyphenat}
\def\BibTeX{{\rm B\kern-.05em{\sc i\kern-.025em b}\kern-.08em
    T\kern-.1667em\lower.7ex\hbox{E}\kern-.125emX}}
\usepackage[none]{hyphenat}
\DeclareUnicodeCharacter{2212}{-}

\usepackage{cite}
\ifCLASSINFOpdf
\else
\fi
\begin{document}
\title{ \huge Physical Adversarial Attacks For Camera-based Smart Systems: Current Trends, Categorization, Applications, Research Challenges, and Future Outlook}
\author{Amira~Guesmi,~
        Muhammad~Abdullah~Hanif,~
        Bassem~Ouni,~
        and~Muhammed~Shafique~%\IEEEmembership{Life~Fellow,~IEEE
\IEEEcompsocitemizethanks{\IEEEcompsocthanksitem A. Guesmi, M. A. Hanif, and M. Shafique are with eBrain Lab, Division of Engineering, New York University (NYU) Abu Dhabi, UAE.\protect\\
%E-mail: see http://www.michaelshell.org/contact.html
\IEEEcompsocthanksitem B. Ouni is with the AI and Digital Science Research Center, Technology Innovation Institute (TII), Abu Dhabi, UAE.}}% <-this % stops a space
%\thanks{Manuscript received July 17, 2023; revised , .}}


\markboth{Journal of \LaTeX\ Class Files,~Vol.~14, No.~8, August~2015}%
{Guesmi \MakeLowercase{\textit{et al.}}: Physical Adversarial Attacks For Camera-based Smart Systems: Current Trends, Categorization, Applications, Research Challenges, and Future Outlook}



\IEEEtitleabstractindextext{%
\begin{abstract}
Deep Neural Networks (DNNs) have shown impressive performance in computer vision tasks; however, their vulnerability to adversarial attacks raises concerns regarding their security and reliability. Extensive research has shown that DNNs can be compromised by carefully crafted perturbations, leading to significant performance degradation in both digital and physical domains. Therefore, ensuring the security of DNN-based systems is crucial, particularly in safety-critical domains such as autonomous driving, robotics, smart homes/cities, smart industries, video surveillance, and healthcare. In this paper, we present a comprehensive survey of the current trends focusing specifically on physical adversarial attacks. We aim to provide a thorough understanding of the concept of physical adversarial attacks, analyzing their key characteristics and distinguishing features. Furthermore, we explore the specific requirements and challenges associated with executing attacks in the physical world. Our article delves into various physical adversarial attack methods, categorized according to their target tasks in different applications, including classification, detection, face recognition, semantic segmentation and depth estimation. We assess the performance of these attack methods in terms of their effectiveness, stealthiness, and robustness. We examine how each technique strives to ensure the successful manipulation of DNNs while mitigating the risk of detection and withstanding real-world distortions. Lastly, we discuss the current challenges and outline potential future research directions in the field of physical adversarial attacks. We highlight the need for enhanced defense mechanisms, the exploration of novel attack strategies, the evaluation of attacks in different application domains, and the establishment of standardized benchmarks and evaluation criteria for physical adversarial attacks. Through this comprehensive survey, we aim to provide a valuable resource for researchers, practitioners, and policymakers to gain a holistic understanding of physical adversarial attacks in computer vision and facilitate the development of robust and secure DNN-based systems.
\end{abstract}

\begin{IEEEkeywords}
Machine Learning Security, Physical Adversarial Attacks, Smart Systems, Camera-based vision system robustness, Security, Trustworthy AI, Computer Vision, Image Classification, Object Detection, Person Detection, Vehicle Detection, Semantic Segmentation, Monocular Depth Estimation, Face Recognition, Person Re-Identification, Optical Flow, Stealthy Physical Attacks, Patch-based attacks, Sticker-based attacks, Camouflage Techniques, Light Manipulation-based attacks, Physical Adversarial Prints, Imaging Device Manipulation, Expectation Over Transformation, Adversarial T-shirts, Adversarial Makeup, Adversarial Eyeglasses, Adversarial Masks. 
%challenges, commun trends, applications, open research problems, system, tools
\end{IEEEkeywords}}

\maketitle

% is not selected <OR> if conference mode is selected - because compsoc
% conference papers position the abstract like regular (non-compsoc)
\IEEEdisplaynontitleabstractindextext

% \fi
\IEEEpeerreviewmaketitle
\input{introduction}   
\input{preliminaries}
\input{attacks}
\input{classification}
\input{detection}
\input{facerecognition}
\input{semanticSeg}

\input{depth}

%\input{other_tasks}
\input{discussion}
\input{conclusion}

\ifCLASSOPTIONcaptionsoff
  \newpage
\fi



%\bibliographystyle{IEEEtran}
%\begin{thebibliography}{1}

%H.~Kopka and P.~W. Daly, \emph{A Guide to {\LaTeX}}, 3rd~ed.\hskip 1em plus

%\end{thebibliography}
\bibliographystyle{IEEEtran}
%\singlespacing
\bibliography{sample}

%\end{IEEEbiography}


%\newpage



%\vfill


\end{document}

%% file: introduction.tex
\section{Introduction}
\label{introduction}
\IEEEPARstart{T}{he} advent of deep learning (DL) has sparked revolutionary changes across various sectors, including computer vision (CV) \cite{simonyan2014deep, redmon2016yolo9000}, natural language processing (NLP) \cite{deng2018deep}, robotics \cite{pierson2017deep}, autonomous driving \cite{al2017deep}, and healthcare \cite{miotto2018deep}. These technologies have already found their way into numerous products and systems, leveraging the vast amounts of available data. %However, they are not immune to adversarial attacks, which pose a threat to their integrity and reliability.
With the ever-expanding utilization of Deep Learning (DL) models, there arises an inherent and pressing need to confront an alarming threat - that of adversarial attacks. The growing prevalence of Deep Neural Networks (DNNs) in real-world applications, particularly safety-critical ones, heightens the urgency to tackle this critical issue with utmost seriousness and diligence. Adversarial attacks represent a distinct threat to the reliability and integrity of DNNs, rendering their vulnerability an issue of paramount concern in the field of artificial intelligence.

Adversarial attacks are cunningly designed manipulations introduced to deceive DNNs, resulting in incorrect outputs, thus challenging their integrity and jeopardizing their reliability. These attacks exploit the vulnerabilities in DNNs, often leading to catastrophic consequences, especially in safety-critical applications where the stakes are incredibly high. Imagine an autonomous vehicle relying on a DNN for real-time decision-making during a potentially life-threatening situation. In the face of an adversarial attack, the vehicle might fail to recognize a stop sign~\cite{b0,fgsm}, causing a disastrous collision. Similarly, in healthcare, a misclassification by an attacked DNN could lead to an incorrect diagnosis, compromising patient well-being.

The alarming reality is that adversarial attacks are not only theoretical concepts but have manifested in real-world scenarios, exposing the urgent need to fortify DNNs against such threats. Ensuring robustness becomes paramount, as it not only safeguards the applications themselves but also protects the individuals and communities that rely on their accurate functioning. In fact, adversarial examples have proven to be effective even in real-world conditions~\cite{physical}. For instance, when an adversarially crafted image is printed out, it can still deceive classifiers under different lighting conditions and orientations, and can be deployed in safety-critical systems. Thus, it becomes imperative to comprehend and mitigate these attacks in order to develop intelligent systems that are safe and trustworthy.

Recognizing the significance of adversarial attacks lead to a surge of interest and efforts in this field. Numerous algorithms have been developed to generate adversarial examples, such as the Fast Gradient Sign Method (FGSM), Projected Gradient Descent (PGD), and Carlini and Wagner attack~\cite{fgsm, pgd, C&W}. These algorithms are often designed to bypass proposed defenses that aim to protect against adversarial attacks~\cite{distillation_SP, DA}. In this scenario (i.e., Digital attack setting), the attacker has the flexibility to arbitrarily alter the input image of a victim model at the pixel level. A more robust and realistic threat model would presume that the attacker exclusively has access and control of the system’s external environment or external objects, rather than its internal sensors and data pipelines (i.e., Physical attack setting).

%Currently, there is a significant amount of work focusing on digital adversarial attacks while less on physical adversarial attacks because performing attack in the physical space is more challenging due to the physical constraints, e.g., spatial deformation, illumination, and camera resolution, etc. Meanwhile, physical adversarial attacks are a greater threat to society because of their operability in the real world.
At present, a majority of research efforts are concentrated on digital adversarial attacks, while comparatively less attention has been given to physical adversarial attacks. This imbalance is attributed to the inherent difficulties associated with conducting attacks in the physical space, which involves overcoming constraints such as spatial deformation, illumination variations, and camera limitations. However, it is crucial to address physical adversarial attacks as they pose a more substantial and realistic threat to society due to their potential for real-world exploitation and impact, and even without having an inside access of the Artificial Intelligence (AI)/Machine Learning (ML) system. %Recognizing the operational nature of physical attacks underscores the need for developing robust defense mechanisms to mitigate their adverse consequences.

%Understanding the techniques employed by adversarial generation algorithms and their potential to circumvent existing defenses is critical. By comprehending the underlying mechanisms, researchers can develop more robust defenses against adversarial attacks. The ultimate goal is to create intelligent systems that are resilient to manipulation and capable of making accurate and trustworthy decisions in the face of potential attacks.

%Few are the surveys that focus on physical adversarial attacks, and the ones published doesn't cover all the attacks and are not up to date

\textit{There is a scarcity of comprehensive surveys that specifically address physical adversarial attacks.} Furthermore, the existing surveys often lack coverage of the full spectrum of attacks and may not reflect the latest developments in the field. As a result, there is a need for more extensive and up-to-date surveys that encompass the diverse range of physical adversarial attacks. 
Existing surveys on physical adversarial attacks have limitations in terms of coverage and scope. For example, Wei et al. \cite{wei2022physical_survey} focus on a subset of 41 physical adversarial attacks, while Fang et al. \cite{fang2023stateoftheart} specifically address optical-based physical adversarial attacks. Similarly, Wang et al. \cite{wang2022survey} primarily survey the tasks of image recognition and object detection, with some coverage of physical adversarial attacks in object tracking and semantic segmentation. However, their coverage is limited to a total of 71 attacks, with 38 attacks related to image recognition and 33 attacks related to object detection. These surveys provide insights into specific aspects of physical adversarial attacks, but \textit{a more comprehensive and up-to-date survey is needed to encompass the broader range of physical adversarial attacks across various computer vision tasks.}

In our comprehensive survey, we have thoroughly examined a total of 94 adversarial attacks, which are presented in Table \ref{all_attacks}. Our survey provides a comprehensive analysis of different attacks on different computer vision tasks. These tasks include Classification, Traffic Sign Recognition, Object Detection, Sign Detection, Person Detection, Vehicle Detection, Infrared Person Detection, Face Recognition, Person Re-identification, Depth Estimation, Optical Flow Estimation, and Semantic Segmentation. %, and various other tasks. 
By considering a wide range of computer vision tasks, our survey provides a comprehensive overview of the existing adversarial attacks and their implications in different domains.

\noindent
\textbf{In summary, the \textit{contributions} of this article are: }
\begin{itemize}

\item We comprehensively review a diverse array of physical adversarial attacks within various computer vision tasks.

\item We propose a taxonomy to categorize physical adversarial attacks based on their real-world form.

\item We analyze and elucidate techniques employed to augment the effectiveness, stealthiness, robustness, and transferability of physical attacks.

\item Detailed insights into physical adversarial attacks are presented, including their specific characteristics: attack goals, strategies for perturbation placement, consideration of changing viewpoints, testing in the physical world, transferability across models, attacker's knowledge level, robustness strategies, stealthiness approaches, physical test types, and operational space.

\item We provide a comprehensive information regarding the datasets employed, evaluated networks, and accessible open-source code for reproducibility.

\item We discuss research challenges and potential avenues of exploration within physical adversarial attacks.

%, including the exploration of multi-model attacks and the investigation of attacks on new tasks.

%\textcolor{red}{put a couple of points regarding tables and what kind of categorization and comparative discussion we provide}
%\item Through tables and discussions, we offer a comparative analysis of different attack techniques based on their effectiveness, robustness, stealthiness, and performance across multiple computer vision tasks.
\end{itemize}
%Comprehensive Overview: We provide a comprehensive overview of the field of physical adversarial attacks in computer vision tasks
%Categorization of Attacks: We propose a taxonomy to categorize physical adversarial attacks based on their characteristics, including attack goals, patch placement strategies, consideration of changing viewpoints, testing in the physical domain, and transferability to other models. 
%Detailed Attack Information: We present detailed information on adversarial attacks, such as the attacker's knowledge level, robustness techniques, stealthiness techniques, physical test types, and space of operation.
%Benchmarking and Datasets: We provide information on the datasets used in various computer vision tasks, the networks evaluated, and links to open-source code for reproducibility.
%Comparative Analysis: Through tables and discussions, we offer a comparative analysis of different attack techniques based on their effectiveness, robustness, stealthiness, and performance across multiple computer vision tasks.
%Identification of Challenges: We identify the challenges in physical adversarial attack research, such as the lack of standardized evaluation metrics for stealthiness and the difficulty of quantitatively evaluating attacks. 
%Future Research Directions: We discuss potential research directions in physical adversarial attacks, including the exploration of multi-model attacks and the investigation of attacks on new tasks
Paper Organization: The structure of the remaining article is organized as follows. In Section \ref{background}, we provide a comprehensive introduction to the preliminaries, which encompass essential topics and concepts necessary for a thorough understanding of the subsequent sections. Section \ref{forms} focuses on a detailed discussion of different forms of physical adversarial attacks. Further contributions of this paper are presented in Sections \ref{classification_}, \ref{detection}, \ref{face}, \ref{semantic_segmentation} and \ref{depth}, where we extensively highlight the recent advancements in physical adversarial attacks. In Section \ref{discussion}, we engage in comprehensive discussions on the findings derived from the reviewed literature and identify potential future research directions and opportunities within the field. Finally, in Section \ref{conclusion}, we provide a succinct summary, encapsulating the key insights and contributions presented throughout this article.

%% file: preliminaries.tex
\section{Background}
\label{background}
%-------------------------------------
\subsection{Camera-based Vision Systems}
%-------------------------------------
The field of environment perception has witnessed significant advancements driven by both industry and the research community. This progress has been particularly notable in the development of deep learning (DL)-based solutions for applications like autonomous robots and intelligent transportation systems. A key focus in this area has been on designing robust DL models for reliable recognition systems, which play a crucial role in achieving high-performance environment perception. In the context of vision-based perception modules, such as those used in automotive cameras, the accurate detection and classification of objects are fundamental for gathering reliable information about the environment. Object detectors and image classifiers serve as the foundation for these perception modules, enabling autonomous vehicles (AVs) to make crucial decisions that ensure safe driving. However, the challenge lies in designing recognition systems that are dependable and robust against different types of adversarial attacks.

Despite the significant efforts invested in developing recognition systems, various studies/investigations have demonstrated that these systems are vulnerable to adversarial attacks \cite{fgsm, C&W}. Adversarial attacks exploit the weaknesses and limitations of DL models by introducing carefully crafted perturbations that can deceive the perception modules. This vulnerability poses a significant obstacle to establishing highly reliable and secure environment perception systems.

Establishing dependable recognition systems remains one of the primary hurdles in achieving high-performance environment perception. Nonetheless, through ongoing research and collaboration, it is possible to develop innovative strategies that enhance the security and reliability of DL-based perception modules. By ensuring the resilience of these modules against adversarial attacks, we can unlock the full potential of autonomous robots and intelligent transportation systems, ultimately leading to safer and more efficient environments for all.

\subsection{Adversarial Attacks}
\begin{figure}[!htp]
\centering
\includegraphics[width=0.25\textwidth, height=3.7cm]{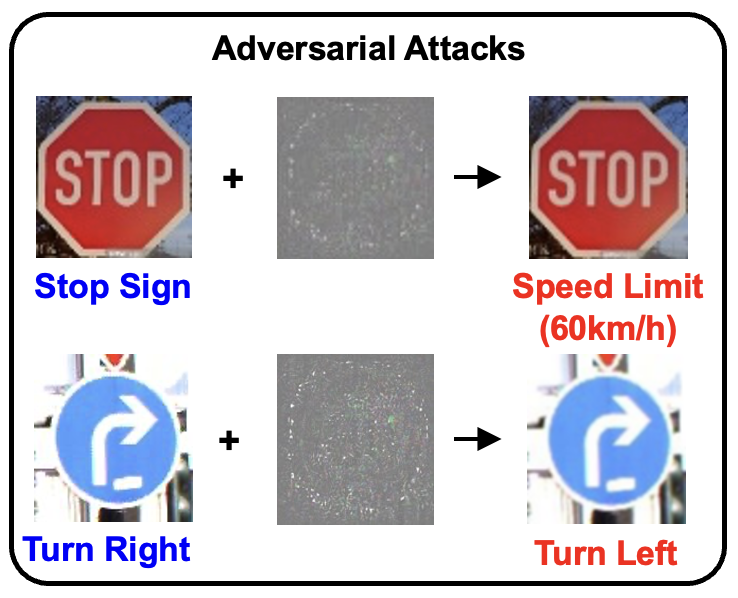} %, height=7cm
\caption{To execute an adversarial attack, an input image is initially taken (left). Subsequently, a deliberately crafted noise vector is applied as a perturbation to the input image (middle). As a consequence, the resulting image becomes an adversarial example (right), which compels the neural network to misclassify the original input image, leading to an erroneous classification outcome. }
\label{adv_attack}
\end{figure}

%adding imperceptible and carefully crafted perturbations..
Adversarial noise, also known as adversarial perturbations or adversarial examples, refers to carefully crafted modifications applied to input data with the intention of misleading or causing errors in machine learning models. Adversarial noise is designed to be imperceptible to human observers while having a significant impact on the model's behavior. As illustrated in Figure \ref{adv_attack}, the addition of a small imperceptible and carefully crafted noise to an image of a traffic sign can result in a significant change in the model's output. In this example, the original image depicts a stop sign, but with the introduction of adversarial noise, the model incorrectly recognizes it as a speed limit sign.

\subsubsection{Problem formulation}
An adversarial example $x^*$ is crafted by adding a small carefully generated perturbation $\delta$ to the clean and correctly classified input image $x$.
The problem of generating an adversarial example $x^*$ can be formulated as a constrained optimization \ref{eq:adv}, given an original input image $x$ and a target classification model $ f(.) $,:

\begin{equation}
\label{eq:adv}
    \min_{\delta} \left\|\delta\right\|_p  s.t. f(x + \delta) \neq f(x)
\end{equation}

Three metrics were proposed to approximate human’s perception of visual difference, namely $L_0$, $L_2$, and $L_\infty$ \cite{carlini2016evaluating}. These metrics are special cases of the $L_p$ norm: 

\begin{equation}
    \left\|x\right\|_p = \left( \sum^{n}_{i = 1} \left |x_i \right | ^{p} \right)^{\frac{1}{p}}
\end{equation}

These three metrics focus on different aspects of visual significance. $L_0$ counts the number of pixels with different values at corresponding positions in the two images. $L_2$ measures the Euclidean distance between the two images $x$ and $x^*$. $L_\infty$ measures the maximum difference for all pixels at corresponding positions in the two images.

Adversarial attacks can be (1) \textit{Digital Attacks}, which attack DNNs by perturbing the input data in the digital space; and (2) \textit{Physical attacks}~\cite{Kurakin2017AdversarialExamplesPhysicalWorld}\cite{Brown2017AdversarialPatch}\cite{Eykholt2018PhysicalWorldAttacks}\cite{Sharif2016FaceRecognitionAttacks}\cite{Man2020GhostImage}, which attack DNNs by modifying the visual characteristics of the real object in the physical world. In contrast to the attacks in the digital world, adversarial attacks in the physical world are more challenging due to the complex physical constraints and conditions (e.g., lighting, distance, camera, etc.), which will limit the attacking ability of generated adversarial perturbations.

\subsection{The Taxonomy of Adversarial Attacks}
According to the threat model, existing adversarial attacks can be categorized based on: \\
\subsubsection{Attacker Knowledge:}
Here, what determines the characteristics of an adversarial attack is the knowledge required by the attacker. Accordingly, attacks can be divided into three main categories: White-box, Black-box, and Grey-box attacks.
\begin{itemize}
    \item \textbf{White-box attacks} refer to adversarial attacks where the attacker possesses complete knowledge of the training and testing data used to train the victim model, as well as the architecture and parameters of the target model. Leveraging this information, the adversary employs various techniques to craft adversarial examples. These examples are carefully designed to exploit vulnerabilities in the target model and deceive it into producing incorrect outputs. By having access to detailed information about the model and its training data, white-box attackers can tailor their attacks to maximize their impact and success rate.
    \item \textbf{Black-box attacks}, refer to adversarial attacks where the attacker has limited or no access to the architecture and parameters of the victim model. In this scenario, the attacker relies solely on the output of the model, such as the predicted label or confidence score, for a given input. By querying the victim model and observing its responses, the adversary aims to reverse engineer the underlying classifier and construct a substitute model. This substitute model, though not an exact replica of the victim model, captures its behavior to a certain extent. Using the substitute model, the attacker can then generate adversarial examples to launch attacks on the victim classifier. Black-box attacks pose significant challenges in terms of limited information and increased uncertainty for the attacker, making it necessary to develop robust defenses against such sophisticated attacks.
    \item \textbf{Grey-box attacks}, also referred to as transferability attacks, involve the use of adversarial examples created to deceive one specific machine learning (ML) model in order to mislead other ML models, even when their architectures differ significantly. The attacker leverages the transferability of adversarial examples, exploiting shared vulnerabilities or similar decision boundaries across different ML models. By crafting adversarial examples that successfully deceive one model, the attacker can effectively fool other models into making incorrect predictions or classifications, even if their internal structures and parameters vary significantly. This highlights the importance of understanding the transferability of adversarial examples and developing robust defenses that can mitigate the impact of such grey-box attacks across diverse ML systems.
\end{itemize}

\begin{figure}[!htp]
\centering
\includegraphics[width=0.5\textwidth]{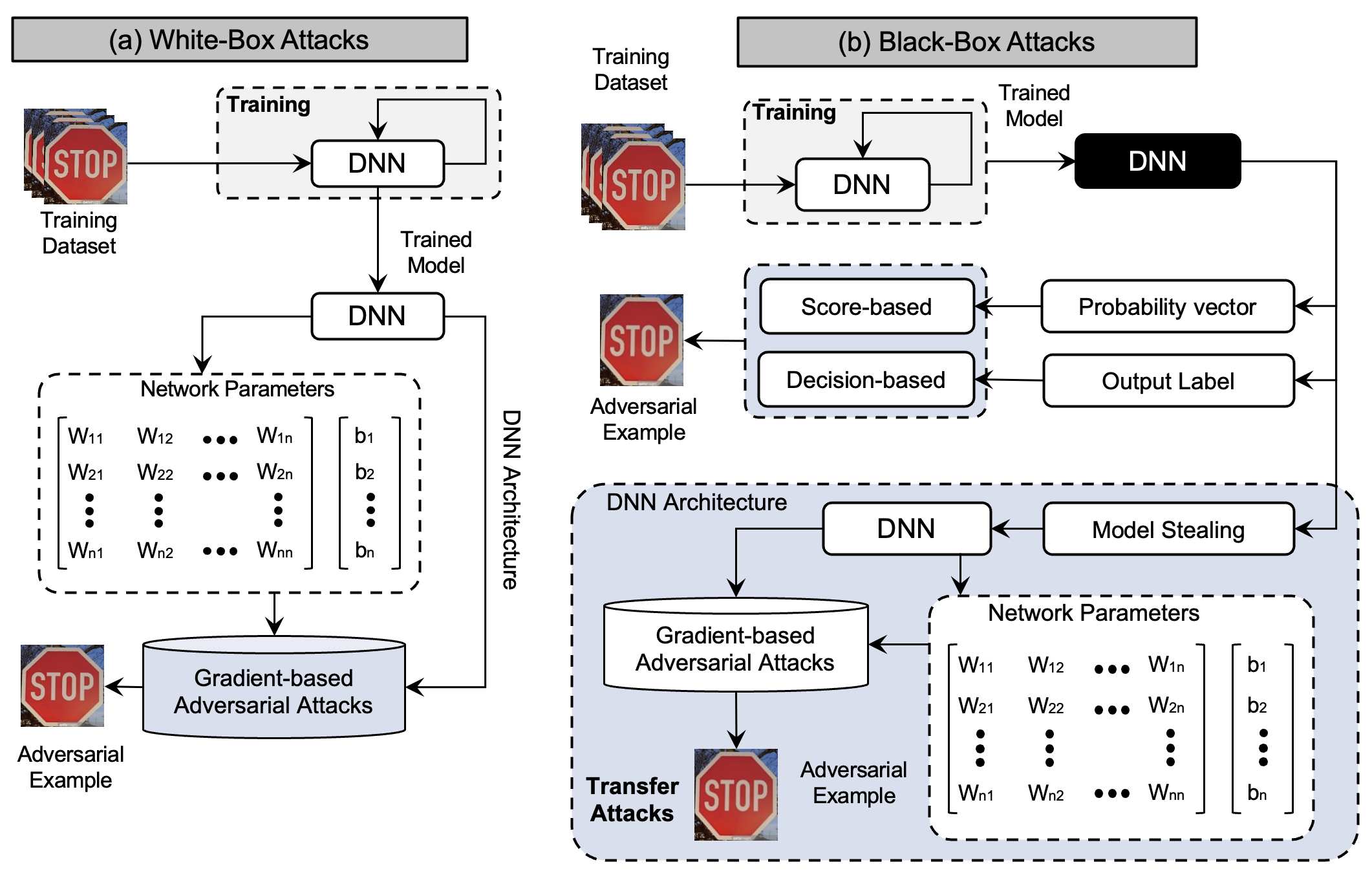} %, height=7cm
\caption{White-box vs Black-box attacks}
\label{setting}
\end{figure}

\subsubsection{Attacker Goal:} 
We can categorize attacks into two classes based on the intent and goals of the attacker. 
\begin{itemize}
    \item \textbf{\textit{Untargeted attack}}, the objective is to make slight modifications to the source image in order to cause misclassification by the target model, without a specific preference for any particular output class. The goal is simply to disrupt the correct classification of the image, see Equation \ref{untargeted}.
    \begin{equation}
       \label{untargeted}
        f(x + \delta) \neq f(x)
   \end{equation}
    \item \textbf{\textit{Targeted attack}}, the attacker aims to make slight alterations to the source image in a way that it is specifically misclassified into a predetermined target class by the target model. The objective here is to steer the classification output towards a specific desired outcome, see Equation \ref{targeted}.
    \begin{equation}
    \label{targeted}
        f(x + \delta) = t
    \end{equation} 
    Where $t$ is the target class.
\end{itemize}
%Understanding the distinction between untargeted and targeted attacks is essential in developing effective defense mechanisms and robust models that can withstand adversarial manipulation.
\subsubsection{Attack Location} 
Attacks in the context of the ML pipeline can be classified into two main categories based on their location and objective. Firstly, \textbf{\textit{poisoning or training attacks}} focus on altering the training process by manipulating the training data. The attacker's goal is to inject malicious samples into the training set, aiming to maximize the classification error of the trained model. These attacks aim to compromise the integrity and performance of the model during the learning phase by introducing biased or misleading data points. Poisoning attacks have been extensively studied and their impact on ML systems has been analyzed in various research works \cite{MuozGonzlez2017PoisoningAttacks,Shafahi2018PoisoningAttacks}.

Secondly, \textbf{\textit{backdoor attacks}} involve the creation of a hidden backdoor in the ML model. The attacker's objective is to implant a specific trigger or pattern in the input data that can activate the backdoor functionality. When the model encounters input instances containing this backdoor trigger, it predictably and consistently classifies them as a target label specified by the attacker. Backdoor attacks pose a significant threat as they can manipulate the model's behavior selectively while maintaining normal performance on other inputs \cite{Chen2017BackdoorAttacks,Gu2019BadNets}.

Additionally, \textbf{\textit{inference or evasion attacks}} target the deployed ML models. These attacks attempt to perturb input instances in such a way that they appear normal to human observers, but are deliberately misclassified by the ML models. The objective is to exploit the model's weaknesses and vulnerabilities to generate adversarial examples that bypass the model's defenses and produce incorrect predictions. Evasion attacks challenge the robustness and reliability of ML models in real-world scenarios where adversarial inputs may be encountered \cite{Biggio2013EvasionAttacks}.

%------------------------------------------------------------------
\subsection{Why do adversarial perturbations work?}
%Reasons for adversarial perturbations
%------------------------------------------------------------------
%Adversarial perturbations have the ability to deceive DNNs for recognition or classification by manipulating clean inputs \cite{zhang2021adversarial}, altering their inherent structures. 
A significant number of researchers have investigated the origins of adversarial perturbations from diverse angles. Although a consensus has yet to be reached on their causes, they are generally classified into three potential categories.
%------------------------------------------------------------------
\subsubsection{ Local over linearization of DNN structure}
\begin{figure}[!htp]
\centering
\includegraphics[width=0.45\textwidth]{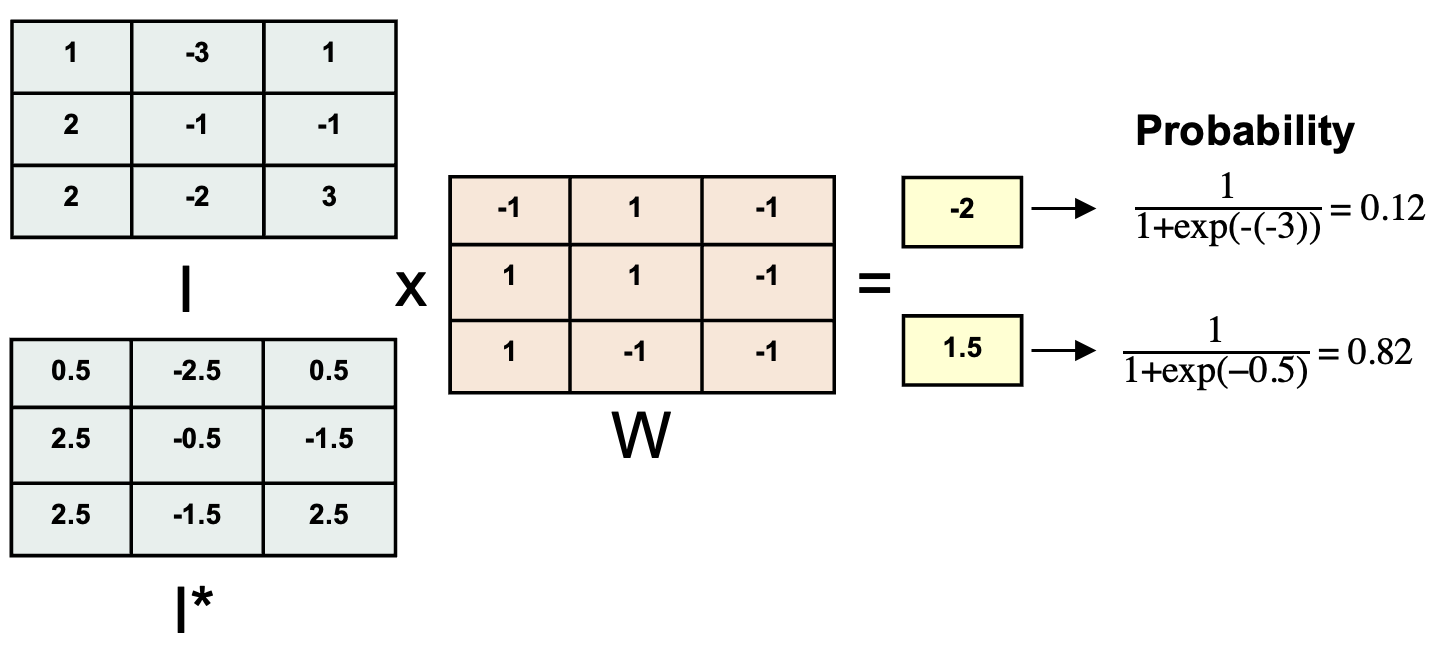} %, height=7cm
\caption{Minimal perturbations are scaled up in the high-dimensional linear classifier. Before the perturbation, the classifier classified the original image into class "A" with a probability of 12\%; by adding or subtracting 0.5 to each pixel of the original image to obtain the adversarial example, the classifier classified the adversarial example into class "A" with 82\% confidence. }
\label{non_linear}
\end{figure}

%minimal perturbations may be amplified in the process of transmission
%------------------------------------------------------------------
According to Goodfellow et al.\cite{Goodfellow2015ExplainingAdversarialExamples}, the cause of adversarial examples is attributable to linear behavior within a high-dimensional space. When utilizing a high-dimensional linear classifier, the tiny perturbations in each dimension are accumulated and amplified by dot product operations. When a nonlinear activation function, such as sigmoid, is applied to compute the classification probability, the probability of the original image I being classified into the "A" class is boosted from 12\% to 82\% after perturbation, as illustrated in Fig \ref{non_linear}. If a linear activation function is employed in the DNN, the original image is more vulnerable to a greater discrepancy in classification probability before and after perturbation.
%------------------------------------------------------------------
\subsubsection{The training data set contains insufficient target features}
\begin{figure}[!htp]
\centering
\includegraphics[width=0.4\textwidth]{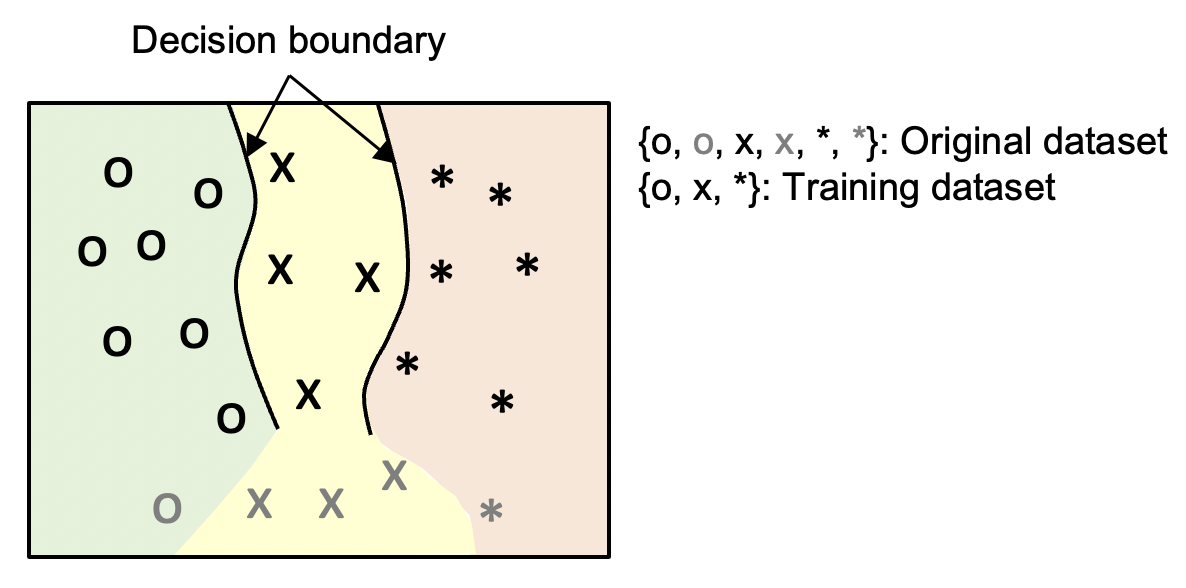} %, height=7cm
\caption{When the training dataset contains insufficient target features, it will cause the decision boundary of the target depth model to stop prematurely. }
\label{original}
\end{figure}

%which causes the decision boundary of DNN to stop prematurely
%------------------------------------------------------------------
Insufficient training data is also posited by some researchers as the root cause of DNN model vulnerability, leading to incomplete training \cite{zhang2021adversarial}. The high dimensionality of datasets, due to the wide range of target features, can result in premature cessation of training and a weak generalization of the decision boundaries of DNN models when the training data is inadequately labeled. Suppose the original target dataset contains 22 features, including `o', `x', `*', `\textcolor{gray}{x}', `\textcolor{gray}{o}', and `\textcolor{gray}{*}', whereas the randomly selected training dataset only comprises `o', `x', and `*', as depicted in Fig \ref{original}. During the training process, if the target model can distinguish the features contained in training dataset proficiently, training might end, and the decision edge will stop evolving. The target model will be unable to correctly classify some features, such as the gray features, that were not included in the training dataset. %To improve the model's robustness, the target depth model must learn more target features by incorporating additional data.
%------------------------------------------------------------------
\subsubsection{The presence of non-robust features in the classifier leads to inaccurate predictions}
\begin{figure}[!htp]
\centering
\includegraphics[width=0.4\textwidth]{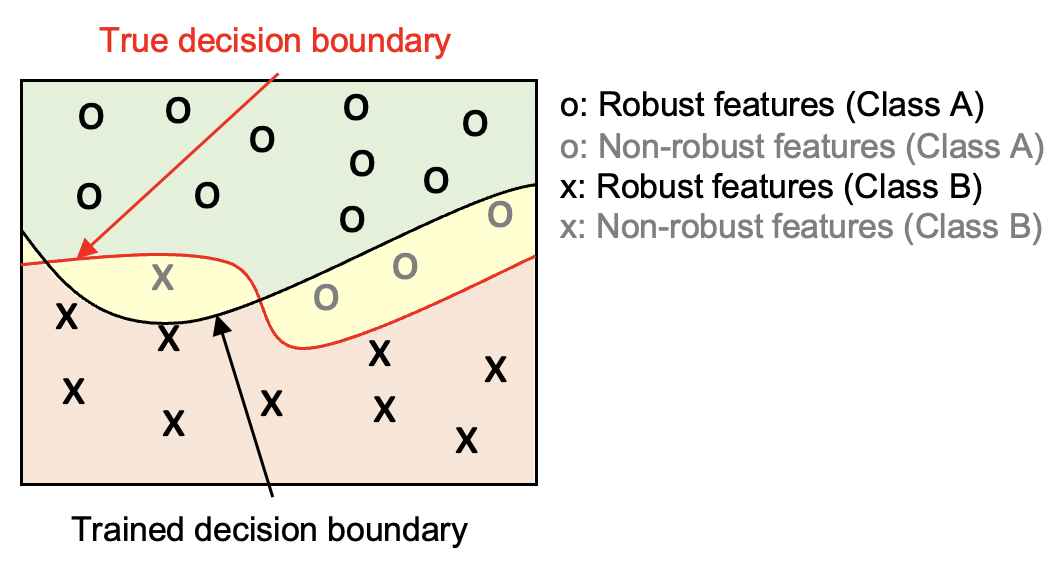} %, height=7cm
\caption{Non-robust features in the dataset are present in the classifier, resulting in inaccurate decision boundaries from the training. }
\label{non_robust}
\end{figure}

%------------------------------------------------------------------
Cubuk et al. \cite{cubuk2017intriguing} argue that the origin of adversarial examples predominantly stems from the inherent uncertainty that neural networks exhibit regarding their predictions. They argue that the uncertainty in the output results of the target model is independent of the network architecture, training method, and data set. Suppose that a classifier contains robust features (`x' and `o') and non-robust features (`\textcolor{gray}{x}' and `\textcolor{gray}{o}'), and the red curve is the true decision boundary of the dataset, which is shown in Fig \ref{non_robust}). The presence of non-robust features in the classifier causes the decision boundary of the trained classifier (black curve in Fig \ref{non_robust}) to not recognize non-robust features in the target well, and therefore, if these non-robust features are added to the input image, it will likely result in the target depth model making false predictions.

%% file: attacks.tex
\section{Physical Adversarial Attacks}
\label{forms}
\subsection{Digital adversarial attack vs Physical adversarial attack}
%Digital adversarial attacks occur in the digital domain, where attackers generate perturbations or modify input data to deceive the model. These attacks often involve carefully crafted modifications to the pixel values or features of digital images or data. The goal is to exploit vulnerabilities in the model's decision-making process and cause misclassifications or incorrect predictions. Digital attacks can be performed with access to the model's architecture, parameters, and training data, and they are typically evaluated using digital inputs and performance metrics.

\begin{figure}[!htp]
\centering
\includegraphics[width=0.5\textwidth]{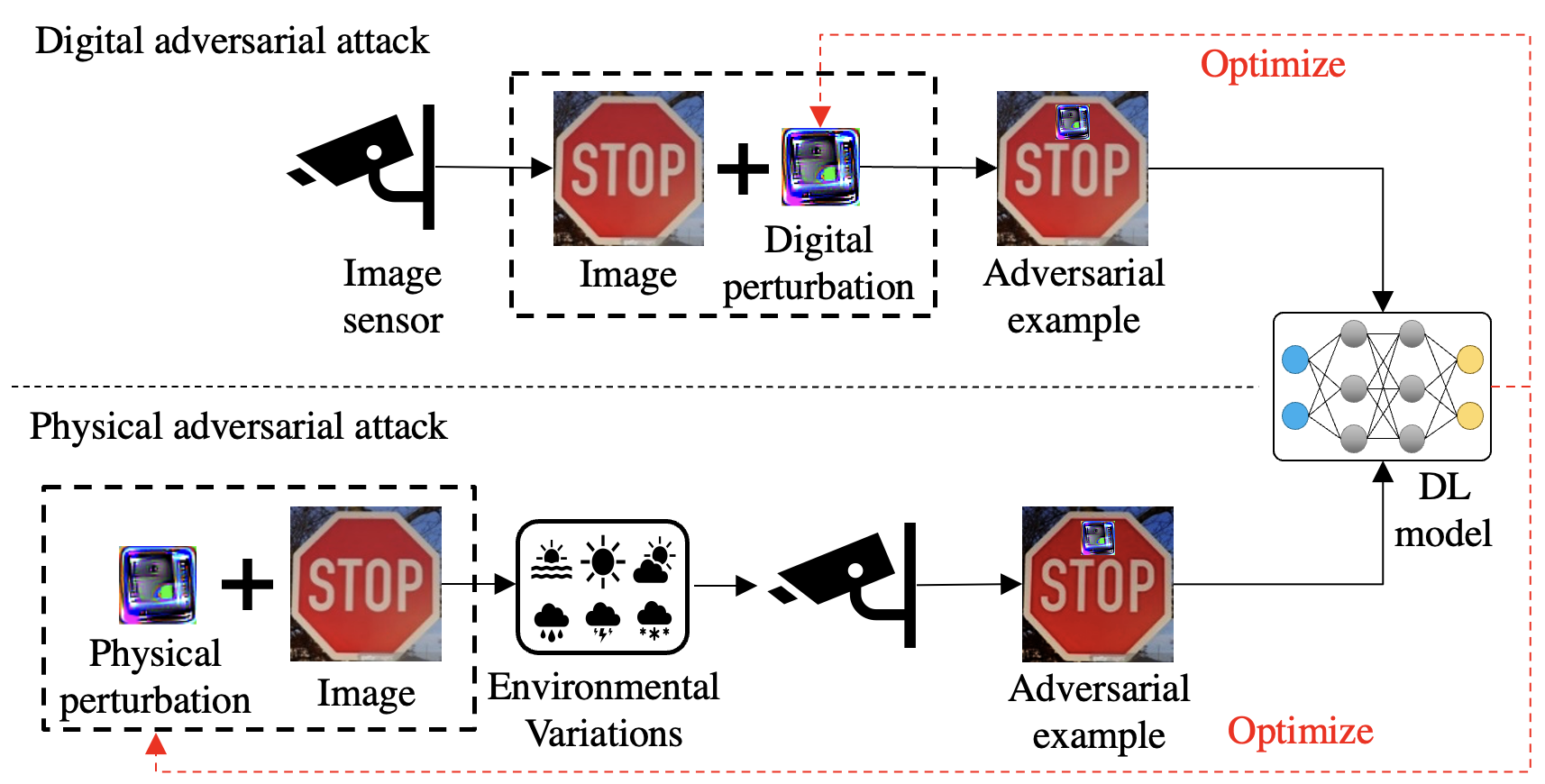} %, height=7cm
\caption{Digital vs Physical adversarial attacks}
\label{digital}
\end{figure}

%On the other hand, physical adversarial attacks target the real-world deployment of machine learning models by introducing modifications to physical objects or signals. These attacks are designed to deceive models when they interact with physical inputs in the real world. Physical adversarial attacks may involve applying stickers, printing patterns, adding noise, or manipulating objects to induce misclassifications or undesired behavior in the model. These attacks aim to exploit the vulnerabilities that arise due to the model's reliance on real-world sensory data. Physical attacks are evaluated by assessing their effectiveness in real-world scenarios, using physical objects, images, or signals.
Adversarial attacks can be categorized into two types based on the setting in which they occur: digital attacks and physical attacks. Figure \ref{digital} illustrates the difference between the two settings.
In a \textbf{digital attack} setting, an attacker has the flexibility to arbitrarily alter the input image of a victim model at the pixel level. Thus, these attacks implicitly assume that the attacker is in control of the DNN's input system (e.g., a camera or input image memory). One of the first adversarial examples was proposed in \cite{vulnerable} which was based on adding small imperceptible noise that tries to push the input image prediction towards the direction of an incorrect class. Since then, a number of digital attacks have been proposed advancing the algorithms for producing adversarial examples \cite{Carlini2017CWAttack, Goodfellow2015ExplainingAdversarialExamples,Chen2019BoundaryAttack, Chen2020HopSkipJumpAttack}.

Inference attacks can be divided into three categories: First, the \textit{Gradient-based attacks} that rely on detailed model information, including the gradient of the loss w.r.t. the input. For example, Fast Gradient Sign Method (FGSM)~\cite{Goodfellow2015ExplainingAdversarialExamples} attack, Projected Gradient Descent (PGD)~\cite{Madry2018ResistentAdversarialAttacks} attack, Carlini \& Wagnar (C\&W)~\cite{Carlini2017CWAttack} attack, DeepFool~\cite{MoosaviDezfooli2016DeepFool} attack. Second, the \textit{Decision-based attacks} that rely only on the final output of the model and minimizing the norm-based adversarial examples such as the One-Pixel-Attack~\cite{Su2019OnePixelAttack}, \cite{Brendel2018decisionbased}, \cite{Brunner2019GuessingSmart}, Caps Attacks~\cite{Marchisio2019CapsAttacks} Boundary Attack~\cite{Chen2019BoundaryAttack}, FADec~\cite{Khalid2020FaDec}, and Hop-Skip-Jump~\cite{Chen2020HopSkipJumpAttack} attack. Finally, the \textit{Score-based attacks.} rely on the predicted scores, such as class probabilities or \textit{logits} of the model. On a conceptual level, these attacks use the predictions to numerically estimate the gradient such as the Local Search Attack (LSA)~\cite{localsearch} attack.

A more robust and realistic threat model would presume that the attacker does not touch the internals of an AI system (e.g., its internal sensors, memory and/or data pipelines), rather can manipulate external environments or external objects. \textbf{Physical Attacks} involve the manipulation of inputs from the physical world, where the DNN-based model exclusively relies on images captured by a camera. In such attacks, the adversary's capability is limited to presenting adversarial images to the camera, which are then processed by the victim model. Unlike digital attacks that can leverage precise and subtle perturbations, physical attacks typically require large or even unrestricted perturbations. This is because in complex real-world environments, small perturbations may go unnoticed or be challenging to capture accurately by the camera.  %Therefore, physical attacks challenge the robustness of DNN classifiers to adversarial inputs in scenarios where the attacker has physical access and control over the visual inputs, emphasizing the need for defenses that can mitigate the impact of such attacks in real-world settings.

\subsection{Physical world challenges}
%Introducing the digital attack into the physical realm poses an additional challenge, as the perturbation must be durable enough to withstand real-world distortions arising from variations in viewing distances and angles, lighting conditions, and camera limitations. Previous studies have revealed that adversarial examples generated via conventional techniques frequently cease to be adversarial after undergoing slight transformations (Luo et al., 2016; Lu et al., 2017).
 
Physical adversarial attacks in the real world present several challenges that need to be considered. Firstly, \textit{environmental conditions} play a crucial role. Factors such as the distance and angle of the camera can affect the quality and clarity of the captured images. In scenarios where the camera is positioned far away or at an unfavorable angle, the attacker may face difficulties in generating effective adversarial perturbations. % that can bypass the model's defenses.

\textit{Spatial constraints} also pose challenges in physical attacks. In certain situations, the attacker may have limitations on manipulating the background imagery. This constraint restricts the attacker's ability to modify the entire scene and forces them to focus their perturbations on specific objects or regions of interest. Such constraints may require the attacker to carefully choose their target and plan their attacks accordingly.

%Moreover, physical limits on \textit{imperceptibility} play a significant role. Perturbations applied to the physical world must adhere to the limitations imposed by the sensing hardware. Imperceptible perturbations are desirable to ensure that the modifications go unnoticed by human observers. However, physical hardware and limitations in the imaging process may introduce constraints on the level of imperceptibility that can be achieved.
Moreover, achieving perturbation \textit{stealthiness} is essential to ensure that the perturbations go unnoticed by both human observers and defenders. The perturbations should blend seamlessly into the environment, making them appear as regular elements, such as an artistic painting or a common object, to avoid raising suspicion.

Furthermore, \textit{fabrication error} is a critical consideration in physical attacks. When generating adversarial perturbations, all perturbation values must correspond to valid colors that can be accurately reproduced in the real world. Any deviations from valid colors due to fabrication error can potentially alert humans or be detected by the system, thereby reducing the effectiveness of the attack.

\subsection{Attack Evaluation}
In the evaluation of physical adversarial attacks, four key aspects are commonly considered: effectiveness in deceiving victim model, robustness under real-world conditions, stealthiness to human observers, and transferability.
\subsubsection{Attack Effectiveness} 
    Attack effectiveness is a critical aspect to consider when assessing the impact of physical attacks. These attacks have demonstrated their effectiveness in significantly degrading the performance of the targeted task, thereby compromising the reliability and accuracy of the victim system. Adversarial manipulations in the physical space can cause misclassifications, incorrect predictions, or erroneous decisions, leading to potentially severe consequences.
    %The attack success rate (ASR) is a measure used to quantify the effectiveness of an adversarial attack. It represents the percentage of successfully misclassified or manipulated examples out of the total number of examples in the evaluation dataset.

    %Furthermore, physical attacks are often designed to be relatively easy to operate, requiring minimal expertise or specialized knowledge. This accessibility increases the potential threat, as individuals with limited technical skills can carry out such attacks. By leveraging the vulnerabilities and limitations of the physical world, attackers can manipulate inputs in a way that the system fails to detect or mitigate the adversarial perturbations effectively.
    
    %Considering the effectiveness and ease of operation of physical attacks is crucial for developing robust countermeasures. Detecting and mitigating these attacks in real-world settings requires comprehensive defense strategies that account for the unique challenges posed by the physical environment. Addressing the effectiveness of physical attacks and developing countermeasures that can effectively neutralize their impact is essential to ensuring the reliability and security of systems operating in the physical space.
\subsubsection{Attack Robustness} 
    Attack robustness is a key factor in evaluating the resilience of physical attacks in dynamic environments. Maintaining attack ability despite variations in the environment is crucial for the sustained effectiveness of adversarial manipulations. One aspect of robustness is being able to maintain attack efficacy across different scenes. This means that the perturbations should remain effective and capable of deceiving the system, regardless of changes in lighting conditions, backgrounds, or other scene-specific factors.
    
    %Additionally, physical attacks need to be robust to physical constraints. These constraints may include limitations on the size or location of the adversarial perturbations due to the physical properties of the objects or the sensing hardware. Robustness to physical constraints ensures that the attacks can still achieve their intended impact even under realistic limitations and practical conditions.
    %Achieving robustness in physical attacks requires thorough analysis, experimentation, and validation in diverse environments. Understanding and addressing potential variations and challenges that may arise in dynamic settings is crucial for developing effective countermeasures. By enhancing the robustness of physical attacks, we can better anticipate and mitigate their potential impact, thereby safeguarding the integrity and reliability of systems operating in real-world scenarios.
\subsubsection{Attack Stealthiness} 
\begin{figure}[!htp]
\centering
\includegraphics[width=0.5\textwidth]{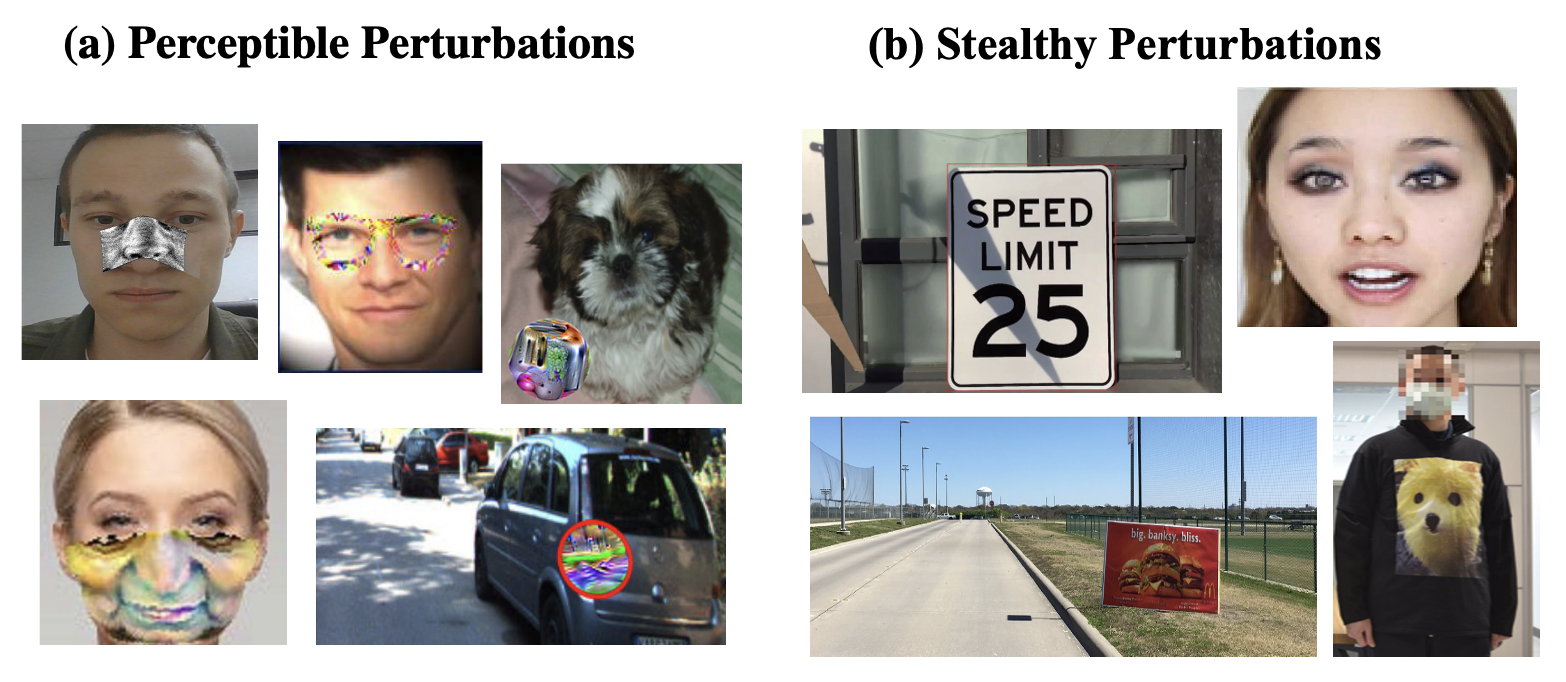} %, height=7cm
\caption{Visualization discrepancy between (a) Stealthy perturbations
    and (b) perceptible perturbations. }
\label{fig:stealthy}
\end{figure}

    Attack stealthiness is a critical characteristic that determines the effectiveness of physical attacks. To be successful, these attacks should ideally go unnoticed by both the observer and the victim, remaining imperceptible to human eyes (See Figure \ref{fig:stealthy}). The ability to maintain stealthiness ensures that the adversary can carry out their attack without raising suspicion or triggering any defensive mechanisms.

\subsubsection{Attack Transferability} 
    Attack transferability refers to the extent to which an adversarial example, initially crafted to cause misclassification on one specific machine learning model, can also induce misclassifications when presented to other, potentially different, models. It demonstrates the generalization capacity of adversarial perturbations across multiple models.
    %If the perturbations introduced by the attacker are easily detectable or noticeable to human observers, there is a higher risk of the attack being discovered either by the victim or the defender before it can have the desired impact. The goal is to make the adversarial modifications blend seamlessly into the surrounding environment, making it difficult for humans to distinguish between legitimate inputs and adversarial examples.
    %Achieving stealthiness requires careful consideration of the physical properties and limitations of the targeted system, as well as an understanding of human perception. By crafting adversarial perturbations that are imperceptible or difficult to discern, attackers can increase the likelihood of a successful attack while minimizing the chances of detection or suspicion.
    
    %Ensuring the stealthiness of physical attacks is essential for their effectiveness and long-term viability. By remaining undetected, these attacks can undermine the trust and reliability of systems, posing significant challenges for defending against such malicious activities. Developing countermeasures that account for stealthy attack techniques and enhancing the robustness of systems to detect and mitigate them are crucial steps in safeguarding against stealthy physical attacks.

\subsection{Evaluation Metrics}

\subsubsection{Attack Success Rate (ASR)} 
    The attack success rate is a metric used to evaluate the effectiveness of an attack in achieving its intended outcome, which is to alter the prediction of a given model and steer it towards a specific target class. This measure quantifies the rate at which the attack successfully manipulates the model's prediction to match the desired target class. By assessing the attack success rate, we can gauge the reliability and potency of the attack technique. A high success rate indicates that the attack consistently achieves its objective, significantly impacting the model's predictions. Conversely, a low success rate suggests that the attack may be less effective or inconsistent in altering the model's output to the desired target class. 
        The attack success rate (ASR) is calculated as:
    \begin{equation}
       ASR  = \frac{Number ~of~ successful ~attacks}{Total~ number ~of~ adversarial ~examples} \times 100
    \end{equation}
    %A high attack success rate indicates that the adversarial attack is effective in deceiving the model, while a low success rate implies that the attack may not be as potent.
\subsubsection{Transfer rate} 
    The transfer rate is a metric used to evaluate the transferability of adversarial examples across different models. It measures the extent to which an adversarial example, originally designed to misclassify a specific model, can also cause misclassifications when presented to other models. By assessing the transfer rate, we gain insights into the generalization of adversarial examples across different models and their susceptibility to similar attacks. A high transfer rate indicates that the adversarial example successfully deceives not only the targeted model but also other models, highlighting the potential widespread impact of the attack. Conversely, a low transfer rate suggests that the adversarial example may be more specific to the targeted model and less effective against other models. 
\subsubsection{Targeted success rate} 
    The targeted success rate is a metric used to evaluate the effectiveness of a targeted attack pattern in achieving its specific objective. It measures the extent to which the attack pattern successfully changes the prediction of the targeted model to the desired target class while leaving other models unaffected. By assessing the targeted success rate, we can quantify the precision and accuracy of the attack pattern in specifically manipulating the targeted model's output without causing undesired effects on other models. A high targeted success rate indicates that the attack pattern consistently achieves its goal of steering the targeted model towards the desired target class, while minimizing any unintended impact on other models. Conversely, a low targeted success rate suggests that the attack pattern may be less effective or more prone to causing disruptions in non-targeted models. %Measuring the targeted success rate helps in evaluating the reliability and specificity of targeted attacks and allows for the development of robust defense mechanisms. Understanding the factors that contribute to a higher or lower targeted success rate enables researchers and practitioners to enhance the security of targeted models and develop countermeasures that can mitigate the impact of such attacks.

\subsection{Adversarial Perturbation Forms} %Forms

\begin{figure*}[!htp]
\centering
\includegraphics[width=0.75\textwidth]{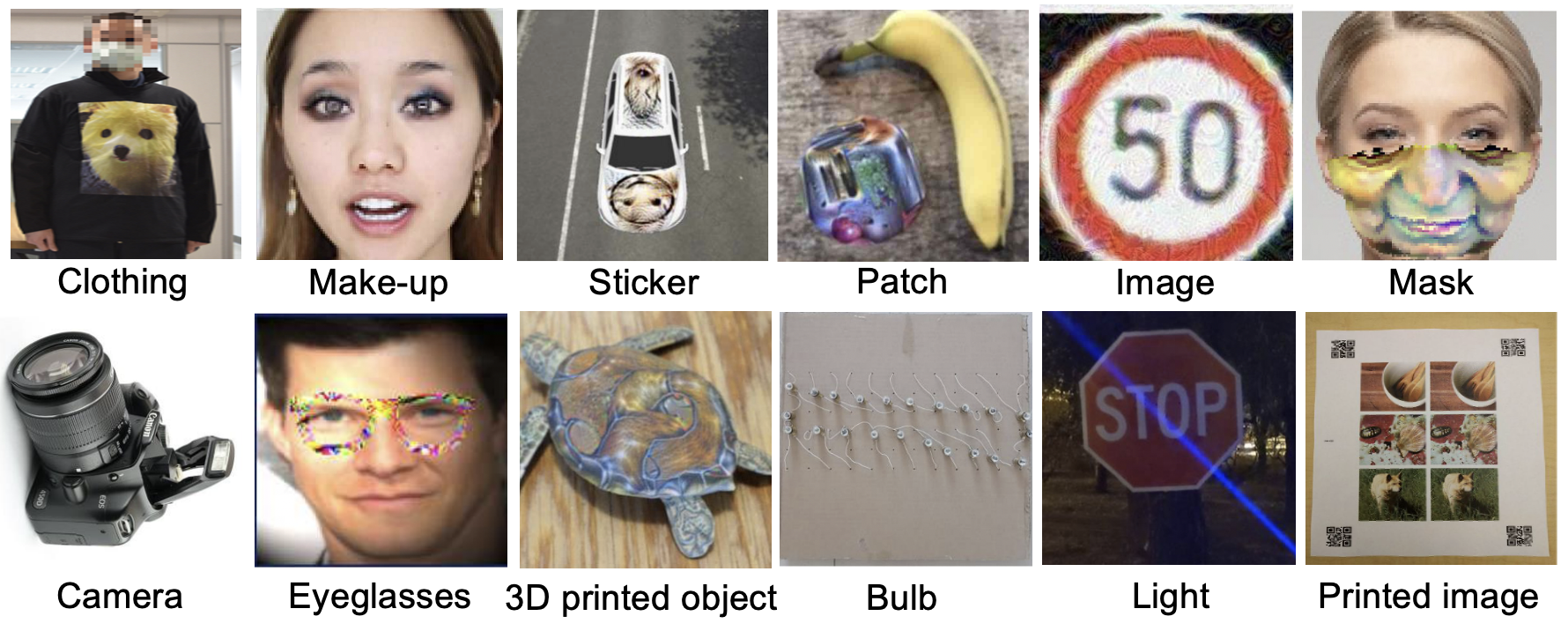} %, height=7cm
\caption{Illustration of different forms of physical adversarial attacks. }
\label{fig:forms}
\end{figure*}

Various physical adversarial perturbation forms have been explored in the realm of adversarial attacks. These forms encompass different techniques and approaches for manipulating physical inputs to deceive machine learning systems. Some notable examples of physical adversarial perturbation forms are presented in figure \ref{fig:forms}
\subsubsection{Patch-based Attacks} % Printed Adversarial Perturbations
These perturbations are carefully designed and printed on physical surfaces, such as images or objects, to exploit the vulnerabilities of computer vision systems. They aim to introduce imperceptible alterations that can lead to misclassifications or incorrect interpretations by the targeted models. This method involves substituting a section of the targeted image with an image patch to impede the performance of DNN-based models. Stickers and patches are comparable in effectiveness, but patches are typically uniform in shape, whereas stickers adapt to non-rigid transformations by being irregular in shape. The adversarial patch is prevalent because it is simple to use and can usually be printed out with ease.

\subsubsection{Sticker-based Attacks} Sticker-based perturbations involve strategically placing physical stickers or decals on objects or scenes. These stickers can contain patterns or symbols that can deceive object detection or recognition systems, causing misclassifications or triggering false interpretations. This method relies on attaching the sticker to the surface of the targeted object, such as cars or individuals, to execute the attack. As it can be affixed over a significant area, its effectiveness and resilience are noteworthy. However, concealing the sticker poses a challenge. Enhancing stealthiness can be achieved by creating a sticker pattern that closely resembles natural images.

\subsubsection{Camouflage Techniques} Camouflage-based adversarial perturbations involve modifying the appearance or texture of an object or scene to blend it with the surroundings, making it difficult for computer vision systems to detect or recognize the object accurately.
\begin{itemize}
    \item \textbf{\textit{Clothing:}} The technique involves incorporating physical perturbations into clothing items such as t-shirts, cloaks, or pants, referred to as adversarial clothing. The wearable nature of this approach greatly enhances its stealthiness. It is widely utilized to deceive person detectors in order to evade surveillance systems.
    \item \textbf{\textit{Makeup:}} This method utilizes special makeup to achieve more stealthy adversarial attacks on face recognition algorithms in the real world. The makeup is designed to blend in with the human face, making this adversarial medium highly stealthy.
    \item \textbf{\textit{Eyeglasses:}} The attacker wears glasses with specific patterns designed to deceive face recognition algorithms, enabling them to avoid recognition or pretend to be someone else. Adversarial glasses are quite effective and resilient, but they lack stealthiness because the unusual and conspicuous patterns make them noticeable.
    \item \textbf{\textit{Masks:}} Face mask-based physical adversarial attacks involve the manipulation of face masks or facial coverings to deceive or manipulate face recognition systems. These attacks leverage the vulnerabilities of face recognition algorithms by introducing specially crafted masks that alter the appearance of the wearer in a strategic manner. In face mask-based physical adversarial attacks, the attacker designs and deploys masks that are intended to either impersonate a different individual or evade recognition altogether. These masks may include various elements such as printed patterns, textures, or shapes that are carefully crafted to exploit the weaknesses of face recognition algorithms. By strategically placing these masks on individuals, attackers can deceive the system into misclassifying identities or failing to recognize individuals altogether.
\end{itemize}

\subsubsection{Light Manipulation} Light-based perturbations exploit the sensitivity of vision systems to changes in lighting conditions. By carefully manipulating the lighting environment, such as adjusting the intensity, direction, or color of light, these perturbations can introduce perceptually subtle changes that affect the model's predictions.
\begin{itemize}
    \item \textbf{\textit{Light:}} This category of technique does not directly alter the object but rather employs a lighting device, such as a laser pointer or projector, to project specific light onto the targeted object. It can rapidly launch potent physical-world attacks and possesses superior stealthiness. However, the attack's potency can decline in an environment with strong lighting as the light carrying the perturbations is not perceptible.
    \item \textbf{\textit{Bulb:}} The attacker can generate special patterns for performing attacks by imaging small glowing light bulbs as a light spot shaped like a Gaussian distribution under the thermal infrared camera. The bulbs can work steadily and continuously, but the method is poorly stealthy.
\end{itemize}

\subsubsection{Physical Adversarial Prints} These perturbations involve printing adversarial patterns or images that are specifically designed to mislead object recognition systems or image classifiers. These prints can be displayed on physical objects or scenes, leading to misclassifications or incorrect interpretations.
\begin{itemize}
    \item \textbf{\textit{Physical Adversarial Objects:}} Adversarial objects are physical three-dimensional objects designed with specific patterns or structures that can manipulate the perception of computer vision systems. %These objects can cause misclassifications, confuse object detectors, or deceive depth estimation algorithms.
    \item \textbf{\textit{Printed Image:}} To misidentify the image, physical perturbations are distributed across the entire image. The primary challenge in this method is to conceal the perturbation effectively while ensuring the attack's effectiveness.
    \item \textbf{\textit{Image:}} Physical perturbations are dispersed throughout the entire image to cause DNN-based models to misclassify the image. The main difficulty in this approach is to successfully conceal the perturbation while guaranteeing the attack's efficacy.
\end{itemize}

\subsubsection{Imaging Device Manipulation} 
\begin{itemize}
    \item \textbf{\textit{Camera:}} This technique executes attacks by altering the camera rather than the targeted object. Two types of this method have been devised to date: the rolling shutter effect of the camera and the camera's image signal processing (ISP). As the modification is carried out on the camera itself, the camera-based approach is stealthy.
    \item \textbf{\textit{Acoustics:}} %This technique refers to an adversarial signal injected into the audio domain to manipulate the behavior of cameras and computer vision algorithms. The attack leverages the fact that cameras often rely on both audio and visual inputs to make decisions, and by manipulating the audio input, the authors demonstrate that they can influence the visual perception of the system.
    This technique refers to an adversarial attack that emits specially crafted acoustic signals that cause subtle perturbations in camera sensor readings, which in turn can cause misclassification or misleading interpretations by DNN-based algorithms.
\end{itemize}

\begin{table}[ht]
  \centering
%%\small
  \caption{Example of transformation distribution used in \cite{saam}.}
  \label{transformations}
  \begin{tabular}{lcc}
    \hline
    \textbf{Transformations} &  \textbf{Parameters} & \textbf{Remark} \\
    \hline
    Rotation    & $\pm20^\circ$ & Camera Simulation   \\
    Cropping    & −0.7 $\sim$ 1.0 & Photograph/Occlude Simulation   \\
    Affine & $0.7$ & Perspective/Deformed Transforms  \\
    Scale    & $[0.25, 1.25]$ & Distance/Resize \\
    Random Noise & $\pm0.1$ & Noise   \\
    Brightness  & $\pm0.1$ & Illumination   \\
    Contrast    & $[0.8, 1.2]$ & Camera Parameters   \\
  \hline
\end{tabular}
\end{table}

\begin{table*}[!htp]
\centering
  \caption{Main physical adversarial attack methods in Computer Vision: Form, Application and Venue.}
  \label{all_attacks}
  \begin{tabular}{lccc}
    \toprule
       \textbf{Method}  & \textbf{Form}  & \textbf{Application}  & \textbf{Venue}  \\
    \midrule 
            GoogleAp \cite{googleap} & Patch & Classification & NIPS 2017\\
            PAE \cite{PAE} & Printed images   & Classification &  AISS 2018   \\
            EOT \cite{EOT} & 3D-printed object & Classification & PMLR 2018\\
            LaVAN \cite{lavan} & Patch & Classification & ICML 2018\\
            LightAttack \cite{LightAttack}  & Light &  Classification & AAAI-S 2018 \\
            D2P \cite{D2P} & Printed images & Classification & AAAI 2019\\
            ACS \cite{ACS} & Sticker & Classification & PMLR 2019\\
            ProjectorAttack \cite{projectorattack} &   Light  & Classification & S\&P 2019  \\
            Adversarial ACO  \cite{ACO} & Patch & Classification & ECCV 2020\\
            Adv-watermark  \cite{advwatermark} & Patch & Classification & ACM MM 2020\\
            ABBA  \cite{abba} & Printed image &  Classification &  NeurIPS 2020   \\
            ViewFool \cite{viewfool} &  Position    & Classification &  NeurIPS 2020  \\
            SLMAttack  \cite{slmattack}     & Light &  Classification  & ArXiv 2021     \\
            Meta-Attack \cite{meta-attack} & Image & Classification &ICCV 2021 \\
            %DAS  & \cite{DAS} & Sticker & Classification & CVPR & 2021\\
            Invisible perturbations \cite{Invisible-Perturbations} & Camera & Classification & CVPR 2021\\
            Adversarial ISP  \cite{ISP} & Camera & Classification & CVPR 2021\\
            AdvLB \cite{AdvLB} & Light & Classification & CVPR 2021 \\
            Adversarial ACO2 \cite{ACO2} & Patch & Classification & IEEE TIP 2022\\ 
            TnT attack\cite{tntattack} & Patch & Classification & TIFS 2022 \\
            Copy/PasteAttack  \cite{pasteattack} & Patch &  Classification &  NeurIPS 2022 \\
            AdvCF \cite{advcf} & Sticker & Classification & Arxiv 2022 \\
            SPAA \cite{spaa}   & Light &  Classification  &  VR 2022 \\ 
            FakeWeather \cite{fakeWeather} & Sticker & Classification & IJCNN 2022\\
            AdvRain  \cite{advrain} & Sticker & Classification & Arxiv 2023\\
             %%%           
            $RP_2$ \cite{RP2} & Patch & Traffic Sign Recognition & CVPR 2018\\
            DARTS \cite{darts} & Image & Traffic Sign Recognition  & Arxiv 2018\\
            RogueSigns \cite{roguesigns}& Printed images & Traffic Sign Recognition & Arxiv 2018 \\ %%%
            PS-GAN \cite{PSGAN} & Patch & Traffic Sign Recognition  &AAAI 2019\\
            AdvCam \cite{advcam} & Image & Traffic Sign Recognition  & CVPR 2020 \\
            OPAD \cite{OPAD} & Light & Traffic Sign Recognition & ICCV 2021\\
            Adversarial Shadow \cite{shadow} & Light &  Traffic Sign Recognition & CVPR 2022\\
            %%%
            PhysGAN \cite{physgan} & Image & Steering Model& CVPR 2020\\
            %%%%%%%%%%%%%%%%%%%%%%%%%%%%%%%%%%%%%%%%%%%%%%%%%%%%%%%%%%%%%%%%%%%%%%%%%%%%%%%%
            TPatch \cite{tpatch} & Acoustics & Detection \& classification & Arxiv 2023\\  
            
            DPATCH \cite{dpatch} & Patch & Object detection & AAAI 2019\\
            Dpatch2 \cite{dpatch2} & Patch & Object detection & ArXiv 2019  \\
            Object Hider \cite{objecthider} &  Patch  & Object detection   &   ArXiv 2020\\
            LPAttack \cite{lpattack} & Patch &  Object detection & AAAI 2020 \\
            SwitchPatch \cite{switchpatch}& Patch & Object detection & ArXiv 2022\\
            Extended RP2 \cite{extendedrp2} & Patch & Sign detection & USENIX 2018\\
            ShapeShifter \cite{shapeshifter} & Image & Sign detection & ECML PKDD 2018\\
            NestedAE \cite{nestedae} & Patch & Sign detection    &  CCS 2019    \\
            Translucent Patch \cite{translucent-patch} & Sticker & Sign detection & CVPR 2021\\
            SLAP \cite{SLAP} & Light & Sign detection & USENIX 2021\\  
            Adversarial Rain \cite{adversarialrain} & Sticker & Sign detection & Arxiv 2022\\ 
            AdvRD \cite{advrd} & Sticker & Sign detection & Arxiv 2023 \\

            Invisible Cloak \cite{invisiblecloak} & Clothing & Person detection & UEMCON 2018\\  
            Adversarial YOLO  \cite{Adversarialyolo} & Patch & Person detection & CVPR 2019\\
            UPC \cite{UPC} & Clothing & Person detection & CVPR 2020\\
            Adversarial T-shirt \cite{adversarialtshirt} & Clothing & Person detection & ECCV 2020 \\ 
            Invisible Cloak2 \cite{invisiblecloak2} & Clothing & Person detection & ECCV 2020\\
            NAP \cite{NAP} & Clothing & Person detection & ICCV 2021\\
            LAP \cite{LAP} & Clothing & Person detection & ACM MM 2021\\  
            %TC-EGA \cite{TC-EGA} & Clothing  & Person detection & CVPR 2022 \\    
            AdvTexture \cite{advtexture} & Clothing &  Person detection & CVPR 2022   \\
            AdvART \cite{advart} & Patch & Person detection & Arxiv 2023 \\  
            Patch of Invisibility \cite{PatchOI}  & Patch & Person detection & Arxiv 2023 \\
            DAP \cite{dap} & Clothing & Person detection & Arxiv 2023 \\              
            Adversarial Bulbs \cite{bulb} & Bulb & Infrared Person detection & AAAI 2021\\
            QRAttack \cite{qrattack} & Clothing & Infrared Person detection & CVPR 2022\\    
            HOTCOLD \cite{hotcold} & Clothing & Infrared Person detection & Arxiv 2022\\
            AIP \cite{AIP}  & Clothing & Infrared Person detection & Arxiv 2023 \\
            AdvIB \cite{AdvIB}  &  Clothing  &   Infrared Person detection      &  Arxiv 2023 \\
            %MeshAdv  & \cite{meshadv}  &    &         &  CVPR  & 2019 \\
            %FIR/ERG  & \cite{FIR} & Patch & Sign detection & CCS & 2019 \\
            CAMOU \cite{camou} & Sticker & Vehicle Detection & ICLR 2019\\
            ER Attack \cite{ERattack} & Sticker & Vehicle Detection & Arxiv 2020\\
            ScreenAttack  \cite{dynamicpatch} & Patch & Vehicle Detection & Arxiv 2020\\
            PG \cite{PG} & Acoustics & Vehicle Detection & S\&P 2021\\

            %%%%%%%%%%%%%%%%%%%%%%%%%%%%%%%%%%%%%%%%%%%%%%%%%%%%%%%%%%%%%%%%%%%%%%

  \bottomrule
\end{tabular} %}
\end{table*}

\begin{table*}[!htp]
\centering
  %\caption{Physical Adversarial Attacks.}
  %\label{ssim}
  \begin{tabular}{lccc}
    \toprule
       \textbf{Method}   & \textbf{Form}  & \textbf{Application}  & \textbf{Venue}   \\
    \midrule 
            %%%%%%%%%%%%%%%%%%%%%%%%%%%%%%%%%%%%%%%%%%%%%%%%%%%%%%%%%%%%%%%%%%%%%%
            DAS \cite{DAS} & Sticker & Vehicle Detection  & CVPR 2021\\
            DTA \cite{DTA} & Sticker & Vehicle Detection & CVPR 2022\\
            FCA \cite{FCA} &Sticker &Vehicle Detection &AAAI 2022\\
            CAC \cite{CAC} & Sticker & Vehicle Detection & IJCAI 2022\\  
            AdvEyeglass \cite{adveyeglass} & Eyeglasses & Face recognition & ACM SIGSAC 2016\\
            AdvEyeglass2 \cite{adveyeglass2} & Eyeglasses & Face recognition & TOPS 2019\\
            %AGN  & \cite{AGN} & Eyeglasses & Face recognition & S\&P & 2019\\  
            CLBAAttack \cite{clbaattack}& Eyeglasses &   Face recognition & BIOSIG 2021 \\
            AdvArcFace  \cite{advarcface} & Patch & Face recognition & SIBIRCON 2019 \\  
            TAP \cite{tap} & Sticker  & Face recognition & CVPR 2021 \\
            ALPA \cite{ALPA} & Light & Face recognition & CVPR 2020\\
            AdvHat \cite{advhat} & Sticker& Face recognition & ICPR 2021\\  
            AdversarialMask \cite{AdversarialMask} & Mask & Face recognition & Arxiv 2022\\
            ReplayAttack \cite{replayattack} & Image & Face recognition & CVIU 2020 \\
            AdvSticker \cite{advsticker} & Sticker & Face recognition  & TPAMI 2022 \\
            Adv-Makeup \cite{advmakeup} & Makeup & Face recognition & IJCAI 2021 \\  
            SSOP \cite{PPattack} & Sticker & Face recognition & Arxiv 2022 \\   
            Face Adv \cite{faceadv} & Sticker & Face recognition & Arxiv 2021\\
            AdvPattern \cite{advpattern} & Clothing & Person re-identification & ICCV 2019\\ 
            %%%%%%%%%%%%%%%%%%%%%%%%%%%%%%%%%%%%%%%%%%%%%%%%%%%
            Adversarial patch \cite{adversarial-patch} & Patch & Depth Estimation& Access 2020\\   
            OAP \cite{OAP} & Patch & Depth Estimation & ECCV 2022 \\
            APARATE \cite{aparate} & Patch & Depth Estimation & Arxiv 2023 \\ 
            SAAM \cite{saam} & Patch & Depth Estimation & Arxiv 2023\\ 
            Projection-based attack \cite{projection} & Light & Depth Estimation & IEICE 2023\\    
            %%%%%%%
            FlowAttack \cite{flowattack} &Patch & Optical flow estimation &ICCV 2019\\
            %%%%%%%%%%%%%%%%%%%%%%%%%%%%%%%%%%%%%%%%%%%%%%%%%%%%%%%
            IPatch \cite{ipatch} & Patch & Semantic segmentation & Arxiv 2021\\ 
            SS Attack \cite{ssattack} & Patch & Semantic segmentation & WACV 2022\\
            %%%%%%%%%%%%%%%%%%%%%%%%%%%%%%%%%%%%%%%%%%%%%%%%%%%%
            PAP \cite{pap} & Patch & Crowd counting & ACM CCS 2022\\
            AB \cite{AB} & Image & Autonomous driving & ITSS 2022 \\ 
            Adversarial Markings \cite{advmarkings} & Sticker & Lane Detection & USENIX 2021\\ 
            %  & \cite{} & & & & \\   
            %  & \cite{} & & & & \\
            %  & \cite{} & & & & \\ 
  \bottomrule
\end{tabular} %}
\end{table*}
\subsection{Techniques For Enhancing The Robustness Of Physical Attacks}
%=========================================================================

\subsubsection{Expectation Over Transformation (EOT)}
%Transformation Augmentation: Input transformation augmentation techniques are used to simulate and account for real-world distortions that may be encountered during physical attacks. These transformations can include changes in lighting conditions, perspective distortions, rotation, scaling, or noise. By incorporating these transformations during the optimization process, the attacks become more robust and can withstand variations in the physical environment.
EOT is a general framework for improving the adversarial robustness of physical attack on a given transformation distribution $T$ \cite{EOT}. Essentially, EOT is a data augmentation technique for adversarial attacks, which takes potential transformation in the real world into account during the optimization, resulting in better robustness. 
EOT is to add random distortions in the optimization to make the perturbation more robust. An example of transformation distribution is presented in Table \ref{transformations}.
\begin{itemize}
    \item \textit{\textbf{Spatial transformations:}} The patch was randomly scaled to a size that is nearly proportional to its actual size in the scene as part of the geometric changes. Rotate the patch at random about the center of the bounding boxes. The aforementioned replicates printing size and location inaccuracies. Perspective change and cropping can also be used to simulate deformed transforms and occlusion.
    \item \textit{\textbf{Appearance Transformations:}} The pixel intensity values are altered by adding random noise, conducting random contrast adjustment of the value, and doing random brightness adjustment to create the color space conversions. 
\end{itemize}

\subsubsection{Spatial Transformer Layer (STL)} 
%Many kinds of Projects imitate the form changes for rectangula adversarial patches after placing it in physical world.
The Spatial Transformer Layer (STL) is a component used in deep neural networks to perform spatial transformations on input data \cite{advhat}. It provides the network with the ability to dynamically and adaptively learn spatial transformations, such as rotation, scaling, and translation, that can be applied to the input data before further processing. In the context of adversarial attacks, the STL can be utilized to project or apply spatial transformations to the input image or perturbation. By leveraging the capabilities of the STL, adversarial attacks can manipulate the appearance or spatial properties of the input in a controlled manner. This can include modifying the position, orientation, or scale of the adversarial perturbation, allowing it to deceive the model and potentially cause misclassification or incorrect predictions.
%-----------------------------------------------
\subsubsection{Total variation norm (TV loss)}
%-----------------------------------------------
%Natural images have representative characteristics of smooth and consistent patches, where color is only gradually within patches \cite{mahendran2015understanding}. Hence, finding smooth and consistent perturbations enhances the plausibility of physical attacks. In addition, due to sampling noise, extreme differences between adjacent pixels in the perturbation are unlikely to be accurately captured by cameras. Consequently, perturbations that are non-smooth may not be physically realizable \cite{Sharif2016FaceRecognitionAttacks}. To address the above issues, total variation (TV) \cite{mahendran2015understanding} loss is presented to maintain the smoothness of perturbation. 

%Numerous experiments have shown that without adding any constraints and focusing only on enhancing the attack effectiveness, the generated adversarial perturbations are visually abrupt and noisy. However, natural images are usually smooth and consistent.
The characteristics of natural images include smooth and consistent patches with gradual color changes within each patch \cite{mahendran2015understanding}. Therefore, to increase the plausibility of physical attacks, smooth and consistent perturbations are preferred. Additionally, extreme differences between adjacent pixels in the perturbation may not be accurately captured by cameras due to sampling noise. This means that non-smooth perturbations may not be physically realizable \cite{Sharif2016FaceRecognitionAttacks}. To address these issues, the total variation (TV) \cite{mahendran2015understanding} loss is introduced to maintain the smoothness of the perturbation.
For a perturbation P, TV loss is defined as

%$L_{tv}$ is the total variation loss on the generated image to encourage smoothness.
%It is defined as:
\begin{equation}
    L_{tv} = \sum_{i,j} \sqrt{(P_{i+1,j} - P_{i,j})^2 + (P_{i,j+1} - P_{i,j})^2}
\end{equation}

Where the subindices $i$ and $j$ refer to the pixel coordinate of the patch $P$.
\subsubsection{Creases Transformation (CT)} %Dynamic Adversarial Patch (DAP)
\begin{figure}[!htp]
\centering
\includegraphics[width=0.5\textwidth]{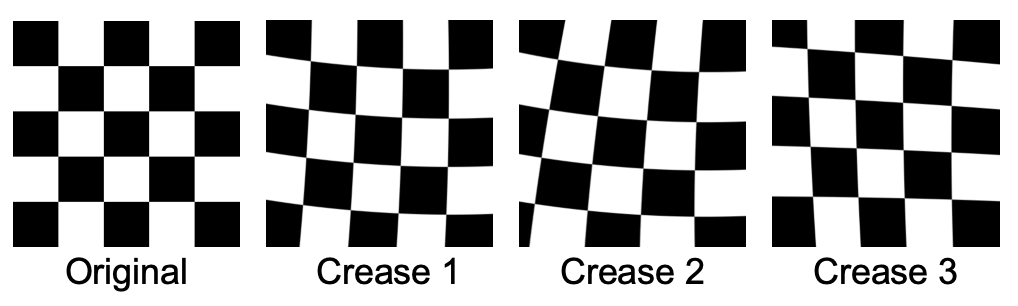} %, height=7cm
\caption{Visual examples of random crease transformations. }
\label{creases_vis}
\end{figure}
Another possible transformation when printing the generated patch on a T-shirt is constantly changing creases in the clothes resulting from a person's movements. To overcome this challenge authors in \cite{dap} proposed the following transformations:
A set of creases can be added to the patch: %Each crease (shown in Figure \ref{creases_vis}) can modeled by randomly selecting a point and a 2D vector ($\pm5$) and moving all pixels in the direction of the vector, with pixels closer to the line formed by the vector extending from the chosen point moving at a greater magnitude. The movement of a point (x, y) when the chosen point is $(x_0, y_0)$ 
Each crease is modeled by randomly selecting a point on the patch, along with a 2D vector representing the crease's direction, which is chosen within a range of $\theta$ degrees to simulate the alignment on the clothing surface. The selected pixels within the patch are displaced in the direction defined by the vector, with varying degrees of movement based on their proximity to the vector's line. This displacement emulates the natural variation in the crease intensity along their length. The displacement of a point (x, y) resulting from the chosen crease point (x0, y0) and the associated 2D vector is calculated as Displacement = Vector x Multiplier, where the multiplier captures the dynamic nature of creases and follows the equation \ref{eq:creases}. This process is performed for each incorporated crease. 
%is the vector multiplied by a multiplier, which follows the equation: 
\begin{equation}
multiplier(x, y) = 1 - \frac{\sin^2{\theta}[(x-x_0)^2 + (y-y_0)^2]}{width^2 + height^2}
\label{eq:creases}
\end{equation}
Where $\theta$ is the angle between the direction of the (x, y) from $(x_0, y_0)$ and the chosen vector, and \textit{width} and \textit{height} are the dimensions of the patch.

\subsubsection{Thin Plate Spline (TPS)}
In the context of adversarial attacks, Thin Plate Spline (TPS) can be used to generate smooth and visually realistic transformations on images, which can deceive machine learning models and introduce adversarial perturbations. TPS-based techniques have been employed to generate adversarial examples by manipulating the pixel values or spatial transformations of images \cite{adversarialtshirt}. By selecting control points and optimizing the TPS parameters, an attacker can create subtle deformations or perturbations that are difficult for a machine learning model to detect, but can cause misclassification or incorrect behavior.

%NPS-Loss, Transformation (e.g., scaling, rotation, TPS mapping, etc.) and Ensemble are beneficial for robustness.}
\begin{figure}[!htp]
\centering
\includegraphics[width=0.3\textwidth]{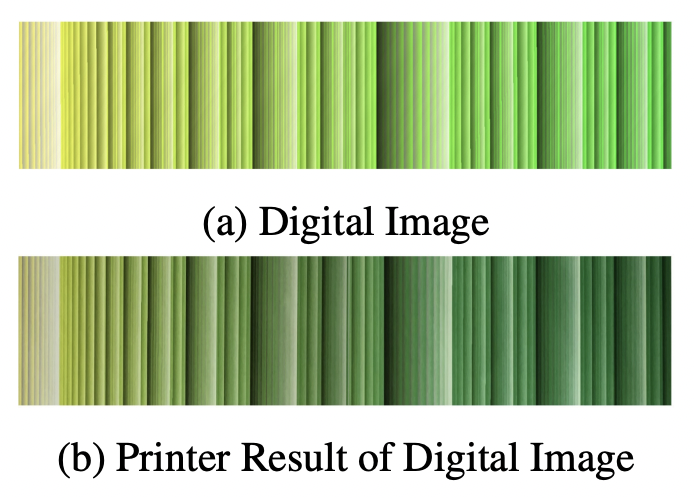} %, height=7cm
\caption{Visualization discrepancy between the digital image (a) and its printed images
(b). Figure adapted from \cite{extendedrp2}. }
\label{nps}
\end{figure}
%==============================
\subsubsection{Non-Printability Score (NPS)}
%==============================
\begin{figure*}[!htp]
\centering
\includegraphics[width=0.65\textwidth]{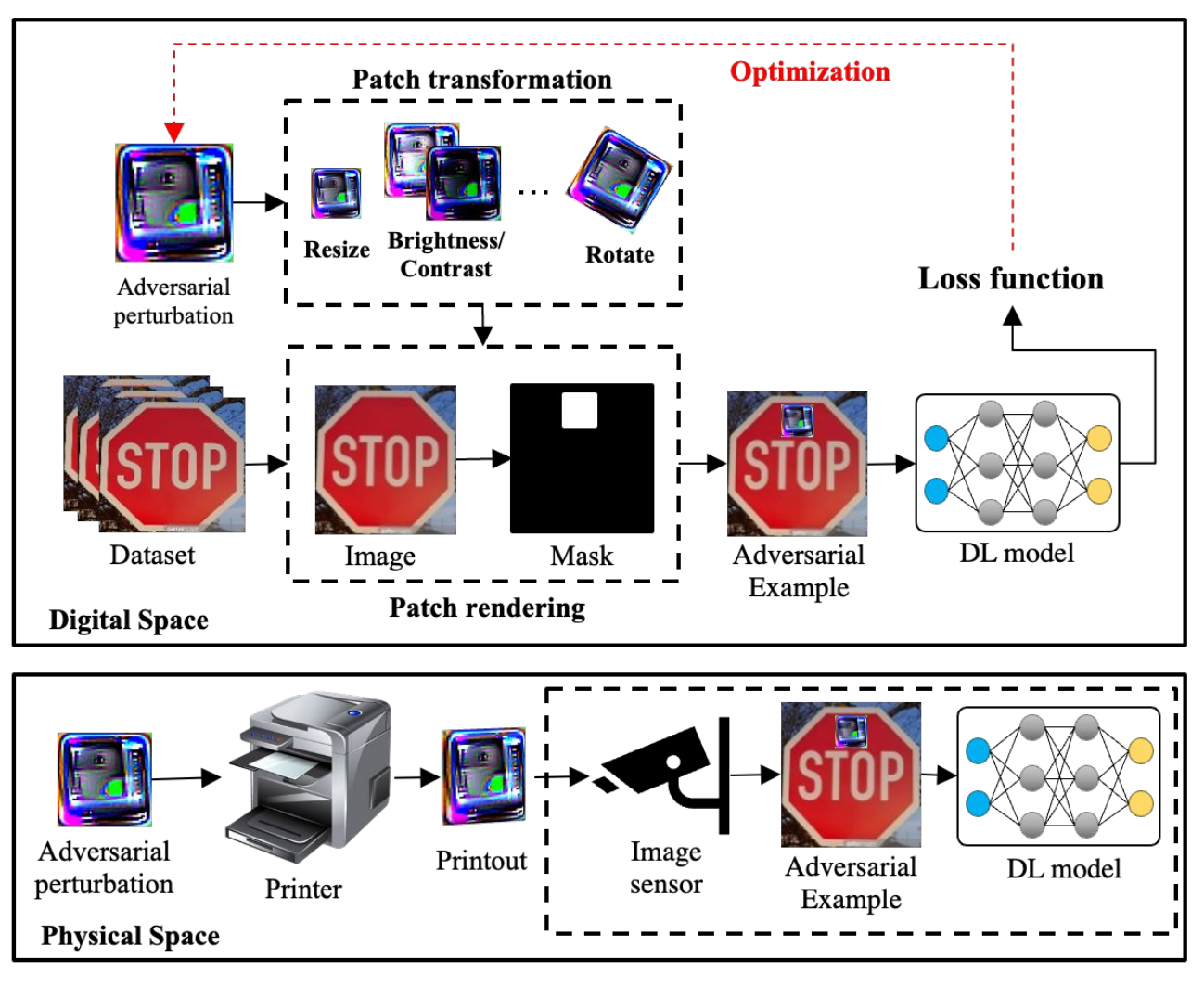} %, height=7cm
\caption{Physical adversarial attacks on the classification task typically involve two stages: optimizing the adversarial perturbation in the digital space and testing the effectiveness of the adversarial perturbation in the physical domain. }
\label{classification}
\end{figure*}
%One necessary way to perform physical attacks is to print the adversarial perturbation. Thus, the perturbation will inevitably have some distortion caused by the printer, resulting in attacks failing in some cases. For instance, Figure \ref{} illustrates the digital color and its printed result. To address this issue, nonprintability score (NPS) loss \cite{Sharif2016FaceRecognitionAttacks} is devised as a metric to measure the color distance between optimized adversarial perturbation and the common printer. In general, the NPS loss can be represented as the following loss:
Printing adversarial perturbations is a crucial method for performing physical attacks, but the printing process can cause distortions that may lead to attack failures. An example of this is shown in Figure \ref{nps}, which demonstrates the difference between a digital color and its printed version. 
To overcome this challenge, researchers have developed the non-printability score (NPS) loss \cite{Sharif2016FaceRecognitionAttacks} as a metric to measure the color distance between the optimized adversarial perturbation and the typical printer. The NPS loss is represented by the following formula:

\begin{equation}
    L_{nps} = \sum_{c_{perturb} \in p} \min_{c_{print}\in C} ||c_{perturb} - c_{print} ||
\end{equation}

\subsubsection{Randomly Transformed Patch}
\cite{invisiblecloak2}
These transforms are a composition of brightness, contrast, rotation, translation, and sheering transforms that help make patches robust to variations caused by lighting and viewing angle that occur in the real world.

%\subsubsection{Nested-AE}
%Nested-AE \cite{nestedae} contains two or more Adversarial examples inside that for different distances or angles. It significantly improve the robustness of adversarial attack at the various position.

\subsubsection{Digital-to-Physical (D2P)} %train cGANs to model the digital-to-physical transformation
Digital-to-Physical (D2P) \cite{D2P} refers to the process of translating digital representations or information into physical objects or actions. It involves converting digital data, such as images, models, or instructions, into a format that can be perceived, interacted with, or executed in the physical world.

%\subsubsection{SSP: size, shape, position} \cite{hotcold}

%\subsubsection{Transformations: Alignment, Rendering, Bending}

%\subsubsection{topologically plausible projection (TopoProj) }

%\subsubsection{Sticker Projection} \cite{advhat}
%Project the obtained adversarial examples with small perturbations in the projection parameters to make the attack more robust.

%=========================================================================
\subsection{Techniques For Enhancing The Stealthiness Of Physical Attacks}
%=========================================================================
Ensuring stealthiness in physical adversarial attacks is crucial to evade detection and maintain the effectiveness of the attacks. Several techniques have been employed to enhance the stealthiness of physical attacks, including:

%-----------------------------------------------
\subsubsection{Generative Models}
\begin{figure}[!htp]
\centering
\includegraphics[width=0.5\textwidth]{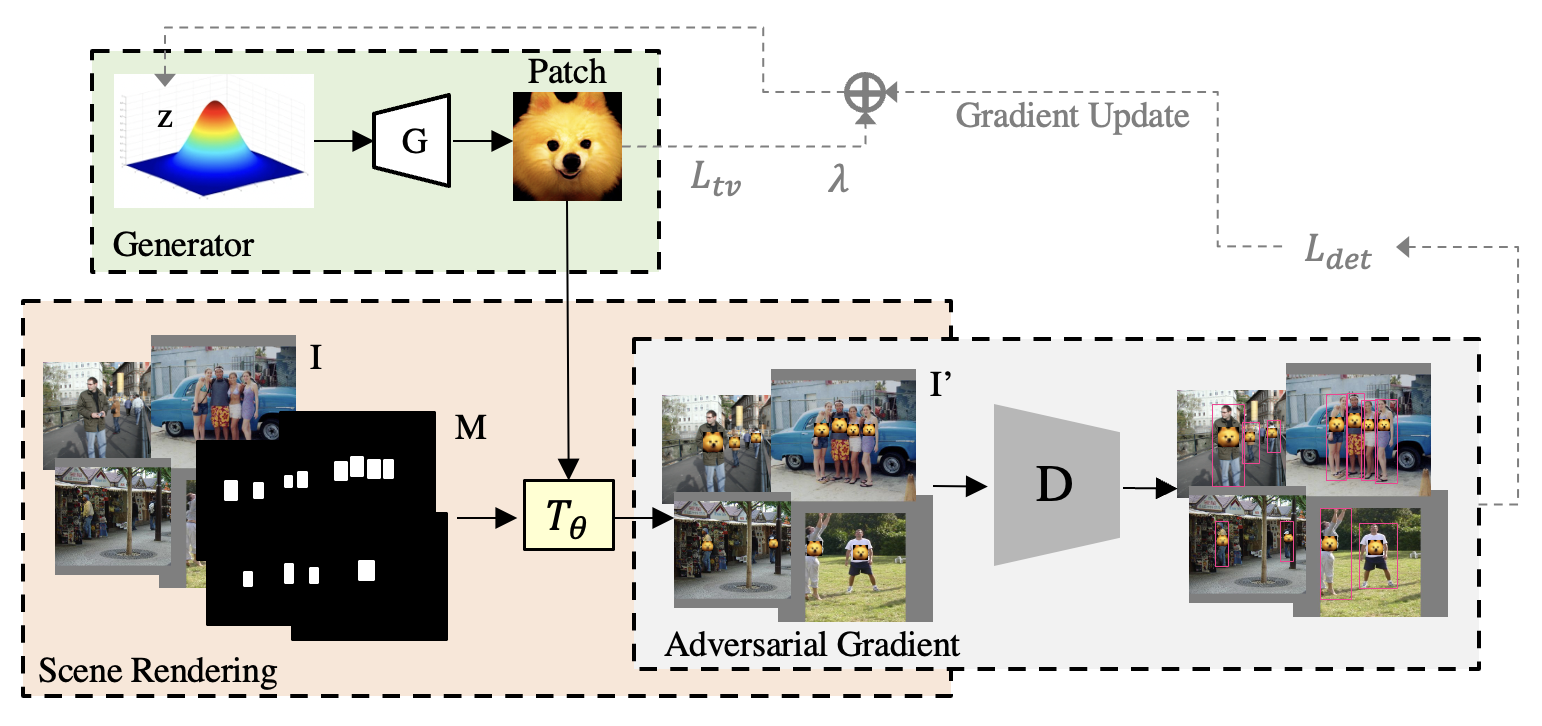} %, height=7cm
\caption{Overview of the naturalistic adversarial patch (NAP) \cite{NAP} generation framework provides an overview of the process used to create patches for object detectors. This framework leverages the knowledge gained from a pretrained Generative Adversarial Network (GAN) trained on real-world images and utilizes the learned image manifold. Through an iterative optimization process, the framework samples the optimal image from the GAN, resulting in the final patch for the object detector.}
\label{NAP_overview}
\end{figure}

%-----------------------------------------------
Generative models are utilized to generate realistic and natural-looking perturbations or adversarial examples. These models, such as generative adversarial networks (GANs) or variational autoencoders (VAEs), can learn the underlying distribution of the input data and generate perturbations that are visually indistinguishable from the original images, ensuring the effectiveness and stealthiness of the attacks.
Generated patches can be natural-looking and look less malicious than noisy perturbation patches in LaVAN \cite{lavan} and AdvPatch \cite{adversarial-patch}. Instead of directly applying perturbations to the input space that leads to noisy adversarial patches \cite{lavan}, \cite{adversarial-patch}, Work like \cite{tntattack} proposed solving the problem of searching for naturalistic patches with adversarial effects by indirectly manipulating the latent space $z$ of a Generative Adversarial Network that has learnt to approach the natural patch distribution.
An illustrative example showcasing the application of this technique is depicted in Figure \ref{NAP_overview}.
%-----------------------------------------------
\subsubsection{Style Transfer}
%-----------------------------------------------
%Style transfer techniques are utilized to transfer the appearance and visual characteristics of the surroundings onto the adversarial patches or perturbations. This ensures that the perturbations inherit the textures, colors, and lighting conditions of the environment, making them visually consistent and less likely to attract attention.
%Neural style transfer is a widely-used technique for transferring the style of one image to another, often treated as an image transformation task \cite{tang2021attentiongan, gu2018arbitrary}.
Style transfer methods play a crucial role in transferring the visual attributes and appearance of the surroundings onto adversarial patches or perturbations. By leveraging these techniques, perturbations can inherit the textures, colors, and lighting conditions of the environment, resulting in visual coherence and reduced likelihood of attracting attention. One widely-used style transfer technique is neural style transfer, which is commonly employed for transferring the style of one image onto another. This process is often treated as an image transformation task \cite{tang2021attentiongan, gu2018arbitrary}.
%-----------------------------------------------
\subsubsection{Semantic constraint}
%-----------------------------------------------
%Semantics (e.g., introducing the semantic features of cartoon pictures to adversarial patch \cite{LAP})

%$L_{sim}$ is the similarity loss between the target natural image $N$ and the adversarial patch $P$. It is defined as:
%\begin{equation}
%    L_{sim} = (\frac{1}{N} \sum_{i,j} (\left | P_{i,j} - N_{i,j} \right |) )^2
%\end{equation}
%project into the onto the surface of $L_p$ norm-balls with radius $\epsilon$ and centered at I. Here we choose I as natural images to ensure the generated camouflage patterns are semantically meaningful.

%$L_{sim}$ is the similarity loss between the target natural image $N$ and the adversarial patch $P$. It is defined based on a similarity metric:\\
%MSE-based similarity loss:
%\begin{equation}
%    L_{sim} = \frac{1}{n} \sum_{i,j} ( P_{i,j} - N_{i,j} )^2 
%\end{equation}
%%where $m$ and $n$ are the dimensions 
%Cosine similarity-based similarity loss:

%\begin{equation}
%    L_{sim} = -\left(\frac{\sum_{i,j}P_{i,j} N_{i,j}}{\sqrt{\sum_{i,j}P^2_{i,j}}\sqrt{\sum_{i,j}N^2_{i,j}}} \right)
%\end{equation}

%The similarity loss term is squared to slow the rate of increase (the slope or the rate of change) and delay the convergence of the similarity metric with respect to the detection loss.

%perceptual color distance $\Delta E$ \cite{spaa}

%SSIM, PSNR, Cosine Similarity

\begin{figure}[!htp]
\centering
\includegraphics[width=0.5\textwidth]{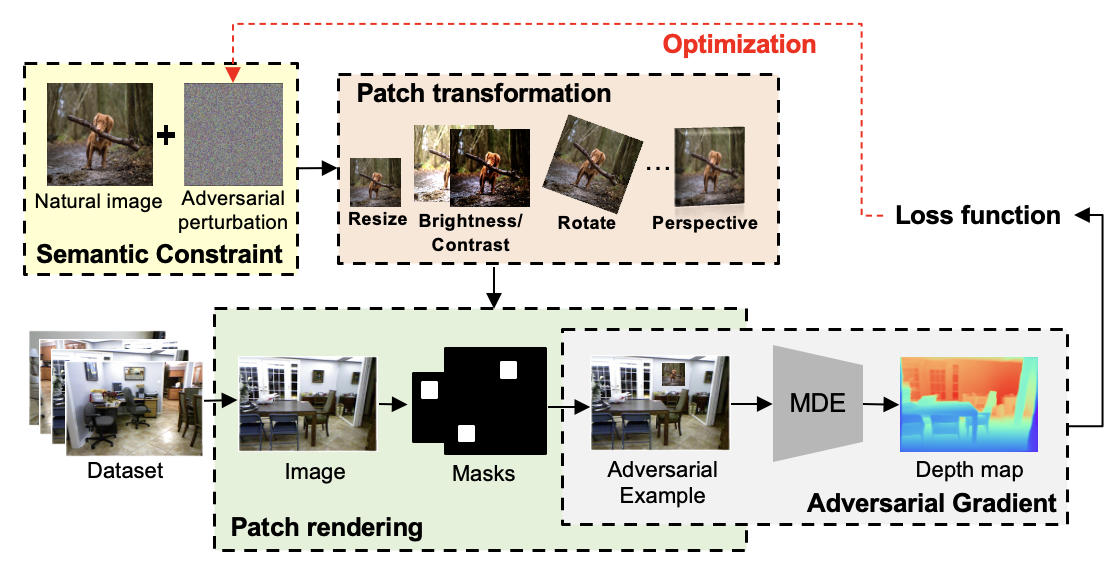} %, height=7cm
\caption{Overview of the stealthy adversarial patch (SAAM) framework: It incorporates a semantic constraint to ensure that the generated adversarial patches remain inconspicuous, making them blend seamlessly into the scene. Additionally, SAAM utilizes a data augmentation technique that considers potential real-world transformations during the optimization process, ensuring the patch's effectiveness across various physical conditions (Figure Adapted from \cite{saam}).}
\label{SAAM_overview}
\end{figure}

%pattern, shape, color control loss functions \cite{nestedae}
Taking inspiration from the imperceptibility constraint commonly employed in Lp-norm based adversarial perturbations, we incorporate a projection function (Equation \ref{semantic_constraint}) to ensure that the generated adversarial patterns maintain visual similarity to natural images throughout the optimization process. By enforcing this constraint, we achieve high-quality semantic patterns that closely resemble a painting on a wall. Empirical results demonstrate the effectiveness of optimizing with this constraint, as it facilitates the creation of visually convincing adversarial patterns that seamlessly blend into their surroundings.
\begin{equation}
    P_{\delta} = N + \delta
\end{equation}
Where $N$ is a chosen natural images to ensure the generated camouflage patterns are semantically meaningful.

\begin{equation}
    \delta^t = Proj_{\infty}(\delta^{(t-1)} + \Delta \delta, N, \epsilon)
    \label{semantic_constraint}
\end{equation}
An illustrative example showcasing the application of this technique is depicted in Figure \ref{SAAM_overview}.

where $\delta^t$ and $\Delta \delta$ denote the adversarial pattern and its updated vector at iteration t, respectively. $Proj_{\infty}$ projects generated pattern onto the surface of $L_{\infty}$ norm-balls with radius $\epsilon$ and centered at N. Here we choose N as natural images to ensure the generated camouflage patterns are semantically meaningful.

%\subsubsection{Style loss}
%Style loss \cite{OAP} is defined as the style distance between target image and adversarial example.

%\subsubsection{Content Loss}
%Content loss is designed to preserve the content of the original image since the style loss could make the adversarial example diﬀerent a lot from the original one.

%\subsubsection{Photorealism Regularization}
%Photorealism Regularization is used to constrain the reconstructed image to be represented by locally affine color transformations of the content image to prevent distortions. 

%\subsubsection{UPC} 
%UPC \cite{UPC} optimization constraint to make generated patterns look natural to human observers.

%=========================================================================
\subsection{Techniques For Enhancing The Transferability Of Physical Attacks}
%=========================================================================
\subsubsection{Ensemble Training} 
%Ensemble training \cite{invisiblecloak2} is used to generate adversarial perturbation that is able to fool an ensemble of detectors/models that were not used for training.
Ensemble training \cite{invisiblecloak2} is a technique utilized to generate adversarial perturbations that can deceive an ensemble of detectors or models that were not involved in the training process. This approach enhances the transferability and robustness of the adversarial perturbations by ensuring their effectiveness across multiple models. %By training the adversarial perturbation on an ensemble of models, it becomes more challenging for the detectors to accurately classify the perturbed inputs, thus increasing the vulnerability of the ensemble to adversarial attacks.
For instance, instead of solely optimizing for the target model, we can introduce an additional objective to maximize the loss of a different model (a non-target model) on the same adversarial perturbation. This encourages the generation of perturbations that are effective on both models.

\subsubsection{Black-Box Attacks}  
Designing adversarial perturbations without assuming any knowledge about the targeted model, thus increasing the likelihood of transfer to other models.
In black-box attacks, the attacker treats the target model as a "black box" and attempts to generate adversarial examples by querying the model and observing its responses without direct access to its internal details.
%Black-box attacks techniques that are designed to be effective even when the attacker has limited or no knowledge of the target model's architecture, parameters, or gradients. 

%% file: classification.tex
%============================================
\section{Physical Attacks on Classification}
%===========================================
\label{classification_}

Physical attacks on classification refer to a class of adversarial attacks that aim to deceive DL-based classifiers by introducing physical perturbations to the input data. Unlike digital attacks that manipulate pixels or features in the digital representation of an image, physical attacks involve making modifications to the physical instance of the input. These attacks exploit the vulnerabilities of these models to subtle changes in the input's physical properties, such as color, texture, shape, or lighting conditions. By strategically manipulating the physical characteristics of an object, an attacker can cause the model to misclassify objects or generate incorrect predictions. %Physical attacks on classification pose a significant threat in real-world scenarios, as they can be executed by introducing imperceptible perturbations to physical objects or scenes. %Defending against physical attacks on classification requires developing robust techniques that can detect and mitigate the impact of physical perturbations, as well as incorporating physical constraints and sensor-based defenses into the classification pipeline. Advancing research in this field is crucial for enhancing the security and reliability of machine learning systems deployed in physical environments.

Physical adversarial attacks on the classification task typically involve two stages: optimizing the adversarial perturbation in the physical space and testing the effectiveness of the adversarial perturbation in the physical domain as illustrated in Figure \ref{classification}.
In the optimization stage, the attacker aims to design an adversarial perturbation that can deceive the classification model when applied to physical inputs. This process often involves iteratively adjusting the perturbation to maximize its effectiveness while considering the constraints imposed by the physical world. Techniques such as optimization algorithms, generative models, or learning-based approaches may be employed to optimize the perturbation's properties. Once the adversarial perturbation is optimized, it is then tested in the physical space to evaluate its effectiveness against the target classification model. This involves applying the perturbation to real-world objects, images, or signals and observing the model's response. The attacker may assess the success rate, misclassification rate, or other relevant metrics to measure the impact of the adversarial perturbation on the model's performance.
By performing these two stages, attackers can exploit vulnerabilities in the physical domain to deceive classification models and potentially manipulate their decisions in real-world scenarios. %Understanding and developing countermeasures against physical adversarial attacks are crucial for enhancing the security and reliability of classification systems in practical applications.

Attacks aimed at compromising this task can take on diverse forms, encompassing various techniques such as patch-based attacks, sticker-based attacks, light manipulation-based attacks, printed images, imaging device manipulation, and 3D printed objects. Throughout the following sections, we will thoroughly examine each of these distinct attack types and categorize them according to their respective forms.

Table \ref{Table:Classification_comparison} provides a comprehensive comparison of various adversarial attack methods on classification task in terms of their attack goals, patch placement strategies, consideration of changing view points, testing in the physical domain, and transferability to other models.
Table \ref{Table:Classification_info} presents detailed information on adversarial attacks, including the attacker's knowledge level, robustness techniques, stealthiness techniques, physical test types, and space of operation.
Table \ref{Table:Classification_dataset} provides information on the datasets used, the networks evaluated, and the links to open-source code for the experiments conducted in the classification task.
%Table \ref{Table:Classification_comparison} includes Attack goal, placement, consider changing view point, test in physical domain, test transferability

%Table \ref{Table:Classification_info} includes Adversarial's knowledge, robustness technique, stealthiness techniques, physical test type, space

%Table \ref{Table:Classification_dataset} includes Dataset, networks, open source code link
%-------------------------------------------------------------------------------------
\subsection{Patch-based Attacks}
%-------------------------------------------------------------------------------------
Patch-based attacks are a specific form of adversarial perturbation that focuses on modifying localized patches or regions within an image with the intention of deceiving machine learning models. These attacks exploit the susceptibility of models to localized perturbations, with the objective of introducing subtle alterations that exert a significant influence on the model's output. By capitalizing on the model's reliance on specific features or patterns, adversaries can generate patches that deceive the model into misclassifying the image or perceiving it in a manner contrary to its intended interpretation. The physical patches employed in such attacks are meticulously engineered to exploit the weaknesses of computer vision systems. % object recognition or detection systems. 
%Through careful manipulation of the appearance and placement of these patches, attackers endeavor to deceive the system by causing it to incorrectly classify or misconstrue the depicted objects or scenes.

\begin{figure*}[!htp]
\centering
\includegraphics[width=0.9\textwidth]{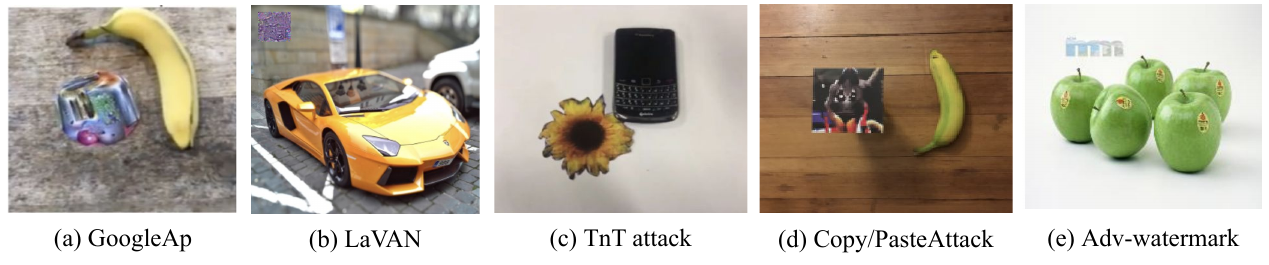} %, height=7cm
\caption{Examples of patch-based physical adversarial examples against image classification: (a) GoogleAp \cite{googleap}, (b) LaVAN \cite{lavan}, (c) TnT attack \cite{tntattack}, (d) Copy/PasteAttack \cite{pasteattack}, and Adv-watermark \cite{advwatermark}. }
\label{AE_Patch_classification}
\end{figure*}
%----------------------------------
%GoogleAp \cite{googleap}
%LaVAN \cite{lavan}
%Advwatermark \cite{advwatermark}
%Adversarial ACO \cite{ACO}
%Adversarial ACO2 \cite{ACO2}
%TnT attack \cite{tntattack}
%Copy/Paste attack \cite{pasteattack}

%$RP_2$ \cite{RP2} (TSR)
%PS-GAN \cite{PSGAN} (TSR)
%----------------------------------
%%%%======= Effectiveness ===========%%%%

The work on adversarial patches was introduced by Brown et al. \cite{googleap} (referred to as \textbf{GoogleAp}), offering a more applicable form of attack for real-world scenarios in comparison to Lp-norm based adversarial perturbations that require object capture through a camera. This approach presents a pragmatic and versatile method for deceiving classifiers in real-world settings. Adversarial patches, often incorporating meticulously designed patterns or textures, are strategically incorporated into specific regions of an image to manipulate the predictions of the model. The fundamental characteristic of patch-based attacks lies in their localized nature, focusing on modifying only a subset of the image rather than altering the entire input. These patches can be strategically positioned in regions of utmost significance or in areas with a higher likelihood of influencing the model's decision-making process. In practical implementation, the patch is strategically inserted into the scene captured by the camera, diverting the classifier's attention away from accurately recognizing the objects present and instead fixating solely on the patch itself. Notably, this targeted attack possesses the capability to deceive the classifier by classifying any object as a toaster or any other desired class envisioned by the attacker. The process of generating the patch imposes minimal constraints, with the primary objective being to maximize the effectiveness of the attack. Consequently, patches containing significant perturbations can be widely disseminated online for other attackers to print and utilize. Furthermore, the attacker does not require prior knowledge of the specific target image. Even when the patches are small in size, they cause the classifiers to disregard other elements within the scene and instead report a pre-determined target class. This research has exhibited remarkable results and has served as a source of inspiration for the development of patch-based adversarial attacks. 

\textbf{LaVAN} \cite{lavan} is a technique for generating localized and visible patches that can be applied across various images and locations. Their approach involved training the patch by iteratively selecting a random image and placing it at a randomly chosen location. This iterative process allowed the model to capture the distinguishing features of the patch across a range of scenarios, thereby enhancing its ability to transfer and its overall effectiveness.

\textbf{Adv-Watermark} \cite{advwatermark} involves the embedding of adversarial watermarks into digital content, such as images or videos. Its objective is to introduce imperceptible modifications within the content that can be detected and authenticated by dedicated algorithms or systems. The adversarial watermarks are specifically engineered to withstand diverse forms of transformations, including compression, cropping, scaling, and other manipulations commonly applied to images. 

Since then, numerous similar approaches have been proposed. One notable example is the work of Eykholt et al. \cite{RP2}, who developed a comprehensive attack algorithm known as Robust Physical Perturbations (\textbf{RP2}). This algorithm demonstrates remarkable success rates in targeted attacks against standard road sign classifiers \cite{stallkamp2012man} in real-world environments, considering diverse environmental conditions. The RP2 method effectively employs specially designed black and white color blocks to generate powerful adversarial perturbations, enabling the manipulation of physical objects and achieving high attack success rates. 

Subsequently, Liu et al. \cite{PSGAN} introduced a distinct approach by framing patch generation as a patch-to-patch translation task, employing the perceptual-sensitive generative adversarial network (\textbf{PS-GAN}). In order to enhance the attack's effectiveness, PS-GAN incorporates visual attention mechanisms to capture perceptually-sensitive information and guide the optimal placement of adversarial patches. Through the utilization of visual attention, PS-GAN ensures that the adversarial patches are strategically positioned to maximize their impact and exploit the vulnerabilities of the targeted system. This approach highlights the potential of employing perceptual-sensitive techniques in the generation of adversarial patches, offering new avenues for exploration in the field.
\begin{figure*}[!htp]
\centering
\includegraphics[width=0.7\textwidth]{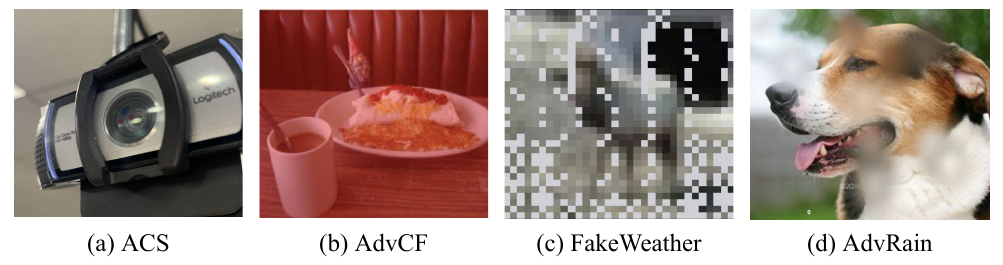} %, height=7cm
\caption{Examples of sticker-based physical adversarial examples against image classification: (a) ACS \cite{ACS}, (b) AdvCF \cite{advcf}, (c) FakeWeather \cite{fakeWeather}, and (d) AdvRain \cite{advrain}. }
\label{AE_sticker_classification}
\end{figure*}

In contrast to previous physical attacks that lacked practical applicability, Liu et al. \cite{wang2021universal, liu2020bias} introduced a novel approach for optimizing a universal adversarial patch specifically targeting real-world automatic check-out systems (\textbf{ACOs}). Their method exploited the perceptual and semantic biases inherent in these systems to optimize the adversarial patch. To exploit the perceptual bias, the authors utilized challenging examples that exhibited strong model uncertainties. From these examples, they extracted a textural patch prior based on style similarities. In order to mitigate the reliance on extensive training data, the authors tapped into the semantic bias by collecting prototypes representing different classes. The universal adversarial patch was then trained using this prototype dataset. Additionally, the authors incorporated Expectation over Transformation (EOT) to enhance the robustness of the adversarial patch. In physical attacks, the printed adversarial patch was affixed to various commodities, and images were captured under different environmental conditions, including different angles (-30°, -15°, 0°, 15°, 30°) and distances (0.3m, 0.5m, 0.7m, 1m). Experimental results demonstrated the ease with which the captured images, containing the adversarial patch, deceived the ACO system. Furthermore, their attack significantly reduced the recognition accuracy of prominent e-commerce platforms (JD and Taobao) on the test set, with degradation rates of 43.75\% and 40\%, respectively. This research sheds light on the vulnerabilities of real-world ACOs and highlights the potential impact of adversarial attacks on e-commerce platforms. %The findings emphasize the need for robust defenses against such attacks to ensure the integrity and security of automated check-out systems in practical settings.

The enhanced version of adversarial ACO (Adaptive Combinatorial Optimization) known as \textbf{Adversarial ACO2} improves upon its predecessor's generalization capabilities, rendering it a universal attack, by leveraging the inherent perceptual and attentional biases of models. Expanding on the foundations of ACO, which involved generating adversarial patches through perceptually biased prior generation and semantically biased prototype generation, the authors of Adversarial ACO2 introduced an attentional-bias-based strategy to enhance its attacking effectiveness. By incorporating attentional bias, the proposed framework places greater emphasis on the generalization and transferability of adversarial patches, thereby strengthening its overall attack capability. This advancement enables Adversarial ACO2 to exhibit improved performance and wider applicability in the context of adversarial attacks.

In a recent investigation, Bao et al. \cite{tntattack} addressed the issue of visual conspicuousness associated with existing physical adversarial patches. To overcome this limitation, the authors proposed a method known as \textbf{TnTattack}. Instead of employing a traditional generative adversarial network (GAN) \cite{xiao2018generating} to generate full-pixel adversarial examples, they developed a generator that maps a latent variable z (random noise) to a realistic image patch suitable for embedding into the target image. In the second stage of their approach, the authors fixed the generator's parameters and employed a gradient-based optimization technique guided by the adversarial loss to determine the optimal latent variable z, which generates a visually naturalistic adversarial patch. The underlying concept revolves around the generator's ability to map a randomly sampled latent variable z to a naturalistic image, where a specific region in the latent space corresponds to the generation of the naturalistic adversarial patch. By reducing the optimization variable to only 100 dimensions of latent variables z, the search space is significantly minimized. In the physical attack scenario, the generated patch image was printed and affixed to the front of clothing items. Subsequently, a 1-minute video was recorded featuring a person wearing the clothes adorned with the adversarial patch. Experimental results demonstrated that over 90\% of the captured images in the video frames effectively deceived the target network, showcasing the efficacy of the TnTattack approach in practical physical attack settings. This technique not only enhances the visual realism of adversarial patches but also validates their effectiveness in real-world physical attack scenarios.

Inspired by the ringlet butterfly's use of adversarial "eyespots" on its wings to startle predators, Casper \cite{pasteattack} postulated that deep neural networks (DNNs) could be deceived by interpretable features in the real world. Motivated by this concept, the author developed a method to generate adversarial perturbations that could exploit vulnerabilities in the victim network, known as \textbf{Copy/Pasteattack}. The proposed approach involved employing a Generative Adversarial Network (GAN) to create a patch that would be pasted onto the target image, causing it to be misclassified as a specific target class by the target model. To enhance the robustness of the adversarial patch, Expectation over Transformation (EOT) was utilized. Additionally, Total Variation (TV) loss was applied to discourage the presence of high-frequency patterns in the patch, ensuring its visual coherence. In physical attacks, the generated adversarial patch was printed and affixed to the object of interest. Evaluation results obtained from recaptured images verified the efficacy of their approach in real-world scenarios. This method not only sheds light on the vulnerability of DNNs to interpretable features but also demonstrates the practical applicability of adversarial attacks in real-world settings. %The Copy/Paste attack technique offers insights into the potential risks and challenges posed by interpretable features in the context of adversarial attacks.

%robustness

Previous studies have revealed that adversarial examples often lose their effectiveness when subjected to real-world image transformations, such as changes in angle and viewpoint \cite{noneed, luo2015foveation}. In order to address this challenge, Athalye et al. \cite{EOT} introduced the \textbf{Expectation Over Transformation (EOT)} algorithm, which serves as a general framework for constructing adversarial examples that remain effective across a selected distribution of transformations. The fundamental concept of EOT involves integrating the modeling of perturbations directly into the optimization process. This approach has demonstrated robustness in conducting adversarial attacks in the physical domain. As a result, subsequent works \cite{shadow, darts, ACO, ACO2} have adopted this framework, aiming to enhance the resilience of attacks and improve overall performance. The utilization of EOT has paved the way for advancements in generating adversarial examples that maintain their effectiveness under various real-world image transformations.

Eykholt et al. \cite{RP2} have identified environmental variability as the primary challenge in generating robust physical perturbations. Existing algorithms designed for digital adversarial perturbations \cite{yuan2021meta} are rendered ineffective when applied in the physical space. Furthermore, several practical challenges arise in this context:
(i) The magnitude of perturbations in the digital domain can be extremely subtle, making it difficult for cameras to detect them due to sensor imperfections.
(ii) Current algorithms generate perturbations that alter the background of an object, which becomes a significant challenge as real-world objects can have diverse backgrounds depending on the viewpoint.
(iii) The fabrication process, such as printing the perturbations, inherently introduces imperfections, thereby adding further complexity to achieving potent attacks.
These challenges highlight the need to address the impact of environmental variability and practical constraints when developing robust physical adversarial perturbation techniques. Eykholt et al.'s work emphasizes the importance of adapting adversarial techniques to real-world conditions in order to overcome these challenges and enhance the effectiveness of physical attacks.

Taking into consideration the aforementioned challenges, Eykholt et al. devised Robust Physical Perturbations (RP2), a method capable of generating perturbations that exhibit resilience to variations in camera distances and angles. To account for synthetic variations, they employed random cropping of the object within the image, adjusted brightness levels, and introduced spatial transformations to simulate various conditions. To address physical conditions, they simulated image capture under diverse scenarios, including changes in distances, angles, and lighting conditions. A related study by Kong et al. \cite{physgan} also implemented color augmentation techniques to enhance image contrast, thereby improving the robustness of adversarial examples against variations in light illumination conditions. These approaches demonstrate the importance of accounting for both synthetic and physical variations in order to develop robust physical adversarial perturbation techniques.

Nevertheless, it is crucial to acknowledge that adversarial patch-based attacks do not consistently deceive neural network classifiers in real-world physical environments. This limitation stems from the inherent two-dimensionality of image patches, which fails to account for variations caused by changes in viewpoints and other natural transformations. The effectiveness of these attacks may diminish when confronted with real-world scenarios that involve three-dimensional objects and dynamic environments. These factors present substantial challenges for the practical application of patch-based attacks outside controlled laboratory settings. It is imperative to consider the constraints imposed by the physical world when evaluating the efficacy and potential limitations of patch-based adversarial attacks.

%stealthiness

In adversarial attacks, ensuring stealthiness, where introduced perturbations are minimally perceptible to human observers while still inducing misclassification, is a fundamental criterion. The assessment of stealthiness often involves quantifying the perceptibility using Lp norms, such as those employed in FGSM \cite{fgsm}, Carlini and Wagner Attack \cite{Carlini2017CWAttack}, and related works \cite{phy9}. These norms are frequently incorporated as additional constraints during the optimization process of the adversarial attack objective, facilitating the generation of imperceptible perturbations. The utilization of Lp norms contributes to the pursuit of stealthy attacks that can effectively deceive machine learning models while minimizing the observable impact on human perception.

Instances of traffic signs marked with scrawls and patches are frequently encountered in the real world. Drawing inspiration from this observation, the Perceptual-Sensitive Generative Adversarial Network (PS-GAN) \cite{PSGAN} adopts a patch-to-patch translation approach within an adversarial framework for patch generation. In this approach, any seed patch is taken as input, and PS-GAN generates an adversarial patch that exhibits a strong perceptual correlation to the targeted image. The adversarial learning process \cite{goodfellow2014generative} is employed, where a generator G generates adversarial patches, and a discriminator D evaluates the perceptual similarity. By effectively leveraging these common phenomena, PS-GAN conceals attacks within the generated patches, utilizing perceptual sensitivity to enhance the adversarial impact.

%-------------------------------------------------------------------------------------
\subsection{Sticker-based Attacks}
%-------------------------------------------------------------------------------------
\begin{figure*}[!htp]
\centering
\includegraphics[width=\textwidth]{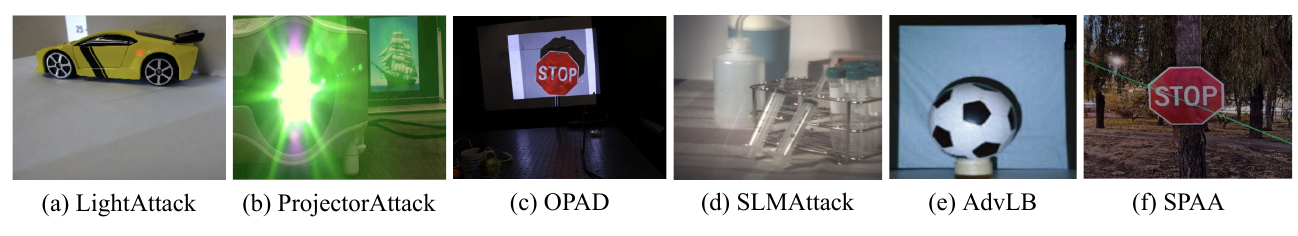} %, height=7cm
\caption{Examples of Light Manipulation-based physical adversarial examples against image classification: (a) LightAttack \cite{LightAttack}, (b) ProjectorAttack \cite{projectorattack}, (c) OPAD \cite{OPAD}, (d) SLMAttack \cite{slmattack}, (e) AdvLB \cite{AdvLB}, and (f) SPAA \cite{spaa}. }
\label{AE_light_classification}
\end{figure*}
Sticker-based physical attacks constitute a distinctive category of physical adversarial attacks wherein physical stickers or decals are affixed to objects or scenes with the intention of deceiving computer vision systems. These attacks exploit the inherent vulnerabilities of object recognition or detection systems by strategically placing stickers in a manner that manipulates the system's perception. The stickers utilized in these attacks are deliberately designed to induce misclassifications or trigger false interpretations by the computer vision system. They may incorporate specific patterns, symbols, or textures that are able to confuse the system's recognition algorithms. Through strategic placement of these stickers on objects, the attacker aims to misguide the system into making erroneous or unreliable judgments. Sticker-based physical attacks pose a tangible real-world threat as they can be executed in diverse settings, encompassing public spaces, surveillance systems, or autonomous vehicles. The physical presence of stickers renders these attacks more challenging to detect, as they seamlessly blend with the environment and may not arouse suspicion among casual observers.
%----------------------------------
%ACS \cite{ACS}
%DAS \cite{DAS}
%AdvCF \cite{advcf}

%FakeWeather \cite{fakeweather}
%AdvRain \cite{advrain}
%Adversarial rain \cite{adversarialrain}
%----------------------------------
Sticker-based attacks can be broadly classified into two categories. The first category entails placing stickers directly on camera lenses. One notable example of this type of attack is the \textbf{adversarial camera sticker (ACS)} \cite{ACS}. ACS introduces perturbations along the optical path between the camera and the object, without modifying the object itself. This technique demonstrates that by strategically positioning a specifically designed translucent sticker over a camera lens, an attacker can induce inconspicuous yet significant perturbations in the resulting captured images. Consequently, the targeted objects may be misclassified as a different expected class due to the influence of these perturbations. The ACS approach exemplifies how perturbations introduced through well-positioned stickers can have a substantial impact on the classification outcomes of computer vision systems. In a similar vein, the \textbf{Adversarial Color Film (AdvCF)} technique, as proposed by Zhang et al. \cite{advcf}, leverages a color film positioned between the camera lens and the target object to facilitate efficient physical adversarial attacks. By manipulating the physical properties of the color film while leaving the target object untouched, AdvCF focuses on generating adversarial perturbations that are effective in both daytime and nighttime settings. In digital simulations, AdvCF achieves an Adversarial Success Rate (ASR) of 82.1\% on a subset of the ImageNet dataset. Moreover, in real-world physical experiments, AdvCF attains ASR values of 80.8\% for indoor testing and 86.7\% for outdoor testing, respectively. The AdvCF approach exemplifies the efficacy of manipulating the color film to introduce adversarial perturbations that can deceive computer vision systems in various environments.

The second type of sticker-based attack involves directly attaching the sticker onto the target object. Wang et al. \cite{DAS} introduced the \textbf{Dual Attention Suppression (DAS) attack}, which generates physical adversarial stickers with the objective of concealing vehicles. The inspiration behind this approach stems from the observation that cerebral activities exhibit similar patterns among individuals when exposed to certain stimulus features \cite{Tricoche2020PeerPE, zatorre1999auditory}. Leveraging this concept, DAS aims to undermine model attention by redirecting the shared equal attention of the model away from the target regions and towards non-target regions. Consequently, the targeted objects within the intended regions receive insufficient attention, leading to misclassification by the threat models. This work capitalizes on the shared attention characteristics among models, highlighting the potential of exploiting these characteristics to deceive computer vision systems through the use of physical adversarial stickers.

%stealthiness
%To ensure the preservation of the camera's imaging integrity, ACS (Adversarial Camera Sticker) \cite{ACS} incorporated the use of the Structural Similarity Index (SSIM) \cite{ssim}. SSIM serves as a metric for quantifying the similarity between two images: the clean image and the perturbed target image. By harnessing the capabilities of SSIM, ACS effectively determined the optimal parameters of the perturbation model, enabling the reconstruction of the observed perturbation while mitigating any potential adverse effects on the camera's imaging capabilities. The utilization of SSIM in ACS demonstrates a meticulous approach to maintaining image quality while introducing adversarial perturbations.
To preserve the integrity of the camera's imaging capabilities, the Adversarial Camera Sticker (ACS) approach, as described in the work of \cite{ACS}, incorporates the Structural Similarity Index (SSIM) \cite{ssim}. SSIM serves as a valuable metric for quantifying the similarity between two images: the original clean image and the perturbed target image. By leveraging SSIM, ACS efficiently determines the optimal parameters of the perturbation model, allowing for the reconstruction of the observed perturbation while minimizing any potential adverse effects on the camera's imaging quality. The incorporation of SSIM in ACS reflects a meticulous and careful strategy for introducing adversarial perturbations while ensuring image fidelity is maintained.

To ensure stealthiness, adversarial patches can be intentionally designed to resemble everyday objects. For example, researchers in \cite{advrain, fakeWeather, adversarialrain} proposed attacks \textbf{AdvRain}, \textbf{FakeWeather}, and \textbf{Adversarial Rain}, respectively, that simulate the visual effects of natural weather conditions, such as raindrops and snowflakes. These attacks involve printing raindrop-like patterns on translucent stickers, which are then affixed externally over the camera lens. This clever approach allows the adversarial patches to blend seamlessly with the surrounding environment, making them inconspicuous to casual observers. By emulating the appearance of natural weather conditions, these adversarial patches enhance their effectiveness in evading detection while still manipulating the captured images to deceive computer vision systems.

%--------------------------------------------------------------------------------------
\subsection{Light Manipulation-based Attacks}
%--------------------------------------------------------------------------------------

%LightAttack \cite{LightAttack}
%ProjectorAttack \cite{projectorattack} *
%OPAD \cite{OPAD} --
%SLMAttack \cite{slmattack}
%AdvLB \cite{AdvLB} --
%SPAA \cite{spaa} *
%Adversarial Shadow \cite{shadow} (TSR) --

Light manipulation-based physical adversarial attacks involve the intentional manipulation of light or visual stimuli to deceive or manipulate visual perception systems. These attacks capitalize on the vulnerabilities of visual processing algorithms with the objective of disrupting or altering the interpretation of visual information by introducing precisely crafted adversarial light patterns. In light manipulation-based physical adversarial attacks, the attacker designs and deploys adversarial light stimuli or patterns with the intention to mislead or deceive visual perception systems, such as image recognition or surveillance cameras. These stimuli are created by modifying or manipulating the intensity, color, or spatial distribution of light within a scene, leading to misclassification or incorrect interpretation of the visual content by the targeted system. These attacks exploit the reliance of visual systems on light cues, emphasizing the need for robustness against adversarial light manipulations to ensure the integrity and reliability of visual perception systems in real-world scenarios.

%--------------------------------------
%OPAD \cite{OPAD}
%ProjectorAttack \cite{projectorattack}
%LightAttack \cite{LightAttack} 
%SPAA \cite{spaa}
%----------------------------------
\textbf{LighAttack}: In contrast to adversarial patch-based attacks, optical attacks offer enhanced stealthiness and control as they do not require direct modification of the target object itself. An important characteristic of optical attacks is the utilization of a projector-camera model, which strengthens the robustness of physical attacks. This model involves projecting an adversarial perturbation into the real world using a projector and capturing it with a camera for evaluation. Taking advantage of this approach, Nichols et al. \cite{LightAttack} constructed a comprehensive training dataset comprising numerous pairs of real projected and captured images. Inspired by the one-pixel attack, they employed the differential evolution (DE) algorithm to optimize the projection point, treating it as a single pixel in low-resolution images (e.g., $32 \times 32$). During physical attacks, the authors projected light onto specific areas of the target object and subsequently recaptured the scene. Through extensive experimentation, they demonstrated the efficacy and effectiveness of these light attacks, showcasing the potential for optical approaches to generate powerful adversarial perturbations in physical settings. The projector-camera model provides greater flexibility and control in the generation of stealthy and impactful adversarial attacks. %, highlighting the importance of considering optical methods as a viable avenue for advancing the field of physical adversarial attacks.
\begin{figure}
    \centering
    \includegraphics[width=0.5\textwidth]{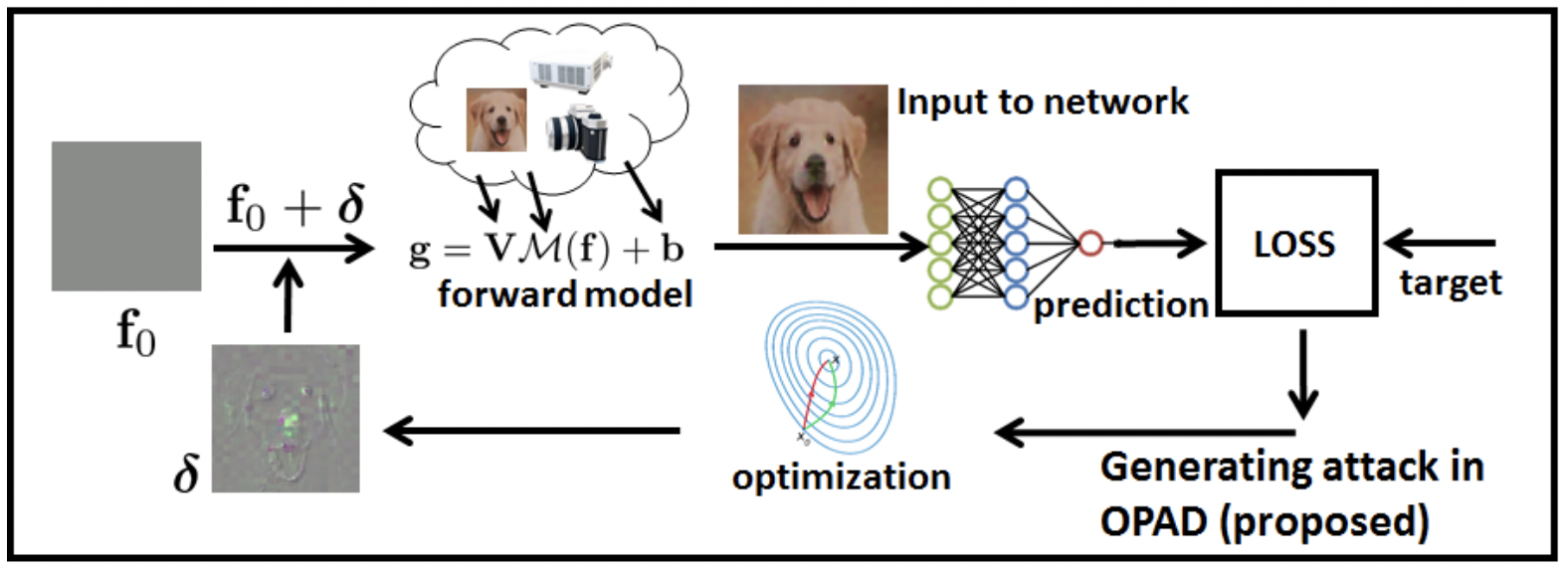}
    \caption{OPAD framework: The starting point, denoted as $f_0$, constitutes the uniform illumination. To introduce perturbation, they incorporate $\delta$ into $f_0$. Subsequently, the interplay of the projector and the scene comes into play, facilitated by the radiometric response and the spectral response, respectively (Figure adapted from \cite{OPAD}).}
    \label{fig:opad}
\end{figure}

\textbf{ProjectorAttack}: Man et al. \cite{projectorattack} conducted physical attacks by employing a light source projected across the entire image. The authors tackled the problem by optimizing the color distribution of the emitted light, specifically focusing on the means and standard deviations of a Gaussian distribution. In real-world scenarios, they adjusted the color values (RGB) of the projector to emit light onto the displayed image and captured photographs for evaluation. Through detailed case studies, the authors provided compelling evidence showcasing the effectiveness of these physical attacks. %This research sheds light on the potential of utilizing projected light as a means of generating impactful adversarial perturbations in practical physical attack scenarios.

By leveraging the inherent properties of light, a light-based approach has been devised to deceive classifiers without requiring any physical contact with the targeted object. Among the notable methods in this category, \textbf{OPAD (Object Physical Adversarial Device)} \cite{OPAD} stands out. OPAD utilizes structured lighting techniques to modify the appearance of the intended object. This attack system comprises a cost-effective projector, a camera, and a computer, facilitating the manipulation of real-world objects in a single shot. OPAD exhibits versatility, enabling the execution of various types of attacks, including targeted, untargeted, white-box, and black-box attacks. The OPAD approach demonstrates the potential of employing structured lighting techniques to generate powerful and effective adversarial perturbations that can deceive classifiers in a contactless manner. OPAD underwent a quantitative experiment to evaluate its robustness against different image transformations, including translation, zoom, and variations in camera ISO settings. The experimental results provide insights into the performance of OPAD under various conditions. It is observed that OPAD exhibits robustness against most image transformations, with the exception of scenarios where the object is significantly zoomed out to a considerable distance or specific camera ISO settings are applied. Notably, when the object is zoomed out extensively or when using high ISO settings, such as 3200 ISO, a significant number of pixels may become saturated. This finding sheds light on the limitations of OPAD when subjected to extreme zoom and high ISO conditions, suggesting potential challenges in maintaining its effectiveness under such circumstances.

\begin{figure*}
    \centering
    \includegraphics[width=0.8\textwidth]{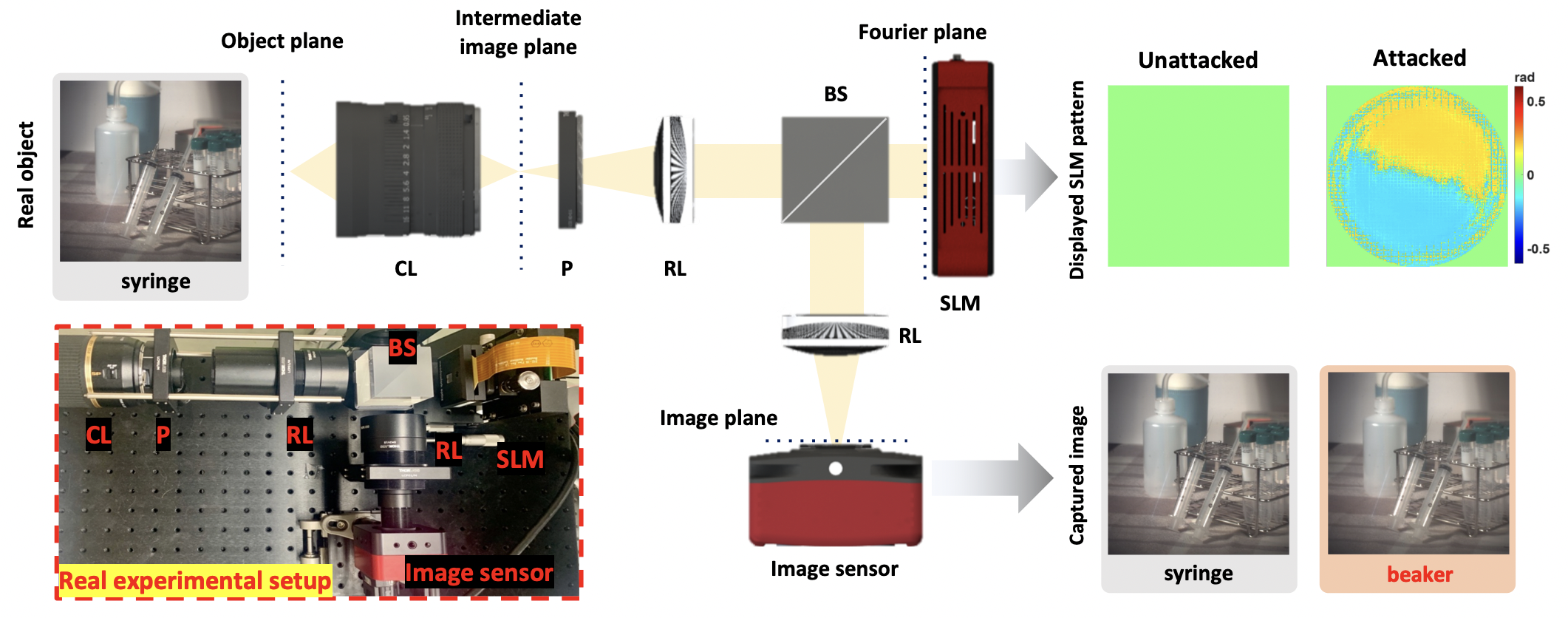}
    \caption{ Illustration of the proposed optical adversarial attack system. It includes a phase modulation module comprising a polarizer, relay lens, beam splitter, and Spatial Light Modulator (SLM) incorporated into the photography system. Two images are obtained: the unattacked image captured without phase modulation and the attacked image captured with adversarial phase modulation. When these acquired images are classified using a deep model, the unattacked image is correctly classified, while the attacked image is misclassified. The system components are labeled as follows: CL: camera lens; P: polarizer; RL: relay lens; BS: beam splitter (Figure adapted from \cite{slmattack})}
    \label{fig:slmattack}
\end{figure*}

In parallel to the OPAD method \cite{OPAD}, Kim et al. \cite{slmattack} proposed the \textbf{SLMAttack} where they employed a spatial light modulator (SLM) to manipulate the phase of light in the Fourier domain, resulting in an adversarial attack. The authors utilized gradient-based algorithms to optimize the parameters of the modulator, enabling them to achieve their desired effect by precisely controlling the phase of the light (See Figure \ref{fig:slmattack}).

Another noteworthy contribution in the field of physical adversarial attacks is \textbf{AdvLB} \cite{AdvLB}, introduced by Duan et al. This approach exploits laser beams as a versatile tool for both digital and physical adversarial attacks, offering the attacker flexibility and rapid execution. The simplicity of using a laser pointer makes AdvLB a potential widespread threat.

Following shortly after, Zhong et al. proposed an effective adversarial attack method known as \textbf{adversarial shadow} \cite{shadow}, which leverages a natural phenomenon. This technique exploits shadows to generate adversarial examples that are both digitally and physically realizable. Instead of relying on gradient-based optimization algorithms, the approach incorporates optimization strategies based on particle swarm optimization (PSO) \cite{488968}. The researchers conducted comprehensive evaluations in simulated and real-world environments, highlighting the potential danger posed by shadows as a viable attack vector. %These contributions expand the arsenal of adversarial attack techniques and emphasize the need for robust defenses against a diverse range of physical attack methods.

\textbf{SPAA (Scene Project-and-Attack)} \cite{spaa} is a projector-based attack technique that manipulates the environment lighting without the need for physically placing adversarial entities within the scene. This attack methodology takes a unique perspective by approximating the real Project-and-Capture process through the utilization of a deep neural network called PCNet. PCNet introduces additional constraints to ensure that the projected adversarial patterns are physically plausible, enhancing the stealthiness and realism of the attack. By leveraging SPAA, adversaries can effectively deceive computer vision systems by subtly modifying the lighting conditions in the scene without the need for direct physical interaction with the objects or the environment.

%stealthiness

%robustness
%OPAD (Object Physical Adversarial Device) \cite{OPAD}, an illumination-based attack technique, 

\begin{figure}[!htp]
\centering
\includegraphics[width=0.35\textwidth]{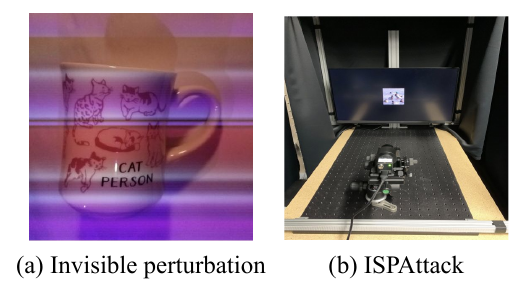} %, height=7cm
\caption{Examples of Imaging device manipulation-based physical adversarial examples against image classification: (a) Invisible perturbation \cite{Invisible-Perturbations} and (b) ISPAttack \cite{ISP}. }
\label{AE_camera_classification}
\end{figure}

Figure \ref{AE_camera_classification} provides visual examples of different attack methods based on the manipulation of imaging devices. While optical attacks offer advantages in terms of stealthiness and control, they are subject to various constraints. Factors such as the surface material of the target object, the saturation of colors, and the intensity of the scene's lighting can limit the effectiveness of optical attacks. These constraints need to be carefully considered when designing and deploying optical attack strategies.
%---------------------------------------------------------------------------------------
\subsection{Imaging Device Manipulation-based Attack} %Attack Imaging pipelines
%---------------------------------------------------------------------------------------

\begin{figure*}[!htp]
\centering
\includegraphics[width=0.7\textwidth]{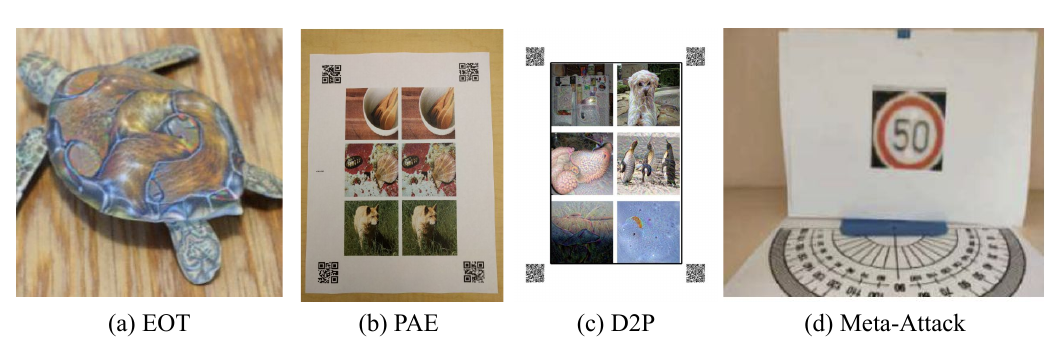} %, height=7cm
\caption{Examples of physical adversarial prints against image classification. (a) EOT \cite{EOT}, (b) PAE \cite{PAE}, (c) D2P \cite{D2P}, and (d) Meta-Attack \cite{meta-attack}. }
\label{AE_prints_classification}
\end{figure*}
%Invisible perturbation \cite{Invisible-Perturbations}
%Adversarial ISP \cite{ISP}

\textbf{Invisible perturbation}: By exploiting the unique characteristics of cameras, camera-based approaches have demonstrated successful adversarial attacks in the physical domain. Sayles et al. \cite{Invisible-Perturbations} made a significant discovery regarding the rolling shutter effect commonly found in consumer-grade cameras, including cell phones \cite{kim2020object}. They found that this effect can generate precise striping patterns that manifest in captured images. Building upon this finding, they developed techniques to modulate visible light in a manner that illuminates an object, leading to misclassification by deep learning camera-based vision classifiers, all while remaining completely imperceptible to human observers. These imperceptible perturbations are commonly referred to as invisible perturbations. The research conducted by Sayles et al. pioneers the exploration of camera-based adversarial attacks, harnessing the rolling shutter effect and utilizing it to generate impactful yet undetectable perturbations that can deceive deep learning-based camera vision classifiers.

\textbf{ISP}: Expanding on the influence of camera optics and the image processing pipeline, Phan et al. \cite{ISP} pioneered a novel method for attacking cameras by targeting specific Image Signal Processing (ISP) configurations. It is worth mentioning that ISPs play a crucial role in converting RAW measurements into RGB images. As hardware ISPs are non-differentiable, the researchers proposed an optimization technique that utilizes a local proxy network to approximate the behavior of the ISP. With this approach, they effectively crafted an adversarial attack that deceived the classifier under a specific camera ISP setting, while leaving other ISPs unaffected. This work highlights the significance of understanding and manipulating ISP configurations as a means to launch targeted adversarial attacks on camera systems, opening up new avenues for exploring the vulnerabilities of image processing pipelines in the context of visual recognition.

%robustness
Invisible Perturbations \cite{Invisible-Perturbations} exploit the rolling shutter effect to conduct physical adversarial attacks. Nevertheless, the efficacy of the rolling shutter effect may be influenced by factors such as camera settings and lighting conditions. To achieve robust attacks in the digital realm, Invisible Perturbations utilizes a differentiable analytical model that emulates the environmental and camera imaging conditions associated with the radiometric rolling shutter effect. This approach enables the accurate representation and manipulation of the rolling shutter effect, thereby facilitating the generation of more reliable and consistent adversarial perturbations. By employing a differentiable model, Invisible Perturbations enhances the ability to optimize and fine-tune the adversarial perturbations, contributing to the robustness and effectiveness of the attack strategy in the digital space.
%------------------------------------
\subsection{Physical Adversarial Prints}
%------------------------------------

%--------------------------
\subsubsection{Printed Objects-based Attacks}
%--------------------------
%EOT \cite{EOT}

%To ensure the effectiveness of physical adversarial attacks from various viewpoints, Athalye et al. \cite{EOT} introduced a groundbreaking approach that utilizes affordable, commercially available 3D printing technology to create 3D physical-world adversarial objects. This pioneering work sheds light on the practical implications of employing 3D adversarial objects in real-world systems based on deep neural networks (DNNs). In their experiments, the researchers fabricated a 3D adversarial turtle that consistently misleads the classifier, causing it to classify the turtle as a rifle, regardless of the angle or viewpoint from which the turtle is observed. This study emphasizes the importance of considering the three-dimensional aspects in physical-world adversarial attacks and underscores the potential vulnerabilities of DNN-based systems in real-world scenarios. The use of 3D printing technology opens up new avenues for exploring adversarial attacks that can have a significant impact on real-world systems, necessitating further research and countermeasures to enhance the robustness of DNNs against such attacks.
Athalye et al. \cite{EOT} have made significant strides in the realm of physical adversarial attacks from multiple viewpoints. They introduced a groundbreaking approach that leverages affordable, commercially available 3D printing technology to create 3D physical-world adversarial objects. This work sheds light on the practical implications of utilizing 3D adversarial objects in real-world systems governed by deep neural networks (DNNs). In their experiments, the researchers constructed a 3D adversarial turtle that consistently misleads the classifier, causing it to mistakenly classify the turtle as a rifle, irrespective of the angle or viewpoint from which the turtle is observed. This study emphasizes the paramount importance of considering the three-dimensional aspects when dealing with physical-world adversarial attacks, thereby highlighting the potential vulnerabilities of DNN-based systems in real-world scenarios. By employing 3D printing technology, this research opens up new avenues for exploring adversarial attacks that can profoundly impact real-world systems. Consequently, further research and the development of robust countermeasures are imperative to enhance the resilience of DNNs against such attacks. %The insights gained from this work underscore the significance of proactive measures to safeguard against physical adversarial attacks, ensuring the reliable and secure operation of DNNs in real-world applications.

\subsubsection{Printed Images-based Attacks}
%-----------------------------
%PAE \cite{PAE}
%D2P \cite{D2P}
%ABBA \cite{abba}

%RogueSigns \cite{roguesigns} (TSR)
%-----------------------------
\begin{figure*}[!htp]
\centering
\includegraphics[width=0.8\textwidth]{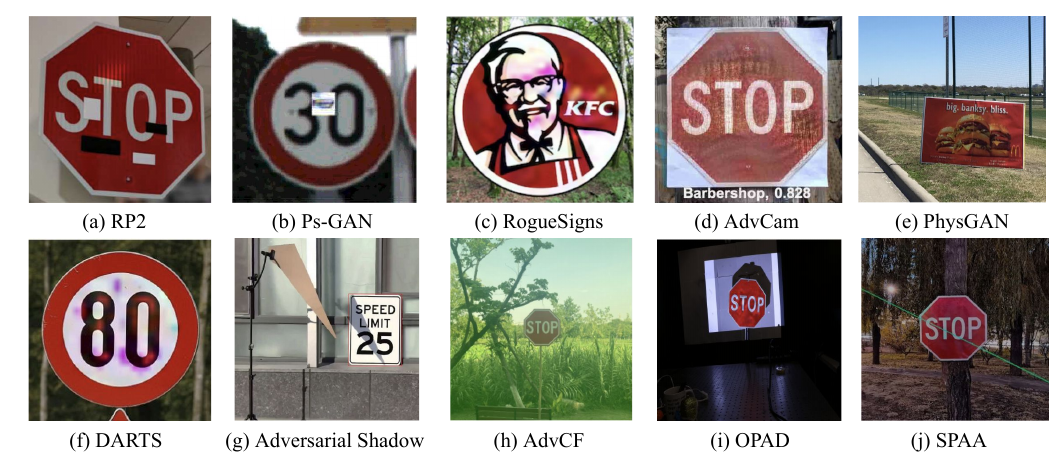} %, height=7cm
\caption{Examples of physical adversarial examples against traffic sign recognition: $RP_2$ \cite{RP2}, Ps-GAN \cite{PSGAN}, RogueSigns \cite{roguesigns}, AdvCam \cite{advcam}, PhysGAN \cite{physgan}, DARTS \cite{darts}, and Adversarial shadow \cite{shadow}. }
\label{AE_TSR}
\end{figure*}
\textbf{PAE}: Kurakin et al. \cite{PAE} conducted an experiment where they employed adversarial images captured using a cell-phone camera as input to deceive the ImageNet Inception v3 image classifier \cite{7780677}. They demonstrated that by selectively modifying a limited number of pixels in the image, the resulting printed and subsequently photographed image could still successfully mislead the classifier. This research highlights the potential vulnerability of image classifiers to adversarial attacks, even when the modifications are made within a restricted pixel space. %It underscores the need for robust defenses against such attacks to ensure the reliability and security of image recognition systems.

\textbf{D2P}: Jan et al. \cite{D2P} conducted a comprehensive analysis to explore the distinctions between digital and physical adversarial images, leading to the development of a novel method called D2P. The researchers initiated their study by curating a large dataset of digital-physical image pairs by printing digital images and recapturing them. They formulated the transformation between the digital and physical domains as an image-to-image translation problem and employed either a Pix2Pix model \cite{isola2018imagetoimage} or a CycleGAN \cite{zhu2017unpaired} to learn the mapping between these domains. To enhance the robustness of the adversarial perturbations in the simulated physical images generated by the models, the authors incorporated Expectation over Transformation (EOT). In physical attacks, the researchers printed the generated adversarial examples and captured them at regular intervals of 15 degrees within the range of -60 to 60 degrees. Experimental results demonstrate that their attack surpasses the performance of EOT and remains effective at various visual angles. However, it is important to note that their approach relies on full-pixel manipulation, which may pose practical limitations in real-world scenarios where fine-grained control over individual pixels may not be feasible.

%Guo \cite{abba} introduced a novel motion-based attack known as \textbf{ABBA}, which leverages the blurring effect caused by object motion to deceive deep neural networks (DNNs). The author optimized specific transformation parameters for both the object and background to achieve the desired blurring effect. The background was extracted using a saliency detection model, resulting in a binary mask that guided the blurring process. To enhance the robustness of the adversarial examples, a spatial transformer network \cite{jaderberg2015spatial} was employed. In physical attacks, real-world images of the object were captured, and the adversarial blur examples were generated using the proposed method. The camera was then moved based on the optimized transformation parameters to capture the resulting blurred image. Experimental results demonstrated that the captured blurred images achieved a success rate of 85.3\%, highlighting the effectiveness of the ABBA attack in exploiting motion-induced blurring for adversarial purposes.
Guo et al. \cite{abba} introduced a motion-based attack called \textbf{ABBA}, which capitalizes on the blurring effect caused by object motion to deceive deep neural networks (DNNs). The author optimized specific transformation parameters for both the object and background to achieve the desired blurring effect. By utilizing a saliency detection model, the background was extracted, resulting in a binary mask that guided the blurring process. To bolster the resilience of the adversarial examples, a spatial transformer network [Reference for Spatial Transformer Network] was employed. In physical attacks, real-world images of the object were captured, and the proposed method was employed to generate adversarial blur examples. Subsequently, the camera was moved based on the optimized transformation parameters to capture the resulting blurred image. Experimental results showcased a success rate of 85.3\%, clearly highlighting the efficacy of the ABBA attack in exploiting motion-induced blurring for adversarial purposes. %This research demonstrates a significant advancement in understanding and utilizing motion-based attacks to deceive DNNs, underscoring the importance of addressing such vulnerabilities in real-world systems.
%-----------------------------
\subsubsection{Image-based Attacks}
%-----------------------------
%Meta-Attack \cite{meta-attack}  (GAN-based)
%DARTS \cite{darts} (TSR)
%Advcam \cite{advcam} (TSR)
%PhysGAN \cite{physgan} (TSR)  (GAN-based)
%-----------------------------
%%%%%%%%%%%%%%%%%%%%%%%%%%%%%%%%%%%%%
\begin{figure*}[!htp]
\centering
\includegraphics[width=0.7\textwidth]{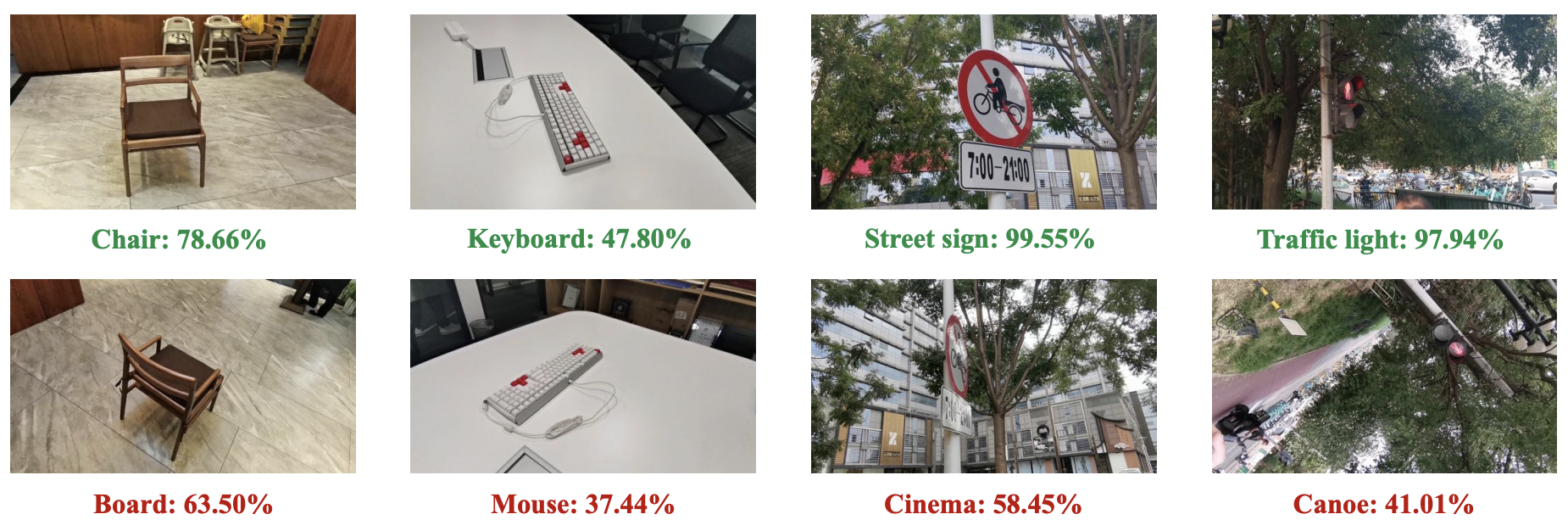} %, height=7cm
\caption{ An illustration of adversarial viewpoints. Top row: images captured from natural viewpoints are shown, and these images are accurately classified. In contrast, Bottom row displays the same objects captured from adversarial viewpoints identified by the ViewFool method. In this case, the images from adversarial viewpoints are misclassified, highlighting the susceptibility of the system to viewpoint-based adversarial perturbations (Figure adapted from \cite{viewfool}).}
\label{fig:viewfool}
\end{figure*}
%%%%%%%%%%%%%%%%%%%%%%%%%%%%%%%%%%%%%
\textbf{DARTS} \cite{darts} introduces a targeted adversarial example generation technique using out-of-distribution images. It explores two variants of the Out-of-Distribution attack: the Logo attack and the Custom Sign attack. In the Logo attack, commonly encountered logos are manipulated to resemble traffic signs, leading to high-confidence detection and classification. The Custom Sign attack involves the creation of custom signs that are adversarial, starting from blank sign templates. By applying masks resembling graffiti or text to these blank signs, adversarial traffic signs can be seamlessly embedded within graffiti-like objects in the environment. Furthermore, the In-Distribution attack modifies images of actual traffic signs using imperceptible perturbations, resulting in misclassification as different traffic signs. These approaches demonstrate the effectiveness of DARTS in generating targeted adversarial examples with practical applications in manipulating traffic sign recognition systems.

%\textbf{AdvCam} \cite{advcam} presents a novel technique for image perturbation that operates in the style space. This approach combines the principles of neural style transfer and adversarial attacks to generate adversarial perturbations that seamlessly integrate into the visual style of an image. By employing AdvCam, perturbations such as rust-like spots can be added to a Stop Sign, allowing them to blend naturally with the surrounding environment. It is important to note that these methods are most effective in static scenes where the context remains relatively constant. However, their efficacy may be limited when applied to dynamic environments or scenes with significant temporal variations, as these conditions may disrupt the consistency and effectiveness of the perturbations.
The \textbf{AdvCam} technique, introduced in \cite{advcam}, introduces a novel approach for image perturbation that operates in the style space. This method combines the principles of neural style transfer and adversarial attacks to generate adversarial perturbations that seamlessly blend into the visual style of an image. For instance, AdvCam can add perturbations such as rust-like spots to a Stop Sign, making them appear natural and inconspicuous in the surrounding environment. It is essential to recognize that the effectiveness of AdvCam and similar methods is most prominent in static scenes where the context remains relatively constant. In such scenarios, the perturbations can maintain their consistency and deception. However, the efficacy of these techniques may be limited when applied to dynamic environments or scenes with significant temporal variations. In such cases, the changing context may disrupt the seamless integration of the perturbations, reducing their effectiveness in deceiving deep neural networks. Despite these limitations, AdvCam represents a notable advancement in the realm of adversarial attacks, providing insights into how image perturbations can be strategically introduced while maintaining visual coherence. As research progresses, addressing the challenges associated with dynamic scenes and temporal variations could lead to more robust and versatile applications of such techniques.

\textbf{PhysGAN} \cite{physgan} introduces a framework based on Generative Adversarial Networks (GANs) to address the challenge of maintaining continuous attack effectiveness in physical-world scenarios. This approach enables the generation of a single physical-world-resilient adversarial example that can continuously deceive a steering model \cite{7410669, chitta2021neat, prakash2021multimodal} throughout an entire trajectory. The main objective of this work is to create a realistic image that incorporates adversarial perturbations and can be physically printed to replace the corresponding original object situated by the roadside. By ensuring the sustained misclassification of the steering model over time, PhysGAN aims to enhance the practicality and effectiveness of physical adversarial attacks in real-world scenarios. The utilization of GANs in PhysGAN allows for the generation of visually convincing adversarial examples that are robust to real-world variations and capable of evading the targeted model's defenses over an extended period of time.

Furthermore, \textbf{Meta-Attack} \cite{meta-attack} addresses physical adversarial attacks by formulating them as few-shot learning problems. This approach introduces a meta-learning algorithm that is both class-agnostic and model-agnostic, aiming to enhance the generalization capability of generative attack models. By leveraging meta-learning principles, Meta-Attack aims to improve the effectiveness and adaptability of physical adversarial attacks across diverse target models, irrespective of their specific classes. This methodology enables the efficient generation of adversarial examples that can deceive multiple models, thereby serving as a valuable tool in the realm of physical adversarial attacks. The meta-learning algorithm in Meta-Attack facilitates the learning of transferable knowledge and the efficient adaptation to new target models, enhancing the overall robustness and versatility of physical adversarial attacks.

%------------------------

PhysGAN \cite{physgan} endeavors to generate a single realistic adversarial example that can be physically printed and used to replace the corresponding original roadside object. The framework employs a generative adversarial network (GAN) architecture, where the discriminator assesses the visual dissimilarity between the generated adversarial roadside sign and the original sign. Simultaneously, the generator is encouraged to produce an example that is visually indistinguishable from the original sign. This strategy aims to make it challenging for individuals to detect the presence of adversarial samples, as they closely resemble the authentic objects in appearance. By achieving a high degree of visual similarity, PhysGAN enhances the stealthiness and effectiveness of physical adversarial attacks in real-world scenarios.

%stealthiness

%Neural style transfer is a widely-used technique for transferring the style of one image to another, often treated as an image transformation task \cite{tang2021attentiongan, gu2018arbitrary}. Similarly, the generation of adversarial patches can be considered as an image transformation problem. Building upon this notion, AdvCam \cite{advcam} introduces a flexible adversarial camouflage approach specifically tailored for crafting and concealing adversarial examples. By leveraging the principles of neural style transfer, AdvCam enables the generation of substantial perturbations that enhance the effectiveness of adversarial attacks. These perturbations are strategically disguised as natural elements, such as snowflakes, to seamlessly blend into the surrounding context, particularly in the context of attacking road signs. This camouflage technique significantly enhances the stealthiness of the adversarial examples, rendering them more difficult to detect and mitigate.
Neural style transfer is a widely-utilized technique for transferring the visual style of one image to another, often employed as an image transformation task \cite{tang2021attentiongan, gu2018arbitrary}. Building on this concept, AdvCam [\cite{advcam} introduces a flexible adversarial camouflage approach explicitly designed for crafting and concealing adversarial examples. By leveraging the principles of neural style transfer, AdvCam enables the generation of substantial perturbations that enhance the effectiveness of adversarial attacks. These perturbations are cleverly disguised as natural elements, like snowflakes, allowing them to seamlessly blend into the surrounding context, particularly in the context of attacking road signs. As a result, this camouflage technique significantly enhances the stealthiness of the adversarial examples, making them more challenging to detect and mitigate. AdvCam's strategy represents a significant step forward in the realm of adversarial attacks, offering a novel way to create inconspicuous and potent adversarial perturbations. By mimicking the appearance of natural elements, these perturbations can effectively deceive deep neural networks, posing potential challenges for the security and robustness of real-world systems. The utilization of neural style transfer in crafting adversarial camouflage exemplifies the versatility and creative potential of AI-based techniques for both constructive and potentially harmful purposes. As such, ongoing research is essential to devise robust defense mechanisms against such stealthy attacks and ensure the reliability of deep neural networks in various applications.

%PhysGAN \cite{physgan} 
%Meta-Attack \cite{meta-attack} are GAN-based attacks

%------------------------------------
\subsection{Position-based Attacks}
%------------------------------------
\textbf{ViewFool} presents a distinct approach to generating adversarial perturbations by focusing on the existence of adversarial viewpoints in the real world, where capturing an image from a specific viewpoint is adequate to deceive the model \cite{viewfool}. To discover such viewpoints, the authors consider the camera's coordinates and orientation (yaw, roll, pitch) in the world axis as optimization variables. They utilize NeRF, a neural radiance field model, to render the object based on the searched coordinates and orientation. In physical attacks, the authors capture images from these adversarial viewpoints and assess the attack performance. Experimental findings demonstrate the susceptibility of DNN models to adversarial viewpoints, highlighting the potential security implications in real-world scenarios.
%%%%%%%%%%%%%%%%%%%%%%%%%%%%%%%%%%%%%

\begin{table*}[!htp]
\centering
  \caption{ Comparison of attack methods against \textbf{Classification} tasks. MC – Misclassification; HA – Hiding objects; P – Adding phantom objects; AL - Altering objects. F – Anywhere in the frame; O – On the target object(s); S – On the sensor (i.e., camera); NO – Creating new object to fool the model; B: Blurring the image.}
  \label{Table:Classification_comparison}
  \begin{tabular}{lcccccc}
    \toprule
       \textbf{Attack}   & \textbf{Attack goal}  & \textbf{Placement} & \textbf{Consider changing view point} & \textbf{Test in physical domain} & \textbf{Test transferability}\\
    \midrule 
          $RP_2$ \cite{RP2} & MC &  O  & Trained for different distances/angels   &  \checkmark     & \checkmark   \\  
          EOT \cite{EOT}   &  MC   & O & Trained for different viewpoints & \checkmark  & $\times$\\          
          LaVAN \cite{lavan}  &  MC   & F  &  $\times$ & $\times$  & \checkmark\\
          Adv-watermark \cite{advwatermark} & MC   &  F & $\times$ &   $\times$ &  \checkmark \\
          %                                      &       &  & Trained for different distances/angels &   &\\
          GoogleAp \cite{googleap}   &   MC    & F & $\times$ & \checkmark  & \checkmark \\
           D2P \cite{D2P}    &  MC     & F & Trained for different angels & $\times$   & \checkmark\\ 
           ACS \cite{ACS}    &  MC     & S &  Trained for different angels  & \checkmark  & $\times$\\ 
           DARTS \cite{darts}   & MC   & O & Trained for different distances/angels   &  $\times$ & $\times$\\ 
           ACO Attack\cite{ACO}  &  MC  & O &  Trained for different distances/angels & \checkmark   & $\times$\\ 
           ACO2 Attack\cite{ACO2}  &  MC  & O &  Trained for different distances/angels & \checkmark   & $\times$\\ 
           ABBA \cite{abba}   &   MC    & B & $\times$ & \checkmark  &\checkmark\\ 
           DAS \cite{DAS}    &   MC    & O & Trained for different angels  & \checkmark  &\checkmark\\ 
          %           &       &  &  &   &\\ 
          Meta-attack \cite{meta-attack}  &   MC  & F & $\times$  & $\times$   & \checkmark\\ 
          AdvCam \cite{advcam}    &   MC  & O & $\times$ & \checkmark  & $\times$\\ 
          OPAD \cite{OPAD}         &    MC   & O & $\times$ & \checkmark   & $\times$\\ 
          FakeWeather \cite{fakeWeather}   &   MC    & S & $\times$ & $\times$  & $\times$\\
          AdvRain \cite{advrain}      &   MC    & S & $\times$ & $\times$  & $\times$\\
          AdvLB  \cite{AdvLB} & MC  & O &  $\times$   &  \checkmark    & $\times$  \\ 
          Adversarial Shadow \cite{shadow} & MC  & O &  $\times$   &  \checkmark    & $\times$  \\ 
          RogueSigns \cite{roguesigns}   & MC   & NO & Trained for different angels & \checkmark  & $\times$ \\
          AdvCF \cite{advcf} & MC &  S  & Trained for different angels  & \checkmark  &  \checkmark  \\
          Copy/PasteAttack  \cite{pasteattack} & MC &  O, F  & Trained for different angels  & \checkmark  &  \checkmark  \\
          LightAttack \cite{LightAttack} & MC &   O  & $\times$  & \checkmark   &  $\times$  \\
         ProjectorAttack \cite{projectorattack} & MC &   S  & $\times$  & \checkmark   &  $\times$  \\
         SLMAttack \cite{slmattack} & MC &   S  & $\times$  & \checkmark   &  \checkmark  \\
         SPAA \cite{spaa} & MC &   O  & Trained for different angels  & \checkmark   &  \checkmark\\

  \bottomrule
\end{tabular} %}
\end{table*}

\begin{table*}[!htp]
\centering
  \caption{Physical adversarial attacks against \textbf{Classification} tasks: Attacker's knowledge, Robustness technique, Stealthiness technique, Physical test type, and Space.}
  \label{Table:Classification_info}
  \begin{tabular}{llllll}
    \toprule
       \textbf{Attack}  & \textbf{Attacker’s}  & \textbf{Robustness} & \textbf{Stealthiness}  & \textbf{Physical}  & \textbf{Space} \\
         & \textbf{Knowledge}  & \textbf{Technique} & \textbf{Technique}  & \textbf{test type}  &  \\
    \midrule 
            PAE \cite{PAE}          & White-box  & -   &  Lp norm & Static &  2D \\
            $RP_2$  \cite{RP2}             & White-box  & D2P &  Lp norm  &   Static & 2D \\
            EOT \cite{EOT}                 & White-box  & EOT &  Lp norm  &   Static & 2D\\
            LaVAN \cite{lavan}             & White-box  & -   &  -      &    Static & 2D \\
            Adv-watermark \cite{advwatermark} & Black-box  &  -  & Uses existing logos &  Static & 2D \\
            GoogleAp \cite{googleap}       & White-box  & EOT &  Lp norm  &    Static & 2D \\
            D2P  \cite{D2P}                & White-box  & EOT, D2P     & -  &   Static & 2D \\
            ACS \cite{ACS}                 &  White-box & -   &    -   &    Static & 2D \\
            PS-GAN \cite{PSGAN}            & White-box  & -   & GAN     &   Static & 2D \\
            DARTS\cite{darts}              &  White \& Black-box  & EOT & - &  Static & 2D \\
            ACO Attack\cite{ACO}           & White-box  & EOT & Style loss  &    Static & 2D \\
            ACO2 Attack\cite{ACO2}         & White-box  & EOT & Style loss  &    Static & 2D \\
            ABBA \cite{abba}               & White-box  & EOT & -  &    Static & 2D \\
            AdvCam \cite{advcam}           & White-box  & EOT, TV & Style loss, Content loss  & Static & 2D \\
            PhysGAN  \cite{physgan}        &  White-box & D2P &  GAN & Dynamic & 2D \\
            Meta-Attack \cite{meta-attack} & White-box  & EOT  & GAN &    Static &  2D \\
            DAS \cite{DAS}                 & Black-box  & EOT & Attention evasion loss & Static  & 2D \\
            TnT attack \cite{tntattack}    & White-box  & - &  GAN &   Static & 2D \\
            Invisible perturbations \cite{Invisible-Perturbations} &  White-box & EOT & - & Static & 2D \\
            Adversarial ISP  \cite{ISP}    & White-box  & - & - &   Static & 2D \\
            OPAD  \cite{OPAD}              &  White-box & EOT & - &   Static & 2D \\
            AdvLB  \cite{AdvLB}            &  Black-box & - &  - &   Static & 2D \\
            Adversarial Shadow \cite{shadow} & Black-box  & EOT &  - & Static & 2D \\
            AdvCF \cite{advcf}             & Black-box  & Print loss & - & Static &  2D   \\
            FakeWeather \cite{fakeWeather} &  Black-box & - & -  &   Static & 2D \\
            AdvRain  \cite{advrain}        &  White-box & EOT & Pattern loss  &    Static & 2D \\
            RogueSigns  \cite{roguesigns}  &  White-box & EOT & -  &    Dynamic & 2D \\
            Copy/PasteAttack  \cite{pasteattack}  &  White-box & EOT, TV & GAN  &    Static & 2D \\
            ViewFool  \cite{viewfool} & White-box & - & - & Static & 3D \\
            LightAttack \cite{LightAttack} & Black-box  &  EOT  & -  &  Static   & 2D \\
            ProjectorAttack \cite{projectorattack} & White-box   &  -  &   -  &  Static  & 2D \\
            SLMAttack \cite{slmattack} &   White-box   &  -  &   -    &   Static      &   2D    \\
            SPAA \cite{spaa}   & White-box & D2P & PCNet  & Static   & 3D\\
  \bottomrule
\end{tabular} %}
\end{table*}

\begin{table*}[!htp]
\centering
  \caption{Attacks on \textbf{Classification} task, Datasets, networks, and Code.}
  \label{Table:Classification_dataset}
  \begin{tabular}{llll}
    \toprule
       \textbf{Attack}   & \textbf{Dataset} & \textbf{Network}  & \textbf{Code} \\
    \midrule 
         PAE      &  ImageNet  & Inception v3  &  https://github.com/Harry24k/AEPW-pytorch \\
         RP2      &  LISA, GTSRB  & Lisa-CNN, GTSRB-CNN  &  https://github.com/evtimovi/robust\_physical\_perturbations \\
         EOT      & Imagenet &      Inception v3    & https://github.com/prabhant/synthesizing-robust-adversarial-examples   \\
         LaVAN   &  Imagenet    &  Inception v3   &  -  \\
         Adversarial Patch    &  Imagenet  &  Inception v3, resnet50,   & -  \\
            (googleAp) &    &   xception, VGG16, VGG19   &    \\
        Adv-watermark &  Imagenet  & Alexnet, VGG19, SqueezeNet,   & https://github.com/jiaxiaojunQAQ/Adv-watermark.git  \\
               &        & Resnet101, InceptionV1, &   \\
               &        & InceptionV3 &   \\
         D2P   &  Imagenet  &  Inception v3   &  https://github.com/stevetkjan/Digital2Physical  \\
         ACS   &  Imagenet  & ResNet-50    &  https://github.com/yoheikikuta/adversarial-camera-stickers  \\
         PS-GAN   &  Imagenet, GTSRB  &  VGG-16   &   - \\
         DARTS   &  GTSRB, GTSDB,    &  Multi-scale CNN,    &  -  \\
                 &  ATD, Logo, &  Standard CNN   &    \\
                 &  Custom Sign  &     &    \\
         Adversarial ACO   &  RPC  &  ResNet-152, VGG-16, AlexNet &  -  \\
         Adversarial ACO2    &         &          &   \\
         AdvCam    &  Imagenet  &  VGG-19   &  https://github.com/RjDuan/AdvCam-Hide-Adv-with-Natural-Styles  \\
         PhysGAN     &  Udacity automatic,  &  NVIDIA Dave-2, Udacity Cg23,   &  -  \\
             &  DAVE-2,Kitti   &   and Udacity Rambo   &    \\
         Meta-Attack  & Imagenet, GTSRD   & VGG-16, VGG-19, ResNet-50    &  -  \\
         DAS     & Carla Simulator  &  inception-v3, vgg-19, resnet-152,    & https://github.com/nlsde-safety-team/DualAttentionAttack   \\
              &    &  DenseNet   &    \\
         
         Invisible perturbations    & ImageNet   &  ResNet-101   &  https://github.com/EarlMadSec/invis-perturbations  \\
         Adversarial ISP     &  Imagenet  &   Resnet-101  & -   \\
         %OPAD    & Imagenet   &  VGG-16, Resnet-50   &  -  \\
         OPAD    &   Imagenet, GTSRB &  VGG-16, Resnet-50, GTSRB-CNN  &  -  \\
         AdvLB     &  ImageNet  &  ResNet-50   &  https://github.com/RjDuan/Advlight  \\ 
         Adversarial Shadow     & LiSA, GTSRB   &  Lisa-CNN, GTSRB-CNN   & https://github.com/hncszyq/ShadowAttack   \\
         AdvCF     & ImageNet   &  ResNet-50   &  -  \\ 
         FakeWeather      &  CIFAR-10  &  LeNet-5, Resnet-32, CapsNet   &  -  \\
         AdvRain     &  Imagenet, Caltech-10  &   VGG-19, Resnet34  &  -  \\  
         ABBA     & NeurIPS’17 adversarial   & Inception v3, Inception v4,   &    https://github.com/tsingqguo/ABBA\\  
            &  competition dataset     &   Inception ResNet v2, Xception       &   \\
         RogueSigns        &    GTSRB     &    multi-scale CNN      & https://github.com/inspire-group/advml-traffic-sign  \\
         Tnt attack    & ImageNet   &  VGG-16, Inception-V3  &  - \\
                      &   &  WideResnet50  &  \\
         Copy/PasteAttack   &   ImageNet    &  ResNet50    &    https://github.com/thestephencasper/feature\_level\_adv  \\
        ViewFool & ImageNet    &   ResNet,  &     https://github.com/Heathcliff-saku/ViewFool \\
         &    & vision transformer    &    \\        
        LightAttack &  CIFAR-10   &  ResNet38  &   -  \\
        ProjectorAttack &  CIFAR-10   &   4 Conv layer followed   &   -  \\  
                    &        &   by two FC layers        &    \\
        SLMAttack   &  ImageNet       &   ResNet50, VGG16, MobileNetV3 &    -   \\
        SPAA    &  ImageNet       &    ResNet-18, VGG-16, Inception     & https://github.com/BingyaoHuang/SPAA \\
  \bottomrule
\end{tabular} %}
\end{table*}

%% file: detection.tex
%======================================
\section{Physical Attacks on Detection}
%======================================
\label{detection}

\begin{figure*}[!htp]
\centering
\includegraphics[width=0.7\textwidth]{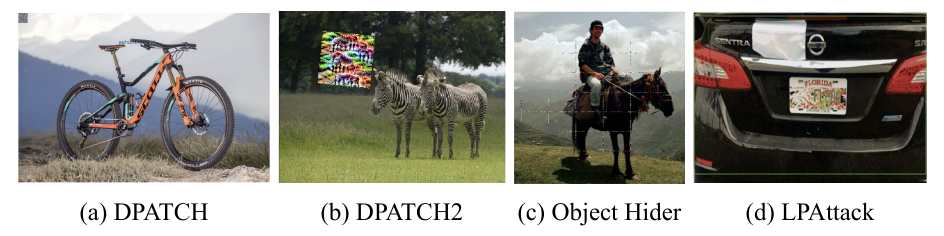} %, height=7cm
\caption{Examples of patch-based physical adversarial examples against object detection: (a) DPATCH \cite{dpatch}, (b) DPATCH2 \cite{dpatch2}, (c) Object hider \cite{objecthider}, and (d) LPAttack \cite{lpattack}.}
\label{AE_OD}
\end{figure*}
Physical attacks on detection encompass a category of adversarial attacks that aim to undermine the resilience of detection systems. These attacks capitalize on weaknesses inherent in the physical characteristics of input data, enabling the manipulation of detection outcomes. By introducing subtle perturbations to the physical properties of the input, an adversary can deceive the detection system and induce incorrect or misleading results. Such attacks pose a significant challenge to the reliability and effectiveness of detection systems in real-world scenarios.

Physical attacks on detection pose significant risks in real-world applications, with potentially severe consequences. In the context of autonomous vehicles, these attacks have the potential to cause dangerous situations or accidents by strategically manipulating road signs or traffic signals, deceiving the vehicle's object detection system. Similarly, in surveillance systems, attackers can exploit physical vulnerabilities to manipulate the appearance of objects or individuals, enabling them to evade detection and carry out malicious activities without being identified. These examples highlight the critical importance of developing robust countermeasures to mitigate the risks associated with physical attacks on detection systems.

%Mitigating the impact of physical attacks on detection necessitates the advancement of resilient algorithms and methodologies capable of effectively countering diverse forms of physical perturbations. One approach involves integrating sensor-based defenses, such as leveraging multi-modal sensing or incorporating temporal consistency, to enhance the system's capability to discern between legitimate and adversarial inputs. Furthermore, techniques like adversarial training, wherein models are trained using adversarial examples, can bolster the system's ability to withstand physical attacks by improving its robustness and adversarial resilience. These strategies collectively contribute to the development of more secure and reliable detection systems in the face of physical adversarial threats.

The initial advancements in deep neural networks for object detection were pioneered by Overfeat \cite{overfeat}, which combined convolutional neural networks (CNNs) with a sliding window technique. This approach, known as Regions using Convolutional Neural Networks (R-CNN), employed a search algorithm to generate region proposals and a CNN to classify each region. However, the region proposal algorithm in R-CNN was computationally slow, limiting its real-time applicability.
To address this limitation, subsequent studies introduced improved algorithms such as Fast R-CNN \cite{fast} and Faster R-CNN \cite{faster}. These approaches replaced the inefficient region proposal algorithm with a CNN, leading to significant speed improvements.

In the field of object identification, these algorithms adopt a two-stage process involving region proposals and subsequent classification. However, recent advancements in "single shot detectors" have emerged, simplifying the process by performing object detection and classification in a single pass. Examples of such detectors include SSD \cite{ssd} and YOLO \cite{yolo} and its subsequent versions YOLOv2 \cite{yolov2}, YOLOv3 \cite{yolov3}, YOLOv4 \cite{yolov4}, YOLOv5 \cite{yolov5}, YOLOv6 \cite{yolov6}, YOLOv7 \cite{yolov7} and YOLOv8 \cite{yolov8}. These networks produce confidence ratings for both object localization and classification simultaneously, allowing for faster processing without sacrificing accuracy.

Table \ref{Table:Detection_comparison} provides a comprehensive comparison of various adversarial attack methods in the detection task. It highlights the attack goals, patch placement strategies, consideration of changing viewpoints, testing in the physical domain, and transferability to other models.
Table \ref{Table:Detection_info} presents detailed information on adversarial attacks, including the attacker's knowledge level, robustness techniques, stealthiness techniques, physical test types, and space of operation.
Table \ref{Table:Detection_dataset} provides information on the datasets used, the networks evaluated, and the links to open-source code for the experiments conducted in the detection task.

%==============================================================================================
\subsection{Object Detection}

\subsubsection{Patch-based Attacks}
%DPATCH \cite{dpatch}
%DPATCH2 \cite{dpatch2} 
%Object hider \cite{objecthider}
%LPAttack \cite{lpattack}

\textbf{DPATCH}: Inspired by the concept of adversarial patch attacks in image classification, Liu et al. \cite{dpatch} expanded this approach to the domain of object detection. Their work aimed to attack the classification output of object detectors using adversarial patches. Through targeted and non-targeted attacks, they successfully demonstrated the vulnerability of object detectors to such patches. However, it is important to note that the effectiveness of their approach in physical adversarial attacks was not specifically evaluated in their study.

\textbf{DPATCH2}: In a related work \cite{dpatch2}, the authors proposed a patch that was trained to be randomly placed within the scene. This approach resulted in all existing objects in the image being missed by the detector, effectively evading detection. Although the implications of this method were discussed, the physical implementation and evaluation of the adversarial patch in real-world scenarios were not explicitly explored.

\textbf{Object Hider}: Zhao et al. \cite{objecthider} presented two novel algorithms for generating adversarial patches in the context of object detection. The first algorithm utilized heatmaps to guide the application of the patch, aiming to conceal specific objects of interest. The second algorithm employed a consensus-based approach, leveraging model ensemble strategies to generate effective adversarial patches. Notably, their proposed algorithms achieved a commendable seventh position in the Alibaba Tianchi adversarial challenge object detection competition. This demonstrates the efficacy of their methods in generating adversarial patches for object detection tasks.

\textbf{LPAttack}: Yang et al. \cite{lpattack} introduced a novel method for generating adversarial license plates to deceive SSD (Single Shot MultiBox Detector) in physical attacks. In contrast to previous patch-based approaches, the authors employed a mask to guide the perturbation region, which was optimized using the C\&W adversarial loss along with color regularization techniques, including NPS (Natural Patch Statistics) and TV (Total Variation) loss. These techniques aimed to ensure the effectiveness of the generated adversarial license plates and eliminate color discrepancies between the digital and physical domains. To enhance physical robustness, the optimization process incorporated EOT (Expectation Over Transformation). In the physical attack scenario, the authors fabricated adversarial license plates and affixed them to real cars. Experimental results demonstrated a significant degradation of 76.9\% in the detection performance of the captured images under both indoor and outdoor conditions. This highlights the vulnerability of SSD to adversarial license plates and the effectiveness of the proposed LPAttack method in physical attack settings.

%==============================================================================================
\subsection{Traffic Sign Detection}
\begin{figure*}[!htp]
\centering
\includegraphics[width=0.6\textwidth]{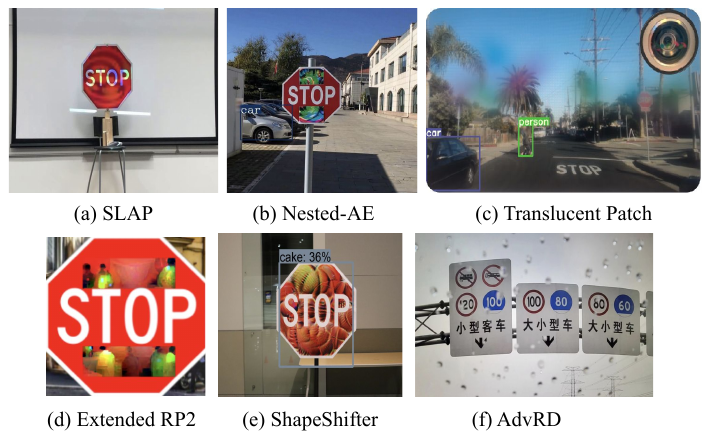} %, height=7cm
\caption{Examples of physical adversarial examples against traffic sign detection: (a) SLAP \cite{SLAP}, (b) Nested-AE \cite{nestedae}, (c) Translucent Patch \cite{translucent-patch}, (d) Extended RP2 \cite{extendedrp2}, (e) ShapeShifter \cite{shapeshifter}, and AdvRD \cite{advrd}. }
\label{AE_TSD}
\end{figure*}

Sign detection has witnessed significant advancements through the application of deep learning models. Convolutional neural networks (CNNs) have proven to be highly effective in accurately detecting and classifying signs in various domains, encompassing traffic signs, street signs, and informational signs. By leveraging the capabilities of deep neural networks, these models can automatically learn discriminative features from sign images, enabling them to generalize well across different lighting conditions, perspectives, and variations in sign appearances. Through training on large-scale datasets with annotated sign images, deep learning models can capture intricate patterns and relationships, endowing them with robustness and reliability in sign detection tasks. This progress in sign detection using deep learning models holds immense potential for enhancing the safety and efficiency of applications such as autonomous driving, smart cities, and intelligent transportation systems.

Traffic sign detection is a pivotal component of autonomous driving technology, as it involves the identification and recognition of road signage, including stop signs, lane lines, and other important traffic indicators, in various driving scenarios. The accuracy and reliability of traffic sign detection are crucial for ensuring safe and informed driving decisions. Incorrect recognition or misclassification of traffic signs can lead to driving violations and potentially result in severe car accidents. Given the significance of this domain, there has been growing interest in the field of adversarial attacks specifically targeted at traffic sign detection. These attacks aim to improve the robustness and resilience of detection algorithms in this critical area, with the goal of enhancing the overall security and performance of autonomous driving systems.

%%---------
\subsubsection{Patch-based Attacks}
%%----------
%Extended RP2 \cite{extendedrp2}
%Nested-AE \cite{nestedae}

\textbf{Extended RP2}: Eykholt et al. \cite{RP2} introduced the RP2 algorithm, which primarily focuses on attacking the task of traffic sign classification. Building upon this approach, Song et al. \cite{extendedrp2} further extended the RP2 algorithm to demonstrate proof-of-concept attacks on object detection networks, introducing a novel attack strategy known as the Disappearance Attack. This attack strategy encompasses two types of attacks: adversarial poster perturbation and adversarial stickers. The adversarial poster perturbation involves introducing perturbations to the entire Stop Sign, while the adversarial stickers method utilizes specially designed patches applied to the Stop Sign. The experiments conducted by the authors employed the YOLOv2 model \cite{yolov2}, resulting in an adversarial poster perturbation achieving an Adversarial Success Rate (ASR) of 85.6\% indoors, and adversarial stickers achieving an ASR of 85.0\% indoors. In outdoor environments, the ASR dropped to 72.5\% for adversarial poster perturbation and 63.5\% for adversarial stickers. Furthermore, the transferability of the attacks was evaluated using the Faster R-CNN object detector \cite{ren2016faster}, yielding an ASR of 85.9\% indoors and 40.2\% outdoors. These results demonstrate the effectiveness of the Disappearance Attack in adversarially manipulating object detection systems for traffic signs.

Certain methods aimed at improving the robustness of deep learning models are applicable to both attacks on object detectors and classifiers. For instance, the Expectation over Transformation (EOT) technique \cite{EOT}, originally proposed for image classification, can be leveraged to enhance the robustness of these models. In the context of physical attacks on traffic sign detection, Eykholt et al. \cite{RP2} adopted the EOT approach to address the variability in environmental conditions. Specifically, they introduced synthetic modeling of two key environmental factors: position (in the X-Y plane) and object rotation (in the Z plane). During the optimization process, in each epoch, the object is randomly placed and rotated to improve the model's robustness to changes in viewing angles. This approach aims to enhance the generalization capability of the model and its ability to handle variations in the physical environment, ultimately contributing to the resilience of the system against adversarial attacks.

%https://drive.google.com/drive/u/0/folders/1I4VLhJEuaINN0icUg-11HoVaIlS122Es

The \textbf{Nested-AE} (Nested Adversarial Examples) algorithm \cite{nestedae} is designed to enhance the robustness of adversarial attacks by incorporating multiple adversarial examples within a single instance. These nested examples are specifically crafted to target different distances or angles, effectively improving the attack performance across various positions. The proposed Nested-AE algorithm aims to enhance the adaptability of the adversarial examples to both long and short distances. Through this approach, the composite scheme achieves successful attacks against popular object recognition algorithms such as YOLO v3 and Faster-RCNN. The attacks are effective within a range of ±60° angle and distances ranging from 1m to 25m. The incorporation of nested adversarial examples in the algorithm enhances its ability to deceive the target models across different positions, demonstrating its potential for robust adversarial attacks.

%%--------------
\subsubsection{Sticker-based Attacks}
%%-------------
%Translucent patch \cite{translucent-patch}
%AdvRD \cite{advrd}
%Adversarial Rain \cite{adversarialrain}

Zolfi et al. \cite{translucent-patch} introduced a universal perturbation called the Translucent Patch, which is designed to deceive object detectors specifically for instances of a particular object class while maintaining the detection of other objects intact. The \textbf{Translucent Patch} is a colored translucent sticker that, when attached to the camera lens, disrupts the imaging process, enabling effective adversarial attacks. To evaluate the effectiveness of the Translucent Patch, the researchers conducted physical adversarial attacks on a Tesla Model X equipped with an advanced driving assistance system (ADAS). In scenarios involving traffic lights and stop signs, the Translucent Patch successfully deceived the ADAS, causing misclassification or failure to detect these critical objects. Additionally, when targeting the YOLOv5 detector, the Translucent Patch resulted in a significant failure rate of 42.27\% in detecting instances of stop signs. These findings highlight the potential risks faced by autonomous vehicles' ADAS, as a malicious attacker could exploit vulnerabilities using a simple sticker placed on the cameras' lenses, compromising the system's reliability and safety.

Enhancing the stealthiness of adversarial attacks on traffic sign detection models presents significant challenges. Physical attacks require noticeable perturbations to be perceptible to sensors like cameras, while traffic signs typically have uniform appearances with single patterns, making any added perturbations highly visible. To address this, researchers have explored novel approaches to achieve covert operations. For instance, Zolfi et al. introduced a camera-sticker-based physical adversarial attack on traffic sign detection \cite{translucent-patch}. Instead of modifying the target object directly, this attack involves affixing a colored translucent sticker onto the camera lens, rendering the attack imperceptible. Another strategy involves conducting the attack within an extremely short timeframe to minimize the exposure duration and reduce the likelihood of detection. These techniques aim to enhance the stealthiness of physical attacks on traffic sign detection models by utilizing unique adversarial forms. % and minimizing the window of vulnerability. Consequently, defending against such attacks becomes even more challenging, requiring the development of robust countermeasures.

\textbf{AdvRD} \cite{advrd} is an adversarial attack method that incorporates raindrops as a form of perturbation to deceive computer vision systems. Raindrops are simulated or physically generated and applied to the input images, aiming to disrupt the performance of object detection or recognition algorithms. The AdvRD attack leverages the visual occlusion and distortion caused by raindrops to manipulate the input images in a way that confuses the underlying computer vision models. By strategically placing raindrops on specific regions of the image, the attacker can introduce perceptible changes that mislead the target system into making incorrect predictions. To achieve this, AdvRD takes into account various factors, such as the size, density, and motion of raindrops, as well as their interaction with the scene's lighting conditions. The attack aims to find the optimal configuration of raindrop placement that maximizes the adversarial effect while maintaining the visual realism of the scene.

%Adversarial Rain \cite{adversarialrain} is a method that aims to manipulate the performance of deep neural network (DNN) perception models by introducing adversarial rain perturbations to input images. The attack leverages the effect of rain on image appearance to deceive the DNN models and cause misclassification or inaccurate predictions. These perturbations are carefully crafted to exploit the vulnerabilities of DNN perception models to rain-related distortions. By strategically placing rain-like patterns in specific regions of the image, the attacker aims to confuse the DNN model and induce incorrect predictions. To defend against the adversarialrain attack, defensive deraining techniques are developed. These techniques aim to remove or mitigate the effects of rain perturbations from the input images before feeding them into the DNN models. By applying deraining algorithms as a preprocessing step, the defensive deraining methods aim to restore the original appearance of the images and enhance the robustness of the DNN models against the adversarialrain attack.
%%--------------
\subsubsection{Light-based Attacks}
%%-------------

Lovisotto et al. introduced a light-based method called \textbf{Short-Lived Adversarial Perturbations (SLAP)} for physical attacks in the context of self-driving scenarios \cite{SLAP}. SLAP leverages a projector to display a specific pattern onto a Stop Sign, resulting in misclassification by the YOLOv3 and Mask-RCNN detectors, effectively deceiving the targeted object recognition. The experiments were conducted in a moving vehicle setting on a designated portion of a private road, where SLAP achieved an adversarial success rate (ASR) of over 77\%. This research highlights the potential vulnerability of self-driving systems to light-based attacks and underscores the importance of developing robust object detection mechanisms in autonomous driving technology. SLAP employs light-based attacks as a means to implement its technique. By leveraging a projector, SLAP is capable of projecting precise adversarial perturbations onto the targeted traffic sign at the speed of light. The instantaneous nature of this attack mechanism poses a significant challenge for human observers to detect the perturbations in real-time, thereby enhancing the overall stealthiness of the attack. SLAP incorporates a series of data augmentation methods alongside the Expectation Over Transformation (EOT) technique to enhance the robustness of its attack. These data augmentation methods include brightness transformations, perspective variations, rotation, aspect ratio adjustments, sign size variations, and grid size restrictions. By applying different brightness transformations, varying the camera angle, adding rotation, resizing the image, adjusting the sign size, and imposing a fixed grid structure, SLAP accounts for various environmental conditions that were previously disregarded. Through extensive empirical evaluations conducted in both indoor and outdoor settings, SLAP demonstrates its effectiveness in causing the targeted object to go undetected across a wide range of environments.
%%--------------
\subsubsection{Image-based Attacks}
%%-------------
%ShapeShifter \cite{shapeshifter}

\textbf{ShapeShifter} \cite{shapeshifter} is a robust physical adversarial attack technique specifically designed for the Faster R-CNN object detector. This method aims to generate adversarial examples that can deceive the detector while maintaining their effectiveness under various real-world conditions. By leveraging the Expectation over Transformation (EOT) technique, ShapeShifter accounts for the variability in the detector's behavior due to changes in environmental factors such as lighting, camera angles, and object orientations.

The attack process of ShapeShifter involves iteratively optimizing the adversarial perturbations to maximize their impact on the Faster R-CNN detector. It employs gradient-based optimization techniques to find the optimal perturbation that can cause misclassifications or false detections by the detector. Additionally, ShapeShifter incorporates data augmentation techniques such as brightness transformations, perspective variations, rotations, aspect ratio adjustments, sign size changes, and grid-based granularity to simulate different real-world scenarios and enhance the robustness of the adversarial perturbations.

\subsection{Lane Detection}

\textbf{Adversarial Markings} \cite{advmarkings} introduces a novel approach to physical adversarial attacks on autonomous driving systems by directly accessing the camera input and lane output images through reverse engineering of the Tesla Autopilot firmware. This unique approach enables a direct connection between the digital and physical domains, enhancing the robustness and effectiveness of attacks in real-world scenarios. By reverse engineering the firmware, Adversarial Markings gains insights into the underlying system and leverages this knowledge to manipulate the input and output images.

One notable feature of Adversarial Markings is the introduction of parameterized perturbations, which enable more efficient optimization of the adversarial examples. These parameterized perturbations provide a flexible and customizable framework for designing perturbations that can maximize the impact on the system's perception and decision-making processes. By optimizing these perturbations, Adversarial Markings identifies the most effective modifications to the camera input and lane output images, leading to successful adversarial attacks.

This reverse engineering-based approach has significant implications for the field of physical adversarial attacks on autonomous driving systems. It bridges the gap between digital simulations and real-world scenarios, allowing for more realistic and robust attacks. The utilization of parameterized perturbations further enhances the adaptability and effectiveness of the attacks, enabling targeted manipulation of the system's perception and decision-making capabilities. %The findings of Adversarial Markings highlight the potential vulnerabilities of autonomous driving systems and underscore the importance of developing robust defenses against such attacks.

%============================================================================================
%=============================================
\subsection{Person Detection} 
\begin{figure}[!htp]
\centering
\includegraphics[width=0.5\textwidth]{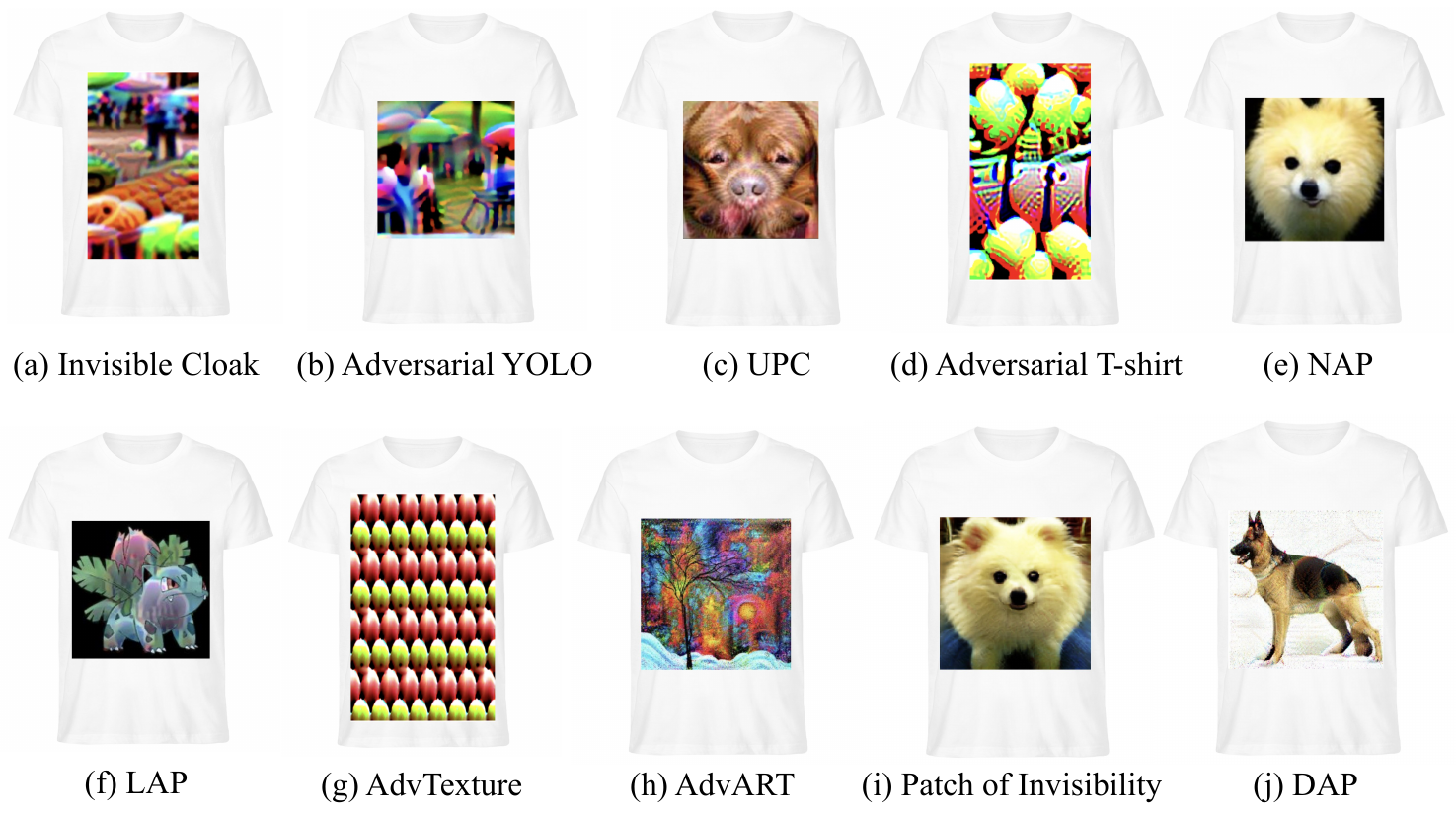} %, height=7cm
\caption{Adversarial T-shirts: (a) Invisible Cloak \cite{invisiblecloak}, (b) Adversarial YOLO \cite{Adversarialyolo}, (c) UPC \cite{UPC}, (d) Advesarial T-shirt \cite{adversarialtshirt}, (e) NAP \cite{NAP}, (f) LAP \cite{LAP}, (g) AdvTexture \cite{advtexture}, (h) AdvART \cite{advart}, (i) Patch of invisibility, and (j) DAP \cite{dap}.}
\label{clothing_PD}
\end{figure}

Person detection has undergone a transformative evolution with the advent of deep learning models, enabling the accurate and efficient identification of individuals in images and videos. Convolutional neural networks (CNNs), in particular, have demonstrated remarkable performance in this domain, leveraging large-scale annotated datasets to learn complex patterns and features associated with human presence. This enables CNN-based models to generalize effectively across diverse scenarios, including varying poses, scales, occlusions, and backgrounds, thus ensuring robustness in real-world applications.

The utilization of deep learning models for person detection holds significant implications in numerous domains, such as surveillance, crowd monitoring, social robotics, and autonomous systems. In the realm of surveillance, deep learning-based person detection systems play a vital role in enhancing security by providing real-time and accurate identification of individuals, assisting in crime prevention and investigation. Crowd monitoring applications benefit from these models by enabling efficient crowd analysis, tracking, and behavior understanding, thereby facilitating effective crowd management and ensuring public safety. In the field of social robotics, person detection enables robots to interact seamlessly with humans, navigate in human-centric environments, and collaborate on various tasks. Moreover, in autonomous systems like self-driving cars, accurate person detection is critical for pedestrian safety and informed decision-making based on their presence.

The adoption of deep learning models has significantly advanced state-of-the-art in person detection, offering robust solutions for identifying individuals across diverse scenarios. Ongoing research and development efforts continue to improve the performance of these models by leveraging advancements in deep learning architectures, dataset collection, and training techniques. As a result, the field of person detection is poised to further enhance safety, security, and situational awareness in a wide range of real-world applications.

The primary goal of physical adversarial attacks in person detection is to devise physical perturbations, such as patches or modified clothing, capable of effectively concealing individuals from automated surveillance cameras in real-world settings.

%------------------------------------
\subsubsection{Patch-based Attacks}
%------------------------------------
%Adversarial YOLO \cite{Adversarialyolo}
%Patch of Invisibility \cite{PatchOI}

The initial approach to conducting physical adversarial attacks on the person detection task is presented in \textbf{Adversarial YOLO} \cite{Adversarialyolo}. This method produces a compact adversarial patch with dimensions of 40cm × 40cm, which, when held by an attacker, can deceive the one-stage detector YOLOv2 \cite{yolov2}. Traditional deep neural network-based detectors possess the ability to predict bounding box positions (Vpos), object probabilities (Vobj), and class scores (Vcls) for a given input image. Adversarial YOLO employs a training procedure that minimizes Vobj and Vcls during the training phase, resulting in the detector disregarding instances of people (the target class). Leveraging the INRIAPerson dataset \cite{inria}, which contains numerous samples of people, Adversarial YOLO generates highly effective adversarial patches capable of achieving successful attacks in real-world scenarios.

At first, the proposed physical attacks aiming at fooling person detectors were generated without any constraints on the patch appearance. Their main focus was on performance and producing effective attacks. For instance, the work in \cite{Adversarialyolo} generates a printable adversarial patch (Adversarialyolo), attached to the person. Able to hide a person from the detector. The produced patch suffers from weak transferability, and the authors did not address issues like robustness to distance/distortions and didn't test detectors beyond Yolov2.

Initially, the early approaches in physical adversarial attacks targeting person detectors were primarily concerned with performance and the generation of effective attacks, without imposing specific constraints on the appearance of the adversarial patch. For example, in the work presented in \cite{Adversarialyolo}, an adversarial patch named Adversarialyolo was created, which could be printed and attached to a person, effectively concealing them from the detector. However, these early approaches exhibited limited transferability, meaning the generated adversarial patches did not generalize well across different detectors, and they did not explicitly address challenges such as robustness to varying distances or distortions. Furthermore, the evaluation of these attacks was primarily limited to the Yolov2 detector, without considering the performance on other person detection models.

In the context of adversarial YOLO, the authors introduced a series of random transformations applied to the adversarial patch before its application to the image. These transformations were designed to enhance the robustness of the generated adversarial patches. The operations included random rotations in both clockwise and counterclockwise directions (up to 20 degrees), random scaling (upscaling and downscaling), random noise addition for introducing subtle pixel variations, and random adjustments to brightness and contrast. By incorporating these random transformations, adversarial YOLO effectively calculated backward gradients and increased the resilience of the generated adversarial patches to variations in the physical environment.

On the other hand, the \textbf{Patch of Invisibility}, proposed by Zhang et al. \cite{PatchOI}, presents a novel approach for conducting naturalistic black-box adversarial attacks on object detectors. Unlike traditional attacks that modify specific regions of an image, the Patch of Invisibility aims to create a universal perturbation that can deceive different images when applied. This is achieved by employing optimization techniques and a pre-trained black-box object detector. The patch is iteratively optimized to minimize its visual impact while maximally deceiving the object detector. The optimization process takes into account the object detector's confidence scores and the patch's visual similarity to the surrounding scene.

The effectiveness of the Patch of Invisibility was evaluated on various object detection models, including Faster R-CNN \cite{ren2016faster} and YOLOv3 \cite{yolov3}. The experimental results demonstrated that the Patch of Invisibility successfully deceived the object detectors, leading to misclassifications and reduced detection performance. This approach showcases the potential of creating universal perturbations that can stealthily evade object detection systems.
%------------------------------------
\subsubsection{Camouflage Techniques}

\textbf{Invisible Cloak \& Invisible Cloak2}: In recent advancements in the field of adversarial attacks, authors in \cite{invisiblecloak2} proposed an approach involving the use of an invisibility cloak to deceive object detectors. Concurrently, Wu et al. \cite{invisiblecloak} introduced a wearable invisibility cloak that has the capability to render objects invisible to detectors, whether the application is digital or physical. To comprehensively evaluate the effectiveness of this approach, the researchers conducted extensive experiments under diverse conditions. They specifically examined the success rates of the attacks and investigated the impact of algorithm and model choices on the cloak's performance. Furthermore, to fully explore the potential of physical attacks utilizing adversarial clothing, they systematically measured the success rates of the attacks in the presence of complex fabric distortions. Additionally, they evaluated the transferability of the attacks across different models, classes, and datasets, providing a comprehensive analysis of the cloak's efficacy.

%Ensemble training has been recognized as an effective technique for enhancing the robustness of adversarial perturbations across multiple detectors. Building on this notion, Wu et al. \cite{invisiblecloak} explored the concept of training adversarial patches to deceive an ensemble of detectors. Specifically, they optimized the patches using an ensemble loss function, enabling the patches to be effective against a variety of detectors. This approach aligns with the strategies employed by other works such as \cite{NAP, TC-EGA, adversarialtshirt, bulb}, which also leverage ensemble training to improve the robustness of adversarial attacks. By leveraging the collective knowledge and diversity of multiple detectors, ensemble training can effectively bolster the resilience of adversarial perturbations, making them more potent against various detection models.
Ensemble training has emerged as a potent technique for enhancing the robustness of adversarial perturbations across multiple detectors. In a study by Wu et al. \cite{invisiblecloak}, the authors further explored this concept by investigating the training of adversarial patches to deceive an ensemble of detectors. The key idea was to optimize these patches using an ensemble loss function, enabling them to be effective against a variety of detectors. This approach aligns with strategies employed in other works, such as \cite{NAP, TC-EGA, adversarialtshirt, bulb}, which also utilize ensemble training to bolster the resilience of adversarial attacks.
By leveraging the collective knowledge and diversity of multiple detectors through ensemble training, adversarial perturbations can be made more potent and robust. This technique effectively improves the performance of adversarial attacks against various detection models, demonstrating its effectiveness in countering adversarial defenses and enhancing the stealthiness of attacks. The utilization of ensemble training to train adversarial patches presents a promising avenue for enhancing the potency and versatility of adversarial attacks in real-world scenarios.

In their quest to conceal individuals from detection systems, Xu et al. \cite{adversarialtshirt} proposed an approach that involves the creation of \textbf{adversarial T-shirts}. Unlike traditional attacks that utilize printed cardboard, the authors developed T-shirts embedded with carefully crafted adversarial patches, which they termed Adversarial T-shirts. To ensure the effectiveness of the attack in real-world scenarios, the Adversarial T-shirt incorporated a TPS-based (Thin-Plate Spline) transformer, which accounted for the temporal deformation of the T-shirt resulting from the pose variations of a moving person. This non-rigid transformation technique enabled the adversarial T-shirt to effectively deceive detection systems in physical environments by adapting to the dynamic nature of the person's movements and preserving the adversarial effect. By combining the principles of adversarial patches and non-rigid transformation, the Adversarial T-shirt represents a novel and promising approach to evading detection in real-world scenarios.

To address the limitation of existing transformations in capturing cloth deformation caused by pose changes in moving individuals, Xu et al. \cite{adversarialtshirt} introduced the use of Thin Plate Spline (TPS) mapping \cite{bookstein1993thin,jaderberg2015spatial}. The TPS mapping incorporates both an affine component and a non-affine warping component. It learns a parametric deformation mapping from an original image to a target image based on a set of control points with predetermined positions. By employing TPS-based transformations, the authors were able to model the temporal deformation of the clothing and improve the robustness of the adversarial T-shirts. This approach enabled the adversarial T-shirts to effectively adapt to the pose changes of moving individuals, enhancing their effectiveness in evading detection systems. The use of TPS-based transformations provides a more comprehensive and accurate representation of cloth deformation, contributing to the success of adversarial clothing in real-world scenarios.

Performing real-world multi-angle attacks poses a significant challenge in the context of adversarial attacks. Existing methods often rely on the assumption that the human body is facing the camera directly, as this allows the camera to capture the adversarial perturbation effectively. However, when the viewing angle changes, the adversarial perturbation may become invisible to the camera, rendering the attack ineffective. To address this limitation and enhance the robustness of attacks against changes in viewing angles, Hu et al. \cite{advtexture} have developed a method for generating \textbf{adversarial textures (AdvTexture)} that can be applied to clothing items with arbitrary shapes. By applying these intricate adversarial textures to clothing such as T-shirts, skirts, and dresses, the camera can capture the adversarial perturbations from any viewing angle. This enables a person wearing such clothing to remain undetected by object detectors, regardless of their body posture or the camera's position. By leveraging the unique characteristics of clothing items and designing adversarial textures that are effective from multiple viewing angles, the proposed method enhances the versatility and robustness of physical adversarial attacks. %This research contributes to the advancement of techniques for generating adversarial perturbations that can deceive object detectors in real-world scenarios, even when facing the challenges posed by changes in viewing angles.
 
All of the aforementioned methods suffer from the drawback of generating patches with easily identifiable conspicuous patterns. To address this limitation, researchers have proposed leveraging the learned image manifold of pre-trained Generative Adversarial Networks (GANs) on real-world images. By utilizing the inherent structure and diversity captured by GANs, these methods aim to generate more natural and inconspicuous patches that can blend seamlessly into the background. This approach allows for the creation of adversarial perturbations that are visually indistinguishable from genuine objects, thereby enhancing their stealthiness and making them more effective in evading detection systems.

In their work, the authors of \cite{UPC} introduced a \textbf{universal camouflage pattern (UPC)} that closely resembles natural images, aiming to enhance the stealthiness of physical attacks. To facilitate fair and comprehensive evaluations of physical attacks, Huang et al. \cite{UPC} proposed the AttackScenes dataset, which provides a standardized collection of 3D environments that accurately simulate real-world conditions. This dataset ensures controllable and reproducible settings, enabling researchers to conduct experiments under fair and comparable conditions.

To deceive both the region proposal network and the classifier, while inducing errors in the regressor, the authors proposed the Universal Physical Camouflage Attack (UPC) method. This approach utilizes adversarial patterns that are strategically designed to evade detection and classification systems. To accommodate non-rigid and non-planar objects, the UPC incorporates various geometric transformations, including cropping, resizing, and affine homography, to mimic the deformable properties of real-world objects.

For human subjects, the UPC includes eight designs printed on masks, T-shirts, and pants, providing practical demonstrations of its effectiveness in physical camouflage. The UPC approach, along with the AttackScenes dataset, offers researchers valuable tools for studying and evaluating physical attacks in controlled and realistic settings.

In order to minimize the detectability of generated adversarial patterns, the UPC approach \cite{UPC} incorporates two constraints: the semantic constraint and the material constraint. The semantic constraint is applied during the optimization process to ensure that the generated adversarial patterns maintain a visual similarity to natural images. This constraint helps to make the perturbations blend in with the surrounding environment, reducing their conspicuousness. Additionally, the material constraint is employed to camouflage the perturbations as texture patterns on human accessories such as T-shirts, pants, and masks. By incorporating texture patterns that are commonly found on these materials, the adversarial patterns become less distinguishable from the regular patterns of the accessories. This material constraint enhances the effectiveness of the camouflage, making the perturbations less noticeable and improving their ability to deceive detection and classification systems. By combining the semantic and material constraints, the UPC approach ensures that the generated adversarial patterns are visually coherent and closely resemble natural images, while also seamlessly blending with the texture patterns of human accessories. These constraints contribute to the overall stealthiness of the physical attacks, making the adversarial perturbations less discernible to human observers and detection algorithms alike.

The \textbf{Legitimate Adversarial Patches (LAP)} approach \cite{LAP} aims to create patches that are capable of deceiving both human observers and detection models in real-world scenarios. To achieve a balance between attack effectiveness and patch rationality, LAP employs a two-stage training process. In the first stage, LAP generates an initial patch by using an original cartoon image as input. This initial patch serves as the starting point for the optimization process in the second stage. In this stage, the input image and initial patch are used to produce the final adversarial patch, which is designed to effectively deceive detection models while appearing visually coherent to human observers. An important consideration in LAP is the application of the adversarial patch to targeted images. In typical real-world applications, the object detection model scales the adversarial sample along with the image to a square shape, which can reduce the effectiveness of the patch. To address this limitation, LAP directly scales the adversarial patch along with the targeted images. This scaling operation introduces additional deformation to the patch, making it more challenging to optimize. Consequently, the attack intensity may be reduced compared to the padding method. However, this approach enhances the physical applicability of the patch, as it aligns the patch with the shape and size of the targeted objects, improving the patch's effectiveness in real-world scenarios. By employing a two-stage training process and incorporating direct scaling of the patch with targeted images, LAP aims to create adversarial patches that are both effective in deceiving detection models and visually plausible to human observers. This approach enhances the applicability and stealthiness of the adversarial patches in real-world settings. LAP takes advantage of cartoon images that possess natural perceptual properties and incorporates a projection function for further optimization. Additionally, LAP conducts a comprehensive evaluation of the rationality of the generated patches, considering color features, edge features, and texture features. By utilizing these indicators of rationality, LAP creates adversarial patches that maintain visual similarity to cartoon pictures.

\textbf{Naturalistic Adversarial Patches (NAP)}: Utilizing the efficiency of generative adversarial networks (GANs) in generating desired samples \cite{wang2020generative}, Hu et al. \cite{NAP} proposed a novel approach for creating physical adversarial patches with the aim of deceiving person detectors (See Figure \ref{NAP_overview}). Their method capitalizes on the learned image manifold of pretrained models, namely BigGAN \cite{biggan} and StyleGAN \cite{stylegan}, which have been trained on real-world images. By leveraging the principles of adversarial YOLO, the authors minimize both the objectness and class probabilities specifically for the person class, thereby creating patches that are effective in evading person detectors. To enhance the diversity of training data and improve the effectiveness of the generated patches, the MPII Human Pose dataset \cite{MPII} is introduced in this work. In a related study by Pavlitskaya et al. \cite{pavlitskaya2022feasibility}, an attempt was made to extend the existing work by introducing a method for suppressing objects near the patch. However, the proposed approach was deemed unrealistic in terms of the positioning of the patch and its size, limiting its feasibility in practical applications.

Drawing upon the remarkable image generation capabilities of advanced generative adversarial network (GAN) models, the authors of NAP (Naturalistic Adversarial Patches) devise a method for generating physically plausible adversarial patches. To maintain a high level of realism, NAP introduces a norm constraint on the latent vector, ensuring that it remains within a specified threshold. This constraint enables a balance between realism and attack performance, as the optimization process navigates the limited latent space of GAN models. However, this trade-off between naturalness and attack effectiveness has been observed in GAN-based approaches \cite{dap}. To address this, the authors of \textbf{AdvART (Adversarial Art)} \cite{advart} propose a novel approach that generates adversarial patches with a naturalistic and inconspicuous appearance resembling artistic paintings, while still maintaining a high level of attack efficiency.
\begin{figure}
    \centering
    \includegraphics[width=0.5\textwidth]{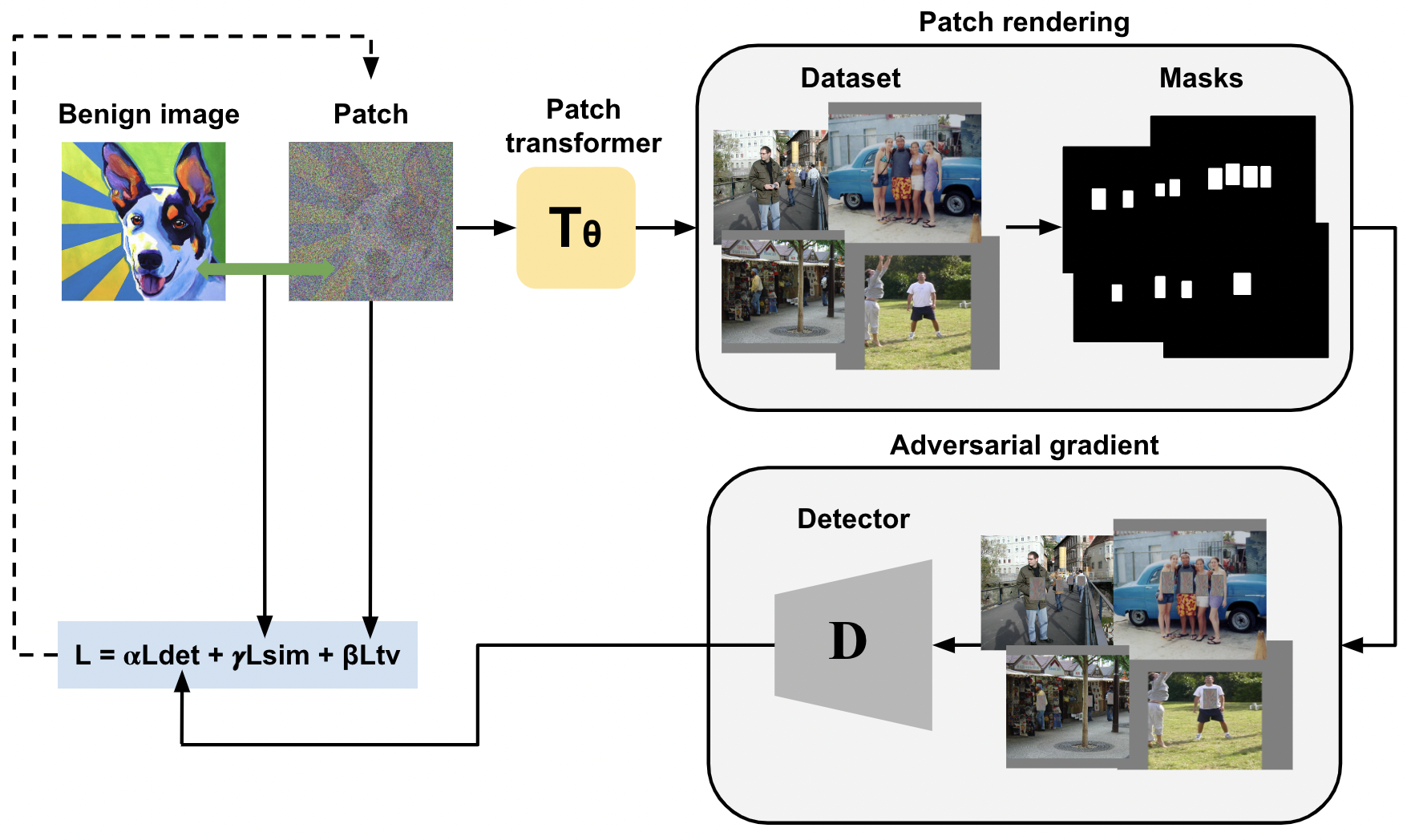}
    \caption{The AdvART approach involves optimizing an adversarial patch, denoted as $P$, to make a target object undetectable by an object detector while preserving a naturalistic and artistic pattern inspired by a given benign image. The process begins by applying geometric transformations to the initial patch using a patch transformer, enhancing its robustness. The transformed patch is then rendered onto the input image using generated masks for seamless integration. Next, the resulting adversarial image is fed to the object detector to compute various loss functions. Based on the computed losses, the gradient of the patch is determined, guiding its iterative update. By leveraging geometric transformations, rendering techniques, and gradient-based updates, the approach seeks to find an effective adversarial patch that conceals the target object from the object detector while appearing visually consistent with the surrounding environment (Figure adapted from \cite{advart}).}
    \label{fig:advart}
\end{figure}
In the optimization process, AdvART introduces an additional loss term to the objective function, which serves as a semantic constraint. This constraint ensures that the generated camouflage pattern holds semantic meaning, preventing the creation of meaningless patterns. By incorporating this semantic constraint, AdvART aims to generate adversarial patches that not only evade detection but also exhibit meaningful and aesthetically pleasing visual characteristics.

\textbf{Dynamic adversarial patch (DAP)}: Moreover, most existing clothing-based physical attacks assume static objects and overlook the potential transformations caused by non-rigid deformations resulting from changes in a person's pose. To overcome this limitation, Guesmi et al. \cite{dap} integrate a "Creases Transformation" (CT) block into their approach. The CT block serves as a preprocessing step, following an Expectation Over Transformation (EOT) block, which generates a diverse set of transformed patches. These transformed patches are incorporated into the training process, enhancing the model's robustness to various real-world distortions such as creases in clothing, rotation, scaling, random noise, brightness and contrast variations, among others. By accounting for non-rigid deformations and incorporating a range of possible transformations, the approach becomes more resilient and adaptable to real-world scenarios.
\begin{figure*}
    \centering
    \includegraphics[width=\textwidth]{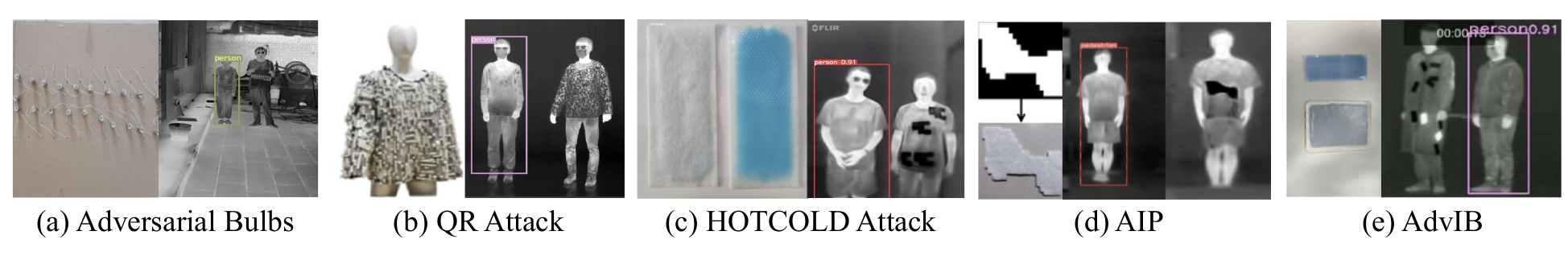}
    \caption{Illustration of different physical infrared attacks: (a) Adversarial Bulbs \cite{bulb}, (b) QR Attack \cite{qrattack}, (c) HOTCOLD Attack \cite{hotcold}, (d) AIP \cite{AIP}, and AdvIB \cite{AdvIB}.}
    \label{fig:infrared}
\end{figure*}
%==================================================================================
\subsection{Infrared Person Detection}

%Infrared person detection involves the utilization of infrared (IR) technology to identify and detect the presence of humans. Infrared radiation, which is a form of electromagnetic radiation, is emitted by objects in the form of heat. Through the use of infrared sensors or cameras, the thermal radiation emitted by objects, including humans, can be detected and captured. This enables the detection of objects even in low-light or no-light conditions. In the context of person detection, infrared sensors or cameras are employed to capture the thermal energy emitted by humans. These sensors can measure the temperature differences between the human body and the surrounding environment. By analyzing the thermal patterns and variations in the acquired infrared images, algorithms can discern human figures from the background or other objects. This allows for the identification and differentiation of humans in various environments. Infrared person detection finds applications in several domains, including security systems, surveillance, and monitoring. It is particularly valuable in situations where visibility is limited, such as during nighttime or in dark areas. In such scenarios, where visual-based detection methods may be ineffective due to low light conditions or camouflage techniques, infrared person detection provides a reliable and effective means of identifying and tracking individuals.
Infrared person detection involves the use of infrared (IR) technology to identify and detect the presence of humans. Infrared radiation, a form of electromagnetic radiation, is emitted by objects as heat. Infrared sensors or cameras are used to detect and capture this thermal radiation. This enables the detection of objects, including humans, even in low-light or no-light conditions.
In the context of person detection, infrared sensors or cameras capture the thermal energy emitted by humans. By measuring the temperature differences between the human body and the surrounding environment, algorithms can discern human figures from the background or other objects. Analyzing the thermal patterns and variations in the acquired infrared images allows for the identification and differentiation of humans in various environments.
Infrared person detection has applications in security systems, surveillance, and monitoring. It is particularly valuable in situations where visibility is limited, such as during nighttime or in dark areas. In such scenarios, where visual-based detection methods may be ineffective due to low light conditions or camouflage techniques, infrared person detection provides a reliable and effective means of identifying and tracking individuals.
%------------------------------------
\subsubsection{Light Manipulation-based Attacks}

The security research in deep neural networks (DNNs) has primarily focused on visible light imaging, leaving the thermal infrared imaging domain relatively unexplored. However, thermal infrared detection systems play a crucial role in various security-related applications, such as autonomous driving, night surveillance, and temperature measurement. Addressing this research gap, Zhu et al. introduced a groundbreaking method known as \textbf{Adversarial Bulbs} \cite{bulb} to launch physical adversarial attacks in the thermal infrared domain. This work represents the first attempt to attack thermal infrared person detectors in real-world scenarios.
The Adversarial Bulbs approach employs a patch-based attack strategy using a set of small bulbs placed on a cardboard surface. These bulbs can be conveniently held in hand, making the attack practical to execute. During the training phase, the position of each small bulb is optimized. Unlike updating individual pixels, the optimization parameter in Adversarial Bulbs is the center point of a two-dimensional Gaussian function, accurately emulating the appearance of bulbs when captured by a thermal infrared camera.
To evaluate the effectiveness of the proposed method, Zhu et al. conducted experiments using the Teledyne FLIR ADAS Thermal dataset, which provides a suitable benchmark for assessing the performance of thermal infrared person detectors. The results of these experiments demonstrate the efficacy of Adversarial Bulbs in deceiving thermal infrared person detection systems, thus highlighting the vulnerability of such systems to physical adversarial attacks in real-world thermal infrared imaging scenarios. %This pioneering research sheds light on the importance of addressing the security concerns associated with thermal infrared imaging and calls for the development of robust countermeasures to mitigate potential adversarial threats in this domain.

Zhu et al. proposed a solution called "infrared invisible clothing" to tackle the challenge of evading detection by infrared detectors. Their approach draws inspiration from the Stefan-Boltzmann law and leverages temperature control materials, including cotton, polyester, thermal insulation tapes such as Teflon and polyimide, and a novel material called aerogel. By designing and using these specific materials, the infrared invisible clothing effectively conceals thermal emissions, rendering individuals wearing this clothing undetectable by infrared detectors from various viewing angles in real-world scenarios. %This research highlights the robustness of the infrared invisible clothing against thermal infrared imaging, enabling the evasion of detection by thermal infrared detectors.
%------------------------------------
\subsubsection{Camouflage Techniques}
%------------------------------------
%QR Attack \cite{qrattack}
%HOTCOLD \cite{hotcold}
%AdvIB \cite{AdvIB}

\begin{figure}
    \centering
    \includegraphics[width=0.5\textwidth]{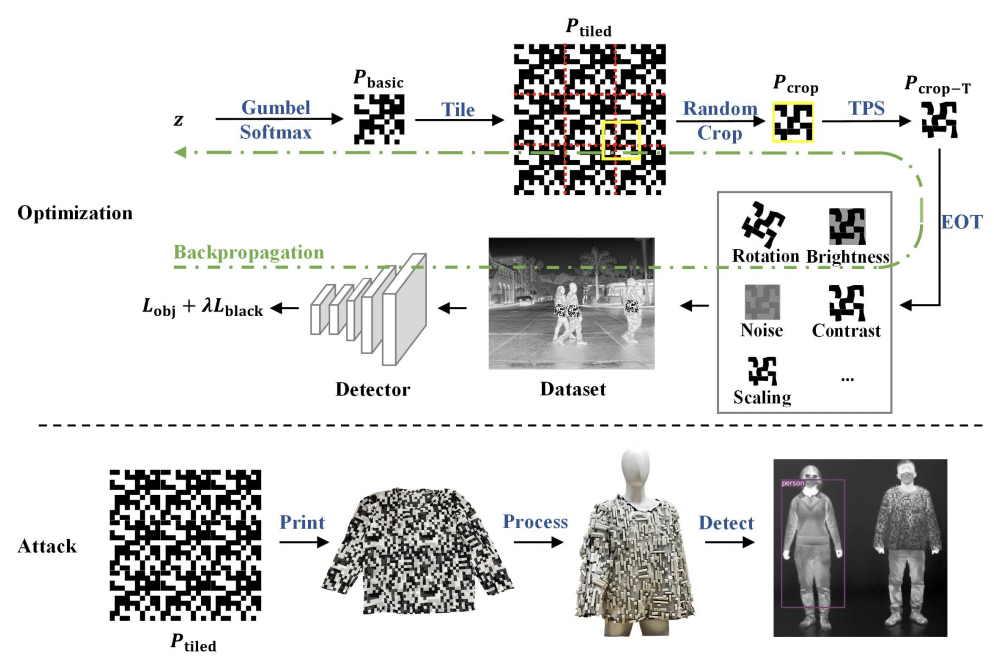}
    \caption{Pipeline of QR attack. Top: attack in the digital world by optimizing a binary pattern. Bottom: attack in the physical world (Figure adapted from \cite{qrattack}).}
    \label{fig:qr}
\end{figure}

\textbf{QR Attack}: Zhu et al. \cite{qrattack} developed a novel approach, referred to as "infrared invisible clothing," to effectively evade person detectors in the context of thermal infrared imaging (See Figure \ref{fig:qr}). Their method utilized a new material called aerogel, which allowed the clothing to remain undetectable by infrared detectors even from multiple viewing angles. The researchers analyzed the specific characteristics of thermal infrared images, including the grayscale nature of the images and the correlation between pixel values and surface temperatures. Leveraging these insights, they designed the cloth pattern of the infrared invisible clothing to resemble a Quick Response (QR) code. Each pixel within the "QR code" was optimized using the Gumbel-softmax technique to maximize the attack effectiveness, ensuring that the clothing effectively concealed the wearer from infrared detectors. This approach demonstrated the potential of using specially designed patterns to create infrared invisible clothing that evades detection in thermal infrared imaging.

In the domain of thermal infrared imaging, researchers have proposed a novel wearable design called the \textbf{HOTCOLD} Block to deceive thermal infrared detectors. This approach, introduced by Li et al. \cite{hotcold}, leverages the characteristics of temperature-sensitive materials and thermal radiation to create a wearable device capable of fooling detectors. The HOTCOLD Block consists of two main components: the "HOT" region and the "COLD" region. These regions are strategically designed to manipulate the thermal signature of the wearer and create misleading heat patterns. The HOT region incorporates materials with a higher thermal emissivity, which enables it to emit thermal radiation at a higher rate compared to the surrounding environment. This leads to a perception of higher temperature in the HOT region by the thermal infrared detectors. In contrast, the COLD region integrates materials with a lower thermal emissivity, resulting in reduced thermal radiation emission and a perception of lower temperature. By incorporating these contrasting regions, the HOTCOLD Block effectively manipulates the thermal signatures of the wearer, causing the detectors to misinterpret the actual body temperature distribution. This wearable design introduces a new dimension to the field of adversarial attacks in thermal infrared imaging, where the focus is shifted towards exploiting the characteristics of temperature-sensitive materials and thermal radiation for deceiving detectors. The HOTCOLD Block demonstrates the potential of wearable devices in undermining the accuracy and reliability of thermal infrared detection systems.
Wei et al. \cite{AIP} introduced a physically feasible method for conducting infrared attacks, referred to as \textbf{AIP (Adversarial Infrared Patches)}. This method takes into consideration the imaging mechanism of infrared cameras, which capture the thermal radiation emitted by objects. The adversarial infrared patches are created by attaching a patch made of thermal insulation materials onto the target object, thereby manipulating its thermal distribution. To enhance the effectiveness of these attacks, the authors propose a novel aggregation regularization technique that facilitates the simultaneous learning of the patch's shape and location on the target object. This allows for the utilization of a simple gradient-based optimization algorithm to solve for the optimal patch configuration. The proposed method enables the generation of infrared patches that can deceive infrared cameras and compromise the accuracy of thermal-based object detection and recognition systems.

\textbf{AdvIB (Adversarial Invisible Block)} \cite{AdvIB} is a novel technique that aims to deceive thermal infrared detectors in the physical world. The objective of this approach is to create a physical object, referred to as the AdvIB, which can effectively camouflage itself from thermal infrared detectors, making it difficult for them to detect its presence. The AdvIB leverages the understanding of the thermal properties and characteristics of objects in the infrared spectrum. By carefully designing the material composition and surface properties of the AdvIB, it can effectively manipulate its thermal signature, causing it to blend seamlessly with the surrounding environment. This camouflage effect enables the AdvIB to go undetected by thermal infrared detectors, even when subjected to close scrutiny. The design of the AdvIB involves optimizing various parameters, such as the material properties, shape, size, and surface texture, to achieve maximum stealthiness. Computational techniques, including numerical simulations and optimization algorithms, are employed to iteratively refine the AdvIB design and ensure its effectiveness in fooling thermal infrared detectors. The AdvIB technique holds significant implications for security and privacy in scenarios where thermal infrared detection is employed, such as surveillance systems, border control, and military applications. By exploiting the vulnerabilities and limitations of thermal infrared detectors, the AdvIB offers a means to circumvent their detection capabilities and potentially evade surveillance or identification. However, it is important to note that the development and deployment of such techniques raise ethical concerns and potential misuse. %As researchers continue to explore and advance the field of adversarial techniques for fooling thermal infrared detectors, it is crucial to consider the broader implications and develop appropriate safeguards to maintain the integrity and effectiveness of thermal infrared detection systems in real-world applications.

%======================================================================================================================================================================****************************************************************************************************************************************************************************====================================================================================================================================================================

\subsection{ Vehicle Detection}

Vehicle detection is a fundamental task in computer vision, with significant applications in traffic monitoring, autonomous driving, and surveillance systems. Deep learning models, particularly convolutional neural networks (CNNs), have emerged as highly effective tools for accurately detecting vehicles in images and videos. CNNs leverage their ability to learn intricate spatial features and patterns, enabling robust vehicle identification across diverse environmental conditions, including variations in viewpoints, lighting, and occlusions. The successful detection of vehicles using deep learning models enables a wide range of applications, such as traffic flow analysis, parking management, object tracking, and collision avoidance systems, contributing to enhanced safety, efficiency, and decision-making in transportation domains.

However, current surveillance and autonomous driving systems heavily rely on deep neural networks for vehicle detection, making them vulnerable to adversarial attacks. Adversarial attacks involve the deliberate application of distinctive patterns or perturbations to a vehicle's body, aiming to render it undetectable by surveillance cameras or disrupt the performance of detection algorithms. The objective of these attacks is to deceive the underlying detection mechanisms employed in these systems, compromising their accuracy and reliability.

%By exploiting the vulnerabilities of deep learning models, adversarial attacks pose a significant challenge to the robustness and security of vehicle detection systems. Researchers are actively investigating techniques to enhance the resilience of these systems against such attacks, including the development of adversarial training methods, defense mechanisms, and anomaly detection algorithms. These efforts aim to ensure the continued effectiveness and reliability of vehicle detection in surveillance and autonomous driving applications, mitigating the potential risks associated with adversarial attacks and maintaining the integrity of these systems in real-world scenarios.

%------------------------------------
\subsubsection{Patch-based Attacks}
%------------------------------------
%ScreenAttack \cite{dynamicpatch}

\textbf{ScreenAttack}: Hoory et al. \cite{dynamicpatch} proposed a dynamic adversarial patch for evading object detection models by utilizing a digital screen. The authors recognized that existing patch-based physical attacks rely on static printed patches, which fail to adapt to the dynamic nature of the physical world, leading to reduced attack success rates. To address this challenge, they introduced the concept of a dynamic adversarial patch that is displayed using a digital screen. In their approach, the authors adopted the loss function from a previous work, Adversarial YOLO \cite{Adversarialyolo}, to train the dynamic adversarial patch. However, they observed that the adversarial patch crafted by the Adversarial YOLO loss function is prone to being misdetected as semantically related categories, making it inconsistent with the background objects. To overcome this issue, the authors introduced a modified loss function that considers the semantically related class (e.g., bus or truck) as the target instead of the original class (i.e., car). The loss function was defined as follows:

\begin{equation}
    J(\delta, x_{adv}, C) = \alpha TV(\delta) + \sum_{y_i \in C} L(f(x_{adv}), y_i)
\end{equation}

Here, $C$ represents the set of semantically related labels, which is used to prevent the adversarial patch from being detected as car-related classes. The authors further divided the training set into multiple subsets and trained an adversarial patch for each subset, enabling improved effectiveness across various environmental conditions. In the physical attack scenario, the authors placed a digital screen on the side and rear of a car. The dynamic adversarial patch, chosen from the patch subsets, was dynamically displayed on the screen in response to changing environmental conditions. The evaluation results on YOLOv2 demonstrated that approximately 90\% of video frames with the adversarial perturbed screen failed to detect the target object (i.e., car), highlighting the effectiveness of the proposed approach. Overall, the ScreenAttack method introduced a dynamic adversarial patch displayed on a digital screen, enabling better adaptation to the dynamic physical world and increasing the success rate of evading object detection models. The modified loss function and subset-based training approach contributed to improved performance and reduced susceptibility to misclassification as semantically related categories, resulting in enhanced consistency with the background objects.

%------------------------------------
\subsubsection{Sicker-based Attacks}
%------------------------------------
%CAMOU \cite{camou}
%ER Attack \cite{ERattack}
%DAS \cite{DAS}
%CAC \cite{CAC}
%FCA \cite{FCA}
%DTA \cite{DTA}
\begin{figure}[!htp]
\centering
\includegraphics[width=0.5\textwidth]{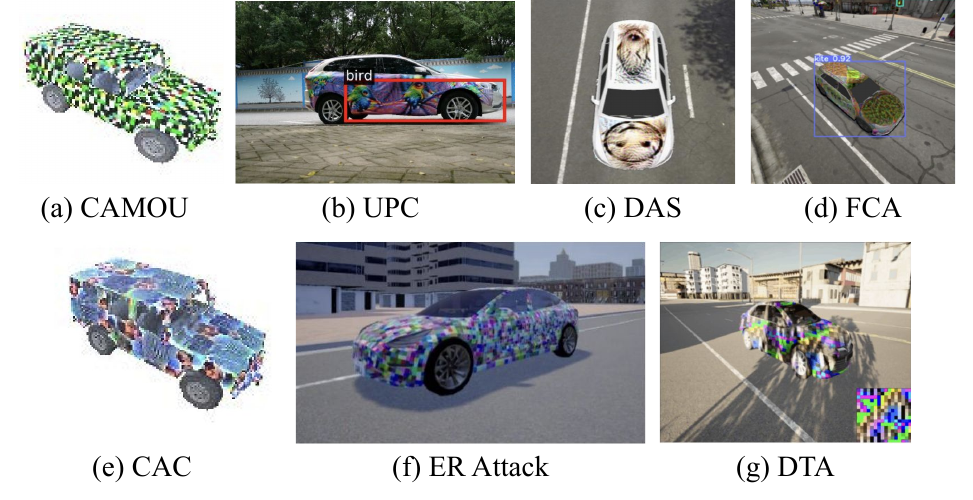} %, height=7cm
\caption{Examples of physical adversarial examples against vehicle detection: (a) CAMOU \cite{camou}, (b) UPC \cite{UPC}, (c) DAS \cite{DAS}, (d) FCA \cite{FCA}, (e) CAC \cite{CAC}, (f) ER Attack, and (g) DTA \cite{DTA}. }
\label{AE_VD}
\end{figure}
To investigate the effectiveness of concealing cars in real-world scenarios, Zhang et al. \cite{camou} conducted a comprehensive set of experiments within a 3D environment, enabling them to simulate the complex transformations induced by the physical surroundings. Due to limitations in terms of cost and time, they opted to utilize the photo-realistic Unreal Engine 4 game engine, which provided them with a versatile platform for experimentation. This engine offered a wide range of configurable parameters, including options for camouflage resolution and pattern, 3D vehicle models, camera and environment settings, and more.
To evaluate the detectability reduction of popular detectors such as Mask R-CNN and YOLOv3, which were pretrained on the MS COCO dataset, the authors employed a clone network. This network aimed to replicate the joint response of the simulator and the detector, enabling the inference of an optimal camouflage for a given 3D vehicle. Through the training process, the network learned effective camouflage strategies that resulted in a significant reduction in the detectability of the vehicles by the detectors.
By leveraging the capabilities of the Unreal Engine 4 game engine and utilizing a clone network to approximate the response of the detectors, Zhang et al. conducted experiments that demonstrated the effectiveness of learned camouflage in concealing cars named \textbf{CAMOU}. The findings of their study shed light on the potential of using virtual environments to simulate real-world scenarios and optimize camouflage strategies for evading object detectors.
\begin{figure*}[!htp]
\centering
\includegraphics[width=\textwidth]{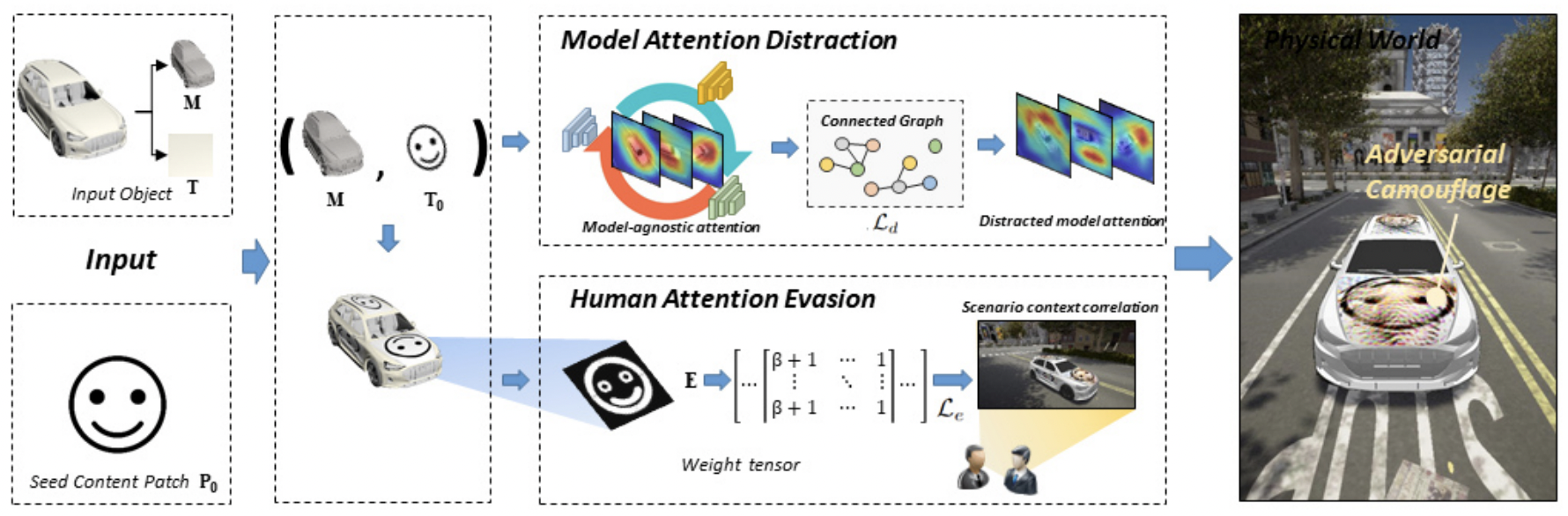} %, height=7cm
\caption{The DAS method is structured as follows: Initially, they address the intrinsic attention characteristic by effectively utilizing the comparable attention patterns across models, thus compelling the displacement of "heat" regions from the target object using loss function Ld. Subsequently, they circumvent the human-specific visual attention mechanism by establishing a correlation between the appearance of adversaries and the contextual scenario. Additionally, they retain the shape information from the seed content image to create visually-natural adversarial camouflage (Figure adapted from \cite{DAS}). }
\label{fig:das}
\end{figure*}
\textbf{ER Attack}: Inspired by prior work, Wu et al. \cite{ERattack} conducted their experiments using the open-source simulator CARLA \cite{dosovitskiy2017carla}, which allowed them to consider simulated physical constraints. Within the CARLA simulation environment, they introduced two important techniques: the Enlarge-and-Repeat (ER) process and a discrete search method. These techniques were employed to generate physically adversarial textures that targeted vehicle detection models. The Enlarge-and-Repeat process enabled the researchers to expand and duplicate a small patch to cover the entire vehicle, thereby creating a visually deceptive texture. This process enhanced the effectiveness of the adversarial attack by exploiting the model's vulnerability to specific patterns or textures. To optimize the adversarial textures, Wu et al. employed a discrete search method that explored different configurations and arrangements of the patch. This method allowed for the identification of the most effective placement and scaling of the patch on the vehicle, maximizing the adversarial impact on the detection model. By utilizing the CARLA simulator, the researchers were able to conduct controlled experiments that incorporated realistic physical constraints. This enabled them to craft and evaluate the performance of the generated adversarial textures in a controlled and reproducible setting, providing valuable insights into the vulnerabilities of vehicle detection models.

\textbf{DAS}: Recognizing a limitation in prior works \cite{camou,ERattack} that overlooked the attention mechanism of the model, Wang et al. \cite{DAS} proposed a method to simultaneously suppress both model attention and human attention, thereby improving transferability of the adversarial attack. To suppress model attention, the authors introduced a technique to disperse the intensity attention map. This was achieved by disconnecting the connected graph generated from the attention map, which effectively disrupted the attention mechanism of the model. For human attention, a seed content patch was used as initialization. The authors encouraged the adversarial texture to remain similar to the seed content, reducing the likelihood of human observers detecting the camouflage. To optimize the adversarial texture, specific part faces of a 3D model were treated as textures and optimized using the neural renderer \cite{kato2017neural}. This approach allowed for fine-grained control over the appearance of the adversarial texture. To evaluate the effectiveness of the proposed method, the authors constructed a dataset sampled from the CARLA simulator, capturing various distances and angles. In physical attacks, the rendered car with camouflage was printed on paper, and the camouflage texture was cropped and applied to a toy car. Experimental results demonstrated a significant reduction in the performance of four different detectors (YOLOv5, SSD, Faster RCNN, and Mask RCNN). On average, the detectors exhibited a performance reduction of 26.05\%, indicating the effectiveness of the proposed method in fooling the object detection models.

The neural renderer has emerged as a valuable tool for performing 2D-to-3D transformations in various domains. One prominent application is the mapping of texture images onto 3D models, which are subsequently rendered into 2D images. This technique has been widely explored in studies by Kato et al. \cite{kato2017neural}, Rematas et al. \cite{rematas2020neural}, and Thies et al. \cite{thies2019deferred}, among others. In the context of adversarial attacks, the neural renderer plays a crucial role in applying adversarial stickers or textures to the surfaces of vehicles. Notably, the \textbf{Full-coverage Camouflage Attack (FCA)} \cite{FCA} takes advantage of the neural renderer to overcome challenges related to partial occlusion and long distances. FCA achieves this by rendering non-planar textures across the entire surface of the target vehicle. By utilizing a differentiable neural renderer, FCA bridges the gap between digital attacks and physical attacks, enabling the generation of realistic adversarial camouflage. Moreover, FCA incorporates a transformation function to transition the rendered camouflaged vehicle into a photo-realistic scenario, enhancing the attack's effectiveness. Comparative evaluations have demonstrated that FCA outperforms other advanced attack techniques, exhibiting superior performance in both digital and physical attack scenarios.

\textbf{CAC (Camouflage Attack via 2D Texture Optimization)}: In contrast to the approach proposed in \cite{FCA}, which focused on optimizing the full coverage faces of the 3D model, Duan et al. \cite{CAC} introduced a method that directly optimizes the 2D texture of the 3D model for adversarial camouflage. The authors achieved this by minimizing the classification output of the Region Proposal Network (RPN) module in the first stage of the Faster R-CNN framework. While considering various physical settings, such as viewing angles and brightness variations, the authors trained the texture using images with a pure color background. In physical attacks, they utilized a 3D printer to print out the complete vehicle model, and their experiments demonstrated the effectiveness of their method in a physical environment with a 360° free viewpoint setting. The CAC technique offers a novel approach to adversarial camouflage by focusing on optimizing the 2D texture of the vehicle model, showcasing its potential in physical attack scenarios.

Despite the remarkable progress in neural rendering techniques, accurately representing the wide array of real-world transformations remains a challenge. This limitation stems from the lack of control over scene parameters compared to traditional photo-realistic renderers. To overcome this challenge, Suryanto et al. \cite{DTA} introduced the \textbf{Differentiable Transformation Attack (DTA)}, a framework specifically designed to generate highly effective physical adversarial camouflage on 3D objects. DTA combines the strengths of a photo-realistic rendering engine with the differentiability of a novel rendering technique. Notably, DTA outperforms previous approaches in terms of both effectiveness and the transferability of the generated adversarial camouflage to other detection models. This significant improvement highlights the potential of DTA to achieve superior performance in generating physical adversarial attacks, making it a promising framework for advancing the field.

The initial works in the field, such as CAMOU \cite{camou} and ER Attack \cite{ERattack}, primarily focused on evading detection models while neglecting to consider human perception. These approaches shared a common limitation: the generated textures were visually attention-grabbing when applied to the surface of a car. Consequently, they were easily detectable by human observers, contradicting the purpose of adversarial attacks.

In an effort to enhance visual stealthiness, Wang et al. \cite{DAS} leveraged the psychological theory that human vision's bottom-up attention mechanism is sensitive to salient objects, such as distortions \cite{connor2004visual}. To evade this attention, the authors focused on generating adversarial camouflage that exhibited high semantic correlation with the surrounding context. They introduced a seed content patch and applied it to the vehicle's surface. To ensure smooth color transitions and prevent noisy images, they incorporated the smooth loss \cite{adveyeglass}, which minimized the squared differences between adjacent pixels during the optimization process. By employing these techniques, Wang et al. aimed to create camouflage that would reduce the likelihood of attracting human attention.

In the field of physical adversarial attacks on vehicles, several approaches have been proposed to address different challenges and enhance the effectiveness of the generated adversarial stickers. CAMOU \cite{camou} employs the Expectation Over Transformation (EOT) algorithm to consider a range of physical transformations, ensuring that the generated stickers can consistently conceal the vehicle from neural detectors across different real-world scenarios. However, this method overlooks the fact that when viewed from different angles, only a portion of the pattern can be captured by the camera. To address this limitation, ER Attack \cite{ERattack} repeats the same adversarial pattern on the vehicle's body, maximizing its adversarial characteristics. While this approach improves robustness against angle changes, it may weaken the overall impact. To tackle the challenges posed by specific physical scenarios, such as multi-view, long-distance, and partially occluded objects, FCA \cite{FCA} treats the adversarial stickers as the texture of the 3D vehicle and applies a transformation function to convert the rendered 3D vehicle into different environmental scenarios, resulting in photo-realistic images. This approach ensures that the generated adversarial stickers remain robust across various physical conditions. Furthermore, existing neural renderers have limitations in accurately blending the foreground object with the background, leading to inaccurate effects such as shadow casting and light reflection. In response, Suryanto et al. \cite{DTA} introduced the Differentiable Transformation Network (DTN), a novel neural renderer that learns object transformations while preserving their original parameters. By extracting the expected transformation of a rendered object and retaining its original attributes, DTN enables the synthesis of photo-realistic images and the generation of robust adversarial patterns. %These advancements demonstrate the ongoing efforts to overcome challenges in physical adversarial attacks on vehicles, ensuring their effectiveness and applicability in real-world scenarios.

%------------------------------------
\subsubsection{Imaging Device Manipulation}
%------------------------------------
%PG \cite{PG} (Acoustics)

\begin{figure}
    \centering
    \includegraphics[width=0.5\textwidth]{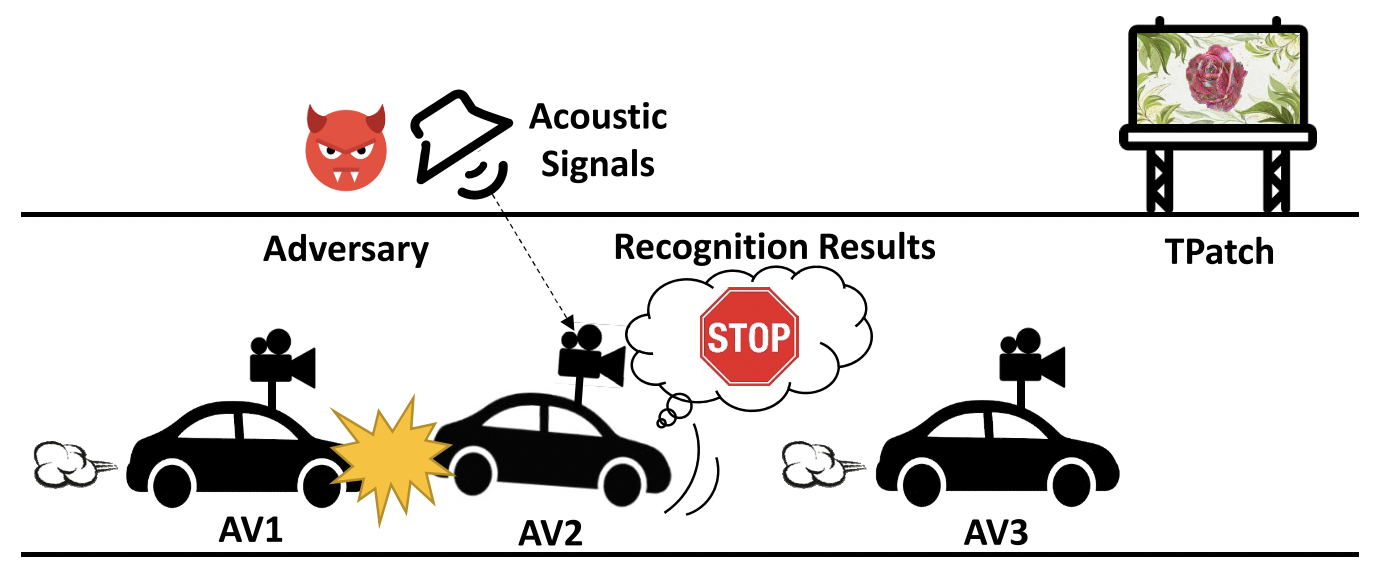}
    \caption{TPATCH attack: a physical adversarial patch is placed at the roadside, seemingly benign to most passing vehicles, such as AV1 and AV3. However, this patch is carefully crafted to exploit the vulnerabilities of a specific targeted vehicle (AV2) under acoustic injection attacks. When AV2 passes by the patch, the adversarial perturbation induced by the acoustic injection causes the vehicle's perception system to mistakenly recognize a non-existing stop sign (Figure adapted from \cite{tpatch}).}
    \label{fig:tpatch}
\end{figure}

\begin{figure}
    \centering
    \includegraphics[width=0.5\textwidth]{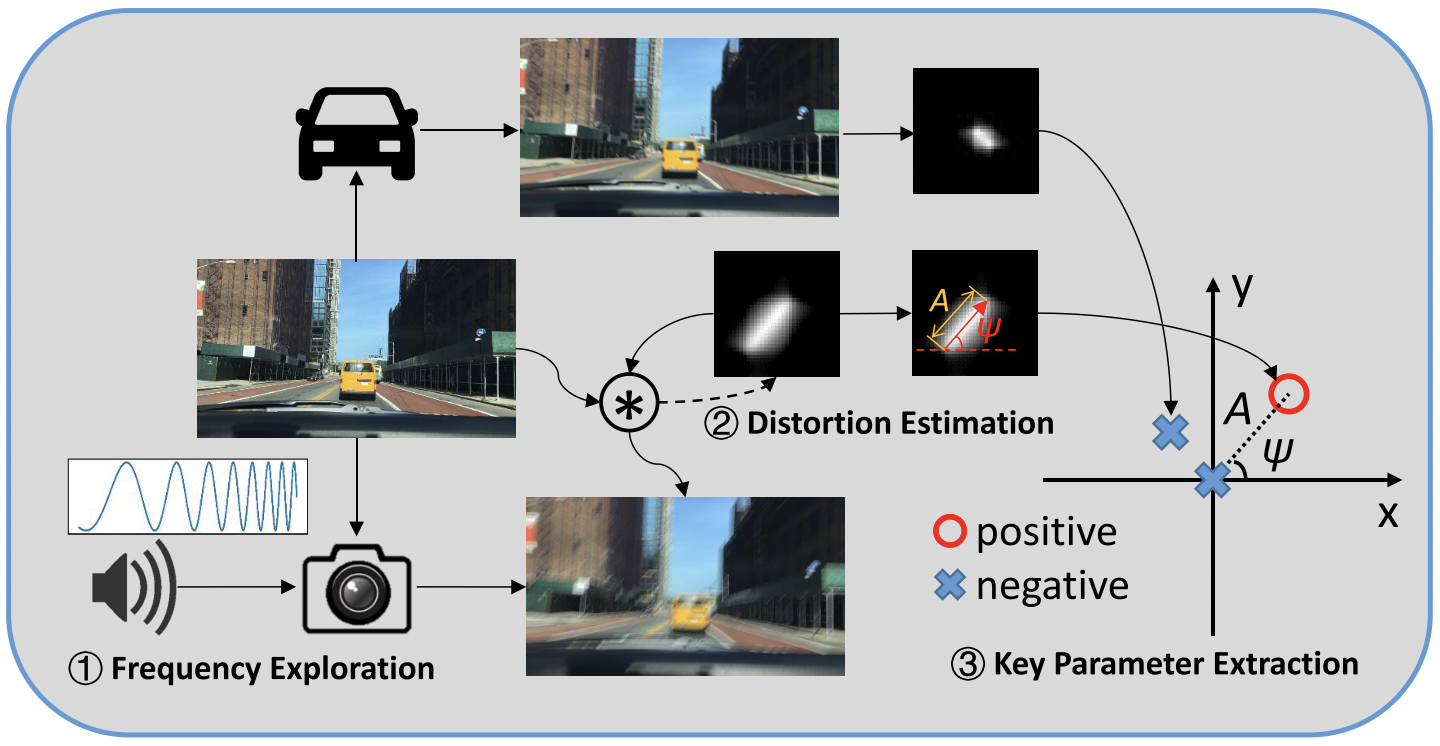}
    \caption{The trigger design process involves the following steps:
    (1) Explore the Resonant Frequency: In this step, frequency-modulated acoustic signals are injected to explore the resonant frequency of the target camera or sensor system. The purpose is to identify the specific frequency at which the camera is most susceptible to image distortion caused by the acoustic signals. (2) Estimate the PSF Kernel: PSF stands for Point Spread Function. In this step, the clear image captured by the camera and the corresponding blurred image induced by the acoustic signals are used to estimate the PSF kernel. The PSF kernel represents the mathematical model of the distortion introduced by the acoustic signals on the image. (3) Extract the Strength and Orientation of Blur: Using the estimated PSF kernel, the strength and orientation of the blur caused by the acoustic signals are extracted. The strength denotes how much the image is distorted, and the orientation indicates the direction of the distortion. (Figure adapted from \cite{tpatch}).}
    \label{fig:tpatch_}
\end{figure}

\textbf{TPATCH} \cite{tpatch} is a unique form of physical adversarial patch triggered by acoustic signals. Unlike traditional adversarial patches, TPatch remains inconspicuous and benign under normal circumstances. However, it can be activated to launch a hiding, creating, or altering attack when exposed to a designed distortion introduced by signal injection attacks directed towards cameras (As illustrated in Figure \ref{fig:tpatch}). To ensure practicality and robustness in the real world, the authors propose two key methods. The first is a content-based camouflage method, which helps TPatch avoid detection by human drivers. The second is an attack robustness enhancement method, which strengthens the effectiveness of the attack against target cameras. The TPatch attack process involves several steps as shown in Figure \ref{fig:tpatch_}. The adversary begins by selecting a physical trigger signal and estimating the image distortion it causes. Positive and negative triggers are then designed based on this information. Using the designed triggers, the adversary trains the TPatch. Finally, the visual camouflage and robustness of the patch are improved to make it more practical in real-world scenarios. Once generated, TPatch can be discreetly attached to various objects, enabling the adversary to launch hiding, creating, or altering attacks on camera systems. This sophisticated approach allows for covert manipulation of visual information captured by the cameras, highlighting the potential risks posed by adversarial attacks in real-world settings. 
\textbf{Poltergeist (PG) }is an approach introduced by Zhao et al. \cite{PG} that aims to launch acoustic adversarial attacks against cameras and computer vision systems. Unlike traditional visual-based attacks, Poltergeist leverages acoustic signals to deceive the target systems. By emitting specially crafted acoustic signals, Poltergeist generates subtle perturbations in the camera's sensor readings, leading to misclassification or misleading interpretations by computer vision algorithms. The effectiveness of Poltergeist lies in the exploitation of the vulnerability in the physical design of cameras, where sound waves can indirectly influence the imaging process. By carefully designing the acoustic signals and considering the unique characteristics of the camera's sensor and image processing pipeline, Poltergeist can introduce imperceptible yet impactful perturbations. These perturbations can cause misclassification of objects, introduce false positives or negatives, or even manipulate the perceived visual content. To generate the adversarial acoustic signals, Poltergeist employs an optimization process that maximizes the attack success while ensuring the imperceptibility of the emitted sound. By formulating an objective function that balances the attack strength and the acoustic signal's audibility, Poltergeist finds the optimal parameters for the acoustic attack. The authors also consider real-world constraints, such as distance and noise interference, to evaluate the practicality and robustness of the attacks. Experimental evaluations demonstrate the effectiveness of Poltergeist in evading various computer vision tasks, including object recognition and tracking. By emitting carefully crafted acoustic signals, Poltergeist can deceive cameras and computer vision systems, highlighting the need for robustness and security considerations in the design of such systems.

%%%%%%%%%%%%%%%%%%%%%%%%%%%%%%%%%%%%%%%%%%%%%%%%%%%%%%%%%%%%
%$L_{det}$: detection loss of object detector of the specific class. 
%$L_{tv}$: total variation loss, to promote smoothness of generated patch.
%$L_{cls}$: classification loss, to promote more-realistic patch generation.
%$\lambda_{tv}$ and $\lambda_{cls}$ are regularization weights.
%%%%%%%%%%%%%%%%%%%%%%%%%%%%%%%%%%%%%%%%%%%%%%%%%%%%%%%%%%%%
\begin{figure*}
    \centering
    \includegraphics[width=0.7\textwidth]{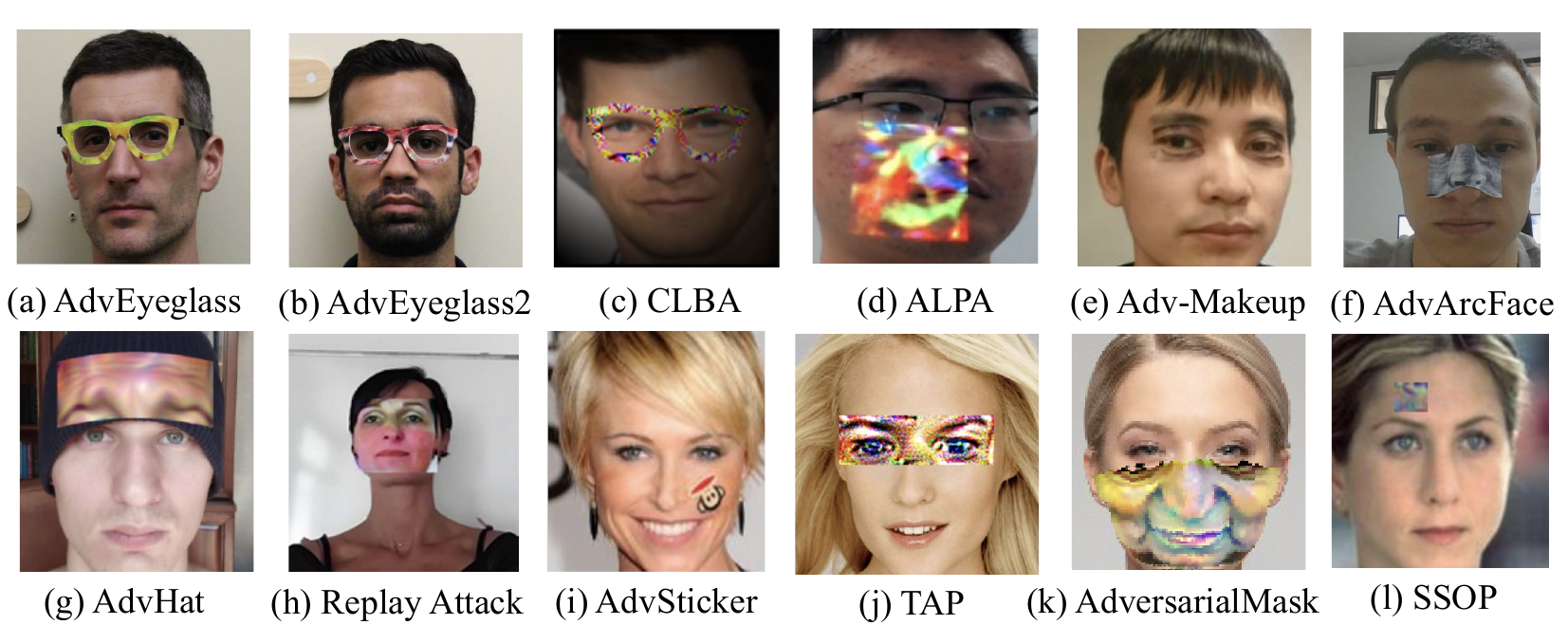}
    \caption{Illustration of different physical attacks on Face recognition task: (a) AdvEyeglass \cite{adveyeglass}, (b) AdvEyeglass2 \cite{adveyeglass2}, (c) CLBA \cite{clbaattack}, (d) ALPA \cite{ALPA}, (e) Adv-Makeup \cite{advmakeup}, (f) AdvArcFace \cite{advarcface}, (g) AdvHat \cite{advhat}, (h) Replay Attack\cite{replayattack}, (i) AdvSticker \cite{advsticker}, (j) TAP \cite{tap}, (k) AdversarialMask \cite{AdversarialMask}, and (i) SSOP \cite{PPattack}.}
    \label{fig:face_recognition}
\end{figure*}
%%%%%%%%%%%%%%%%%%%%%%%%%%%%%%%%%%%%%%%%%%%%%%%%%%%%%%%%%%%%
\begin{table*}[!htp]
\centering
  \caption{ Comparison of attack methods against \textbf{Detection} tasks. MC – Misclassification; HA – Hiding objects; P – Adding phantom objects; AL - Altering objects. F – Anywhere in the frame; O – On the target object(s); S – On the sensor (i.e., camera); NO – Creating new object to fool the model; B: Blurring the image.}
  \label{Table:Detection_comparison}
  \begin{tabular}{lcccccc}
    \toprule
       \textbf{Attack}   & \textbf{Attack goal}  & \textbf{Placement} & \textbf{Consider changing view point} & \textbf{Test in physical domain} & \textbf{Test transferability}\\
    \midrule 

          DPATCH \cite{dpatch}             &  MC-HA-CA  & F & $\times$  & $\times$  & \checkmark \\
          ShapeShifter \cite{shapeshifter} &   MC    & NO &Trained using stop signs   & \checkmark & \checkmark \\
          Object Hider \cite{objecthider}  &  HA     & F & $\times$ & $\times$  &\checkmark \\
          SwitchPatch \cite{switchpatch}   &   MC   & O & \checkmark & \checkmark  &\checkmark \\
          Adversarial T-shirt \cite{adversarialtshirt}   &   HA    & O & Trained on moving people & \checkmark  & \checkmark \\
           Translucent Patch \cite{translucent-patch} &  HA     &  S & $\times$ & \checkmark   & \checkmark \\
          Adversarial YOLO \cite{Adversarialyolo}   &    HA   & O & $\times$ & \checkmark  & \checkmark\\
          TC-EGA \cite{TC-EGA} &  HA & O &  Trained for different viewing angles &  \checkmark   &  \checkmark \\
          CAMOU \cite{camou} & HA   & O  &   Trained for different viewing angles        &  $\times$     & \checkmark     \\
          UPC \cite{UPC} & HA   & O &  Trained for different viewing angles &  $\times$     & \checkmark     \\
          DAS \cite{DAS} & HA   & O &  Trained for different viewing angles &  $\times$     & \checkmark     \\
          FCA \cite{FCA} & HA   & O &  Trained for different viewing angles &  \checkmark     & \checkmark     \\
          CAC \cite{CAC}& HA   & O &  Trained for different viewing angles &  \checkmark     & \checkmark     \\
           DTA \cite{DTA} & HA   & O &  Trained for different viewing angles &  \checkmark     & \checkmark     \\
          Adv. Bulbs \cite{bulb}, & HA & O & $\times$  &  \checkmark     & \checkmark     \\
          QRAttack \cite{qrattack}    & HA   & O   &Trained for different viewing angles     &  \checkmark   & \checkmark\\
           HOTCOLD \cite{hotcold}    & HA   & O   &$\times$  &  \checkmark   & \checkmark\\
            AIP  \cite{AIP}   & HA    &  O   &  $\times$    &   \checkmark    & \checkmark\\
           AdvIB  \cite{AdvIB}   & HA    &  O   &  Trained for different viewing angles/distances     &   \checkmark    & \checkmark\\
          PG\cite{PG}  &   HA-P-AL     & S &   $\times$    &  \checkmark  & \checkmark\\
          TPatch \cite{tpatch}   &   HA-P-AL     & S &   $\times$    &  \checkmark  & \checkmark\\
          AdvART \cite{advart}   &   HA     & O &   $\times$    &  \checkmark  & \checkmark\\
         NAP \cite{NAP}   &   HA     & O &   $\times$    &  \checkmark  & \checkmark\\
         DAP \cite{dap}   &   HA     & O &   \checkmark    &  \checkmark  & \checkmark\\
          NastedAE \cite{nestedae}   &   HA-P     & S &   Trained for different viewing angles/distances    &  $\times$   & \checkmark\\

  \bottomrule
\end{tabular} %}
\end{table*}

%\subsection{clothing/ wearables}

%\subsubsection{Attacks on Person Detectors}

%Person detection, 
%\subsubsection{Attacks on Thermal Infrared Detectors}
%infrared person detection

%\subsection{Camouflage}
%2D vs 3D patch/stickers
%\subsection{Acoustics-based attacks}

\begin{table*}[!htp]
\centering
  \caption{Physical adversarial attacks against \textbf{Detection} tasks. Attacker's knowledge, Robustness technique, Stealthiness technique, Physical test type, and Space.}
  \label{Table:Detection_info}
  \begin{tabular}{llllll}
    \toprule
       \textbf{Attack}  & \textbf{Attacker’s}  & \textbf{Robustness} & \textbf{Stealthiness}  & \textbf{Physical}  & \textbf{Space} \\
         & \textbf{Knowledge}  & \textbf{Technique} & \textbf{Technique}  & \textbf{test type}  &  \\
    \midrule 
            ShapeShifter \cite{shapeshifter}   & White-box  &  EOT & -  & Dynamic &  2D \\
            Extended $RP_2$ \cite{extendedrp2}   & White-box  &  TV,NPS,D2P &  - & Static &  2D \\
            Adversarial YOLO \cite{Adversarialyolo}& White-box  &  EOT,TV,NPS & -  & Dynamic &  2D \\
            NestedAE \cite{nestedae} & White-box & D2P, Alignment & Pattern, Shape and Color control loss  & Dynamic &3D \\
            %    &     &    &    &  \\
            DPATCH \cite{dpatch}  & White-box  & EOT  &  - & Static &  2D \\
            DPatch2 \cite{dpatch2}  & White-box  &  EOT & -  & Static &  2D \\
            LPAttack \cite{lpattack}    & White-box  & EOT,TV,NPS  &  - & Static &  2D \\
            Translucent Patch \cite{translucent-patch} & White-box  & Affine,NPS  & -  & Static &  2D \\
            SwitchPatch \cite{switchpatch} & White-box  & TV  & -  & Static &  2D \\
            Object Hider \cite{objecthider}    & White-box  &  - & -  & Static &  2D \\
            ScreenAttack \cite{dynamicpatch}  & White-box  & TV  & -  & Static &  2D \\
            Invisible Cloak \cite{invisiblecloak} & White-box  & EOT,TV  & -  & Static &  3D \\
            Adversarial T-shirt \cite{adversarialtshirt} & White-box  & EOT,TPS  & -  & Static &  2D \\
            Invisible Cloak2 \cite{invisiblecloak2}   & White-box  & TPS  & -  & Static &  2D \\
            LAP \cite{LAP}  &   White-box &   TV,NPS       &    &   Static        & 2D  \\
            NAP \cite{NAP}&  White-box    &   TV           & GAN  &   Static      &  2D \\
            TC-EGA \cite{TC-EGA}   &  White-box    &   EOT,TPS &  -   &   Static       &  2D  \\
            SLAP \cite{SLAP}   & White-box       &   EOT,TV &  -   &     Static       &  2D \\
            %MeshAdv \cite{meshadv}    &  white-box     &   -     &    -   &   Static      & 3D  \\
             CAMOU \cite{camou}  & White-box   &   EOT   &    -   &    Static     &  2D  \\
             ER attack \cite{ERattack}  & Black-box  & -  & -  &  Static   &  3D \\
             UPC \cite{UPC}  & White-box & EOT, TV   &  $L_\infty$ norm   & Static & 3D \\
             DAS \cite{DAS}  &  White-box &  TV &  Evasion loss  & Static & 3D \\
             FCA \cite{FCA}  & White-box &  TV & -  & Static & 3D \\
             CAC \cite{CAC}  & White-box & EOT & -   & Static & 3D \\
             DTA \cite{DTA}   & White-box   & EOT    &  -   & Static    & 3D\\
             Adversarial Bulbs \cite{bulb} & White-box &  EOT,TV    & -  &   Static &  2D\\
             QRAttack \cite{qrattack}  & White-box &  EOT,TPS &  Material  & Static  & 2D \\
             HOTCOLD \cite{hotcold}  &  White-box   &  SSP    &   Material    &  Static   & 2D  \\
             AIP  \cite{AIP} &  White-box   &  Binary \& Aggregation regularization    &   Material    &  Static   & 2D  \\
             AdvIB \cite{AdvIB}  & Black-box  &  EOT   &  Material    &   Static   & 2D \\
             PG\cite{PG}  & Black-box  &   -  &   -   &  Static    & 2D\\
             %TC-EGA \cite{TC-EGA}  & White-box  &     &    Static  &   2D   & \\
             AdvART \cite{advart} & White-box  & EOT, TV    &  Similarity loss    &  Static    & 2D\\
             DAP \cite{dap} & White-box  & EOT, TV    &  Similarity loss    &  Dynamic    & 2D\\
             TPatch \cite{tpatch}& White \& Black-box  & PDA,TSE    &  Content loss   &  Static    & 2D\\  %verify     
   \bottomrule
\end{tabular} %}
\end{table*}

\begin{table*}[!htp]
\centering
  \caption{Attacks on \textbf{Detection} task, Datasets, networks, and Code.}
  \label{Table:Detection_dataset}
  \begin{tabular}{llll}
    \toprule
       \textbf{Attack}   & \textbf{Dataset} & \textbf{Network}  & \textbf{Code} \\
    \midrule 
            Extended RP2    &  video frames recorded   &  Yolov2, Faster RCNN   &  https://github.com/endernewton/tf-faster-rcnn  \\  
                &  in a laboratory environment  &     &    \\  
            DPATCH    & Pascal VOC 2007   &  Yolo, Faster RCNN   & -   \\  
            ShapeShifter    & MS-COCO   &  Faster R-CNN   & https://github.com/shangtse/robust-physical-attack   \\  
            Adversarial YOLO    & MS-COCO    &  yolov2   &  https://gitlab.com/EAVISE/adversarial-yolo  \\  
            %FIR/ERG    & MS-COCO    &  YOLO V3 and faster-RCNN   &    \\  
            CAMOU    &  MS-COCO  &  Mask R-CNN   &  -  \\  
            UPC    &  AttackScenes  &  FR-RES50-14, FR-RES152-14,   &  https://github.com/mesunhlf/UPC-tf  \\  
                &    &   FR-MN-14, RFCN-RES101-07,   &    \\  
                &    &  SSD-VGG16-0712,Yolov2-14,   &    \\  
               &    &   Yolov3-14 and Retina-14    &    \\  
            Adversarial T-shirt    &  Dataset contains 40 videos   &  Yolov2, Faster RCNN   &  https://github.com/jandress94/adversarial\_tshirt  \\  
                &  (2003 video frames)  &     &    \\  
            Invisible Cloak   &   50 photos with a person standing       &     Tiny YOLO       &   - \\
                            &  in the middle of the camera             &          &   \\
            Invisible Cloak2    &  MS-COCO   & yolov2    & https://github.com/zxwu/adv\_cloak   \\  
            ER Attack    & Carla simulator   &  detectors in Carla simulator   &  -  \\  
            ScreenAttack     &  a dataset for a car with screens   &  YOLOv2 and Fast R-CNN   &   - \\  
             Patch   &   attached from various angles  &     &    \\  
            NAP    & INRIA, MPII Dataset   & YOLOv2, v3, v3tiny , Fast R-CNN,    &  https://github.com/aiiu-lab/Naturalistic-Adversarial-Patch  \\  
                            &    &  v4, v4tiny    &    \\  
            DAP    & INRIA, MPII Dataset   & YOLOv3, v3tiny, v4, v4tiny,   &  -  \\  
                            &    &  Fast R-CNN     &    \\  
            DAS    & Carla simulator   &  Yolo-V5, SSD,Faster R-CNN,   &  https://github.com/nlsde-safety-team/DualAttentionAttack  \\  
                &    &  and Mask R-CNN    &    \\  
            Translucent Patch    &  MS-COCO  &  YOLOv2, faster R-CNN, yolov5   &  -  \\  
            LAP    &  VOC 2007+2012  &  YOLOv2   &  -  \\  
            SLAP    &  LISA, GTSRB  &  Yolov3, Mask-RCNN,   & https://github.com/ssloxford/short-lived   \\  
                &   &   Lisa-CNN, Gtsrb-CNN   & -adversarial-perturbations \\  
            Adversarial Bulbs    & FLIR ADAS dataset v1   & Yolov3    & -   \\  
            QRAttack     & training set of PEOPLE FLIR   &  Yolov3, ensemble models   &  -  \\  
            %Clothing & & & \\
            HOTCOLD  &  FLIR ADAS  &  yolov5   & https://github.com/weihui1308/HOTCOLDBlock   \\  
            AdvIB  &  FLIR ADAS  &  yolov3, DETR, RetinaNet,  &  -  \\
                 &         &      Faster Rcnn, Mask Rcnn, Libra Rcnn           &    \\
            PG    & BDD100K, KITTI   &  YOLO V3/V4/V5 and Fast R-CNN,   &  https://github.com/USSLab/PoltergeistAttack  \\  
                &    &  one commercial detector (Apollo)    &    \\  
            DTA    &  Carla simulator  &  EfficientDetD0 and YOLOv4, eval SSD,   & https://islab-ai.github.io/dta-cvpr2022   \\  
                &    &  Faster R-CNN, and Mask R-CNN    &    \\  
             %TC-EGA   & Inria   &  YOLOv2   &  https://github.com/WhoTHU/Adversarial\_Texture  \\  
             AdvART   &  INRIA, MPII Dataset  &  YOLOv2 , Fast R-CNN, YOLOv3,4, tiny   & -   \\  
             FCA   &  Carla simulator  & YOLOv3, YOLO-V5, Faster RCNN,   &  https://github.com/idrl-lab/Full-coverage-camouflage- \\  
             & &  Mask-RCNN, SSD &adversarial-attack/tree/gh-pages/src  \\
             CAC   &  PascalVOC-2007, 2012,   &  Faster R-CNN with Inception-v2   &  -  \\
                 &   COCO2014         &            &         \\
             Adversarial Rain   &  dataset DEV,               & Inception v3, v4,Inception ResNet v2,     &  -  \\  
                                & MS COCO 2014, KITTI              & Xception, Faster RCNN,     &    \\ 
                                &   & VGG16 (v16), MobileNet (mn), & \\
                                &   & ResNet50, ResNet101, ResNet152 & \\
             AdvRD   & GAN: Rain-drop Removal ,  &  Inc-v3 Inc-v4 IncRes-v2    &  -  \\  
                      & Tsinghua-Tencent 100K   & Res-101 Rob-Res50 IncRes-v2ens    &    \\  
                      &  GTSRB, NIPS-17                      &            &        \\
              TPatch  &  KITTI and BDD100K  &  YOLO V3/V5 and Faster R-CNN & https://github.com/forget2save/TPatch   \\  
              %  &    & VGG-13/16/19, ResNet-50/101/152, Inception v3, and MobileNet v2     &    \\  
              NestedAE &  COCO     &  YOLO V3, Faster RCNN       &   -  \\
              Dpatch2 &  PascalVOC-2007     &   YOLOv3     &   -  \\
              LPattack  & 500 license plates images   & SSD Inception, Faster R-CNN, YOLO & - \\
              Object Hider & dataset provided  &   YOLO, Faster-RCNN     & https://github.com/FenHua/DetDak \\
                       &     by Alibaba Group        &           &      \\
              SwitchPatch  & 3 video clips taken from   &    YOLOv3, v4, v5s     &   -  \\
                   &     vehicle surveillance cameras   &      &     \\
            TC-EGA   &   Inria      &  YOLOv2, YOLOv3,   &  https://github.com/WhoTHU/Adversarial\_Texture    \\
                &      &     Faster R-CNN, Mask R-CNN      &    \\
  \bottomrule
\end{tabular} %}
\end{table*}

%% file: facerecognition.tex
%====================================================================
\section{Physical Attacks on Face recognition \& person re-identification}
%====================================================================
\label{face}
Face recognition and person re-identification are important tasks in computer vision, with the aim of identifying individuals across various images or videos. Face recognition entails the detection and recognition of specific individuals based on their facial characteristics, enabling applications such as identity verification, access control, and surveillance systems. The advent of deep learning models, specifically convolutional neural networks (CNNs), has brought about a revolution in face recognition, showcasing exceptional accuracy and robustness in handling variations in pose, lighting conditions, and facial expressions. These models possess the ability to learn discriminative facial features and map them to distinctive representations, thus facilitating accurate identification and verification of individuals.

Person re-identification, in contrast, focuses on the task of associating individuals across disparate cameras or video frames, even when faced with substantial changes in appearance and variations in viewpoints. Deep learning models have played a pivotal role in propelling the field of person re-identification forward. These models excel at acquiring comprehensive representations that encompass both global and local cues from person images. By harnessing the potential of extensive annotated datasets, these models can effectively learn to extract discriminative features and facilitate accurate matching of individuals across diverse scenes or temporal contexts.

Both face recognition and person re-identification carry substantial implications for security, surveillance, and social applications. They play a pivotal role in enhancing access control mechanisms, facilitating real-time monitoring, and enabling the tracking of individuals across diverse environments. Deep learning models have played a crucial part in elevating the accuracy and scalability of these tasks, opening avenues for advanced applications in fields such as biometrics, law enforcement, and smart cities. Consequently, these advancements contribute to bolstering security measures and promoting public safety.

Physical attacks on face recognition and person re-identification involve the manipulation of physical properties, such as facial appearance or visual characteristics, with the intent to deceive or manipulate the performance of these systems. Face recognition focuses on the identification of individuals based on their facial features, while person re-identification aims to recognize and track individuals across diverse camera views or scenarios.

%Physical attacks on face recognition and person re-identification can manifest in various forms. Attackers may employ tactics such as wearing specially crafted masks or accessories to alter their facial appearance, utilizing makeup or disguises to obscure their identity, or strategically manipulating the physical environment to perplex the perception of these systems.

The ramifications of these attacks are particularly significant in security-sensitive applications, including access control systems, surveillance, and identity verification. By exploiting the vulnerabilities of face recognition and person re-identification systems, attackers can potentially gain unauthorized access, evade surveillance, or impersonate other individuals.
\begin{itemize}
    \item \textbf{Impersonation attack}: the adversary's objective is to be recognized as a specific individual different from their own identity. For instance, the attacker may attempt to discreetly disguise their face in order to be recognized as an authorized user of a laptop or phone that employs face recognition for user authentication. Alternatively, an attacker could aim to deceive law enforcement by simultaneously tricking multiple geographically distant surveillance systems into erroneously identifying their presence in different locations.
    \item \textbf{Dodging attack}: the attacker endeavors to have their face mistakenly identified as any arbitrary face other than their own. From a technical perspective, this category of attack is intriguing because causing a machine learning (ML) system to incorrectly identify a face as any random person should, in theory, be easier to achieve with minimal modifications to the face, as compared to specifically targeting a particular impersonation subject. Apart from malicious purposes, dodging attacks could also be employed by individuals seeking to safeguard their privacy by evading excessive surveillance.
\end{itemize}

Table \ref{Table:Face_recognition_comparison} offers a comprehensive comparison of various adversarial attack methods in the face recognition \& person re-identification tasks. It outlines their attack goals, patch placement strategies, consideration of changing view points, testing in the physical domain, and transferability to other models.
Table \ref{Table:Face_recognition_info} provides detailed information on adversarial attacks, including the attacker's knowledge level, robustness techniques, stealthiness techniques, physical test types, and space of operation.
Table \ref{Table:Face_recognition_dataset} presents information on the datasets used, the networks evaluated, and the links to open-source code for the experiments conducted in the face recognition \& person re-identification tasks.
%%%%%%%%%%%%%%%%%%%%%%%%%%%%%%%%%%%%%
%\subsection{Metrics}

%Impersonate Success Rate (ISR) (Replay Attack)
%Fooling rate (FR)

%%%%%%%%%%%%%%%%%%%%%%%%%%%%%%%%%%%%%

%------------------------------------
\subsection{Camouflage Technique}
%------------------------------------

\subsubsection{Eyeglasses-based Attacks}
%AdvEyeglass \cite{adveyeglass}
%AdvEyeglass2 \cite{adveyeglass2}
%AGN \cite{AGN}

Face Recognition Systems (FRS) are widely used in various applications such as surveillance and access control. Therefore, it is crucial to thoroughly investigate the potential vulnerabilities inherent in FRS. Sharif et al. conducted a study on this matter, as documented in their research \cite{adveyeglass}. They devised a systematic approach to target advanced face recognition algorithms by creating a pair of eyeglass frames. When individuals wear these specially designed adversarial eyeglasses, they can effectively evade recognition or assume the identity of another person. This research underscores the importance of comprehending the risks associated with FRS and the need for robust countermeasures. The researchers demonstrated the capability of an attacker, who lacks knowledge about the internal workings of the system, to carry out inconspicuous impersonation within a commercial Face Recognition System (FRS) \cite{face++}. Additionally, they investigated the dodging attack, in which an assailant can deceive the widely-used face-detection algorithm \cite{viola2001rapid}, resulting in its erroneous identification of the attacker as any arbitrary individual. These findings highlight the existence of deceptive practices within FRS and emphasize the significance of addressing such vulnerabilities in order to ensure the system's integrity and reliability.

Extensive experimental investigations have revealed that adversarial perturbations generated without imposing any constraints, solely focusing on maximizing attack effectiveness, often exhibit visual abruptness and noise. However, natural images possess smooth and coherent characteristics. To address this disparity, Sharif et al. incorporated the total variation loss, denoted as $L_{tv}$ and introduced by Mahendran and Vedaldi \cite{mahendran2015understanding}, as a means to mitigate inconsistencies among neighboring pixels. The inclusion of the total variation loss aimed to enhance visual coherence and smoothness in the generated adversarial perturbations, aligning them more closely with the characteristics exhibited by natural images. The adoption of the total variation loss has subsequently been embraced in numerous subsequent studies \cite{advhat, SLAP, extendedrp2, LAP, Adversarialyolo, DAS, advpattern}, owing to its evident benefits and effectiveness.

In practical settings, devices such as printers and screens have limitations in reproducing the entire RGB color space. Therefore, when crafting adversarial perturbations intended for printing, it is desirable to design perturbations that primarily consist of colors that can be accurately reproduced by the printer. Motivated by this consideration and drawing inspiration from color management technology \cite{giorgianni1998digital}, Sharif et al. \cite{adveyeglass} introduced the concept of the non-printability score (NPS) as a component of the optimization process. The NPS represents how well the colors in the perturbations can be represented by a standard printer.

During the training phase, the algorithm is guided to generate printable perturbations by minimizing the NPS loss function, denoted as $L_{nps}$. The NPS loss $L_{nps}$ quantifies the distance between the generated perturbation vector and a library of printable colors obtained from the physical world. By incorporating the NPS loss, the generated perturbations can be more accurately reproduced in the physical space, enhancing the attack's robustness against hardware limitations and increasing the likelihood of successful realization of the adversarial perturbations using printers.

%---------

\textbf{Eyeglass attack:} Sharif et al. \cite{adveyeglass} conducted research on physical adversarial attacks against face recognition models. In their approach, they utilized facial accessories, specifically eyeglasses, as carriers for adversarial perturbations. To enhance the robustness of the perturbations, they trained them on a set of images, enabling them to perform effectively under various imaging conditions. The smoothness of the perturbations was improved by incorporating the TV loss \cite{mahendran2015understanding}, and the printability of the perturbations was ensured through the introduction of the NPS loss.

The authors formulated an optimization problem to optimize the perturbation, as follows:

\begin{equation}
\text{argmin}_\delta \left( \sum_{x \in X} \text{softmaxloss}(x + \delta, y) + \kappa_1 \text{TV}(\delta) + \kappa_2 \text{NPS}(\delta) \right)
\end{equation}

Here, $\text{softmaxloss}$ represents the adversarial loss for the face recognition system, and $\kappa_1$ and $\kappa_2$ balance the objectives.

In the physical experiment, the authors printed and cropped the adversarial eyeglass frame, which was then affixed to a regular pair of eyeglasses. Participants were instructed to wear the modified eyeglasses and stand at a fixed distance from the camera in an indoor environment. A collection of 30-50 images was captured for each participant. The experimental results indicated that their adversarial eyeglass frame effectively deceived the face recognition system.

Although the eyeglass frame generated by Sharif et al. \cite{adveyeglass} achieves some level of deception in facial recognition, Sharif et al. \cite{adveyeglass2} argue that the appearance of the eyeglass frame remains conspicuous. In order to create an inconspicuous eyeglass frame, the authors propose the use of a generative model to craft the adversarial glasses frame with a more natural pattern. The approach involves collecting a large dataset of real glasses frame images as the training set, followed by training an adversarial generative network (AGN) using the following loss functions: Equation \ref{untargeted_eq} for the untargeted attack and Equation \ref{targeted_eq} for the targeted attack, respectively.

\begin{equation}
\label{untargeted_eq}
\begin{aligned}
    L_G^{untargeted} = \sum_{z \in Z}lg(1 - D(G(z)))
    - \kappa (\sum_{i \neq x}F_{y_i}(x + G(z)) \\- F_{g_t}(x + G(z)))
    \end{aligned}
\end{equation}

\begin{equation}
\label{targeted_eq}
\begin{aligned}
    L_G^{targeted} = \sum_{z \in Z}lg(1 - D(G(z))) 
    - \kappa (F_{y_t}(x + G(z)) \\ - \sum_{i \neq t} F_{y_i}(x + G(z)))
    \end{aligned}
\end{equation}

where the $z$ is sampling from a distribution $Z$, $D(.)$ and $G(.)$ are the output of discriminator and generator, respectively. $F_{y_t}$ and $F_{y_i}$ are the network’s softmax output of the ground-truth label $y_t$ and another label $y_i$. $lg(.)$ is the logarithmic function and $\kappa$ is the weight to balance the loss function. 

Through the minimization of the loss function, the generator effectively learns to generate realistic glasses frames with adversarial characteristics. To further enhance the robustness of the adversarial glasses in physical attacks, the authors introduce three specific methods. Firstly, they propose the utilization of multiple images of the attacker to improve the robustness of the adversarial glasses under diverse conditions. This approach enables the glasses to maintain their deceptive properties across various environmental factors. Secondly, the authors collect a wide range of pose images to bolster the attacks' resilience against changes in pose. By incorporating pose variations during the training phase, the adversarial glasses become more effective in fooling the face recognition system regardless of the subject's pose. Thirdly, the authors employ the Polynomial Texture Maps approach \cite{malzbender2001polynomial} to restrict the RGB values of the eyeglass frames within a specific luminance range. This technique ensures that the adversarial attacks remain effective even under varying illumination conditions. In physical attack scenarios, the authors print out the eyeglass frames and the corresponding adversarial patterns, affix them to the actual eyeglass frames, and then recapture images for evaluation. Experimental results demonstrate that their proposed method outperforms the approach presented in \cite{adveyeglass} by achieving a remarkable 45\% improvement under different physical conditions, such as head poses with a range of 13.01° in pitch, 17.11° in yaw, and 4.42° in roll. Figure X (to be referenced) provides a visual illustration of the printed adversarial eyeglasses generated by both \cite{adveyeglass} and \cite{adveyeglass2}.

%CLBA attack \cite{clbaattack} (On Brightness Agnostic Adversarial Examples) performs non-linear brightness transformations while leveraging the concept of curriculum learning during the attack generation procedure.
\textbf{CLBA attack}: In a recent study, Singh et al. \cite{clbaattack} introduced the concept of curriculum learning to enhance the robustness of attacks against facial recognition models in the presence of changing brightness conditions. During the optimization process, the authors continuously vary the brightness of face images combined with the adversarial eyeglasses. This approach enables their attack to remain effective and robust against variations in brightness levels. The efficacy of the proposed attack strategy under changing brightness conditions was further validated through physical attack experiments.

%----------------
\subsubsection{Mask-based Attacks}
%---------------
%AdversarialMask \cite{AdversarialMask}

\textbf{AdversarialMask:} In the prevalence of the COVID-19 pandemic, Zolfi et al. \cite{AdversarialMask} drew inspiration and developed a mask-like universal adversarial patch, thereby enhancing the stealthiness of the attack. By wearing the specially crafted adversarial mask, adversaries can potentially bypass face recognition systems in certain scenarios, such as airports. In their approach, the authors introduced a mask projection method that transforms both faces and masks into the UV space to emulate the appearance of actual masks being worn. The mask image is optimized using a combination of cosine similarity loss and total variation (TV) loss to ensure adversarial effects and smoothness. Moreover, geometric transformations and color-based augmentations are employed to enhance the robustness of the mask against different environmental conditions. To carry out physical attacks, the authors fabricated the mask using two different materials: regular paper and a white fabric mask. They assembled a group of 15 male and 15 female participants to wear the devised masks, and the experimental results demonstrated a significant decline in the performance of the face recognition model when faced with individuals wearing the adversarial mask (e.g., accuracy decreased from 74.83\% to 5.72\% and 4.61\% for the paper and fabric masks, respectively).
\begin{figure*}
    \centering
    \includegraphics[width=\textwidth]{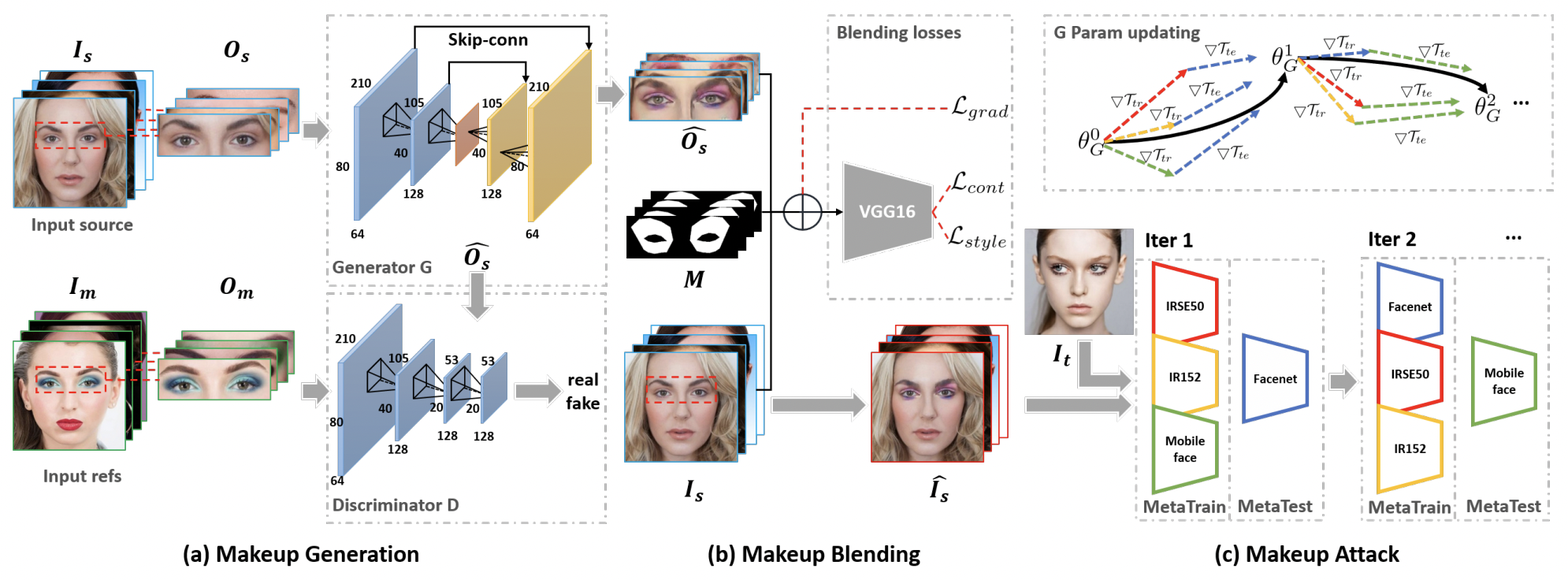}
    \caption{The Adv-Makeup framework: encompassing three integral components: makeup generation, makeup blending, and fine-grained meta-learning adversarial attack. In the makeup generation phase (a), facial images are crafted, featuring eye shadow meticulously applied over the orbital regions. Subsequently, the makeup blending process (b) comes into play, refining the appearance of the eye shadow to seamlessly integrate it with the surrounding context, rendering the generation imperceptible. The framework is further fortified by the inclusion of fine-grained meta-learning adversarial attack (c), a strategic mechanism that facilitates transferable impersonation attacks. This is achieved by leveraging different colors in the 'G param updating' stage to represent various victim face recognition models, thereby enhancing the overall adaptability and potency of the framework (Figure Adapted from \cite{advmakeup}).}
    \label{fig:makeup}
\end{figure*}
%----------------
\subsubsection{Makeup-based Attack}
%----------------
%Adv-Makeup \cite{advmakeup}

Yin et al. \cite{advmakeup} conducted an extensive analysis of existing attacks on Face Recognition Systems (FRS) and put forward a novel and unified method called Adv-Makeup for generating adversarial faces. The \textbf{Adv-Makeup }approach focuses on a practical scenario where makeup is applied to the eye regions in a manner that effectively misleads FRS models, while maintaining a visually imperceptible appearance that resembles natural makeup. The proposed approach to generate this attack is presented in Figure \ref{fig:makeup}.

%Specifically, Adv-Makeup introduces a dedicated makeup generation module that applies natural-looking eye shadow to the orbital region. This approach ensures that the generated makeup is visually consistent with the overall appearance of the human face. Furthermore, a task-driven fine-grained meta-learning adversarial attack strategy is employed to enhance the effectiveness of the generated makeup in evading FRS models. Experimental results demonstrate that Adv-Makeup achieves significantly higher attack effectiveness compared to previous methods such as AdvEyeglass \cite{adveyeglass}. This notable advancement underscores the success of the Adv-Makeup approach in undermining the robustness of FRS systems. The authors also emphasize the concept of enhancing concealment through the choice of the adversarial medium and propose the utilization of makeup as a carrier for perturbations. Given its inherent harmony with the human face, makeup serves as an effective and stealthy adversarial medium, further augmenting the success of the Adv-Makeup approach.
The Adv-Makeup technique introduces a dedicated makeup generation module that applies natural-looking eye shadow to the orbital region of a human face. This approach ensures that the generated makeup blends seamlessly with the overall appearance of the face, making it visually consistent and inconspicuous. Additionally, Adv-Makeup employs a task-driven fine-grained meta-learning adversarial attack strategy to enhance the effectiveness of the generated makeup in evading Face Recognition Systems (FRS).
In comparison to previous methods like AdvEyeglass \cite{adveyeglass}, experimental results demonstrate that Adv-Makeup achieves significantly higher attack effectiveness. This remarkable improvement underscores the success of the Adv-Makeup approach in undermining the robustness of FRS systems. The authors highlight the importance of enhancing concealment through the choice of the adversarial medium and propose the utilization of makeup as a carrier for perturbations. Makeup, being inherently harmonious with the human face, serves as an effective and stealthy adversarial medium, further augmenting the success of the Adv-Makeup approach.
The Adv-Makeup technique represents a noteworthy advancement in adversarial attacks, utilizing makeup as a medium to create inconspicuous and effective perturbations on the human face. This approach raises important considerations in the field of face recognition security and calls for further research in developing robust countermeasures to mitigate such adversarial threats.
%----------------
\subsubsection{Clothing-based Attack} (person re-identification)
%----------------
\begin{figure*}
    \centering
    \includegraphics[width=\textwidth]{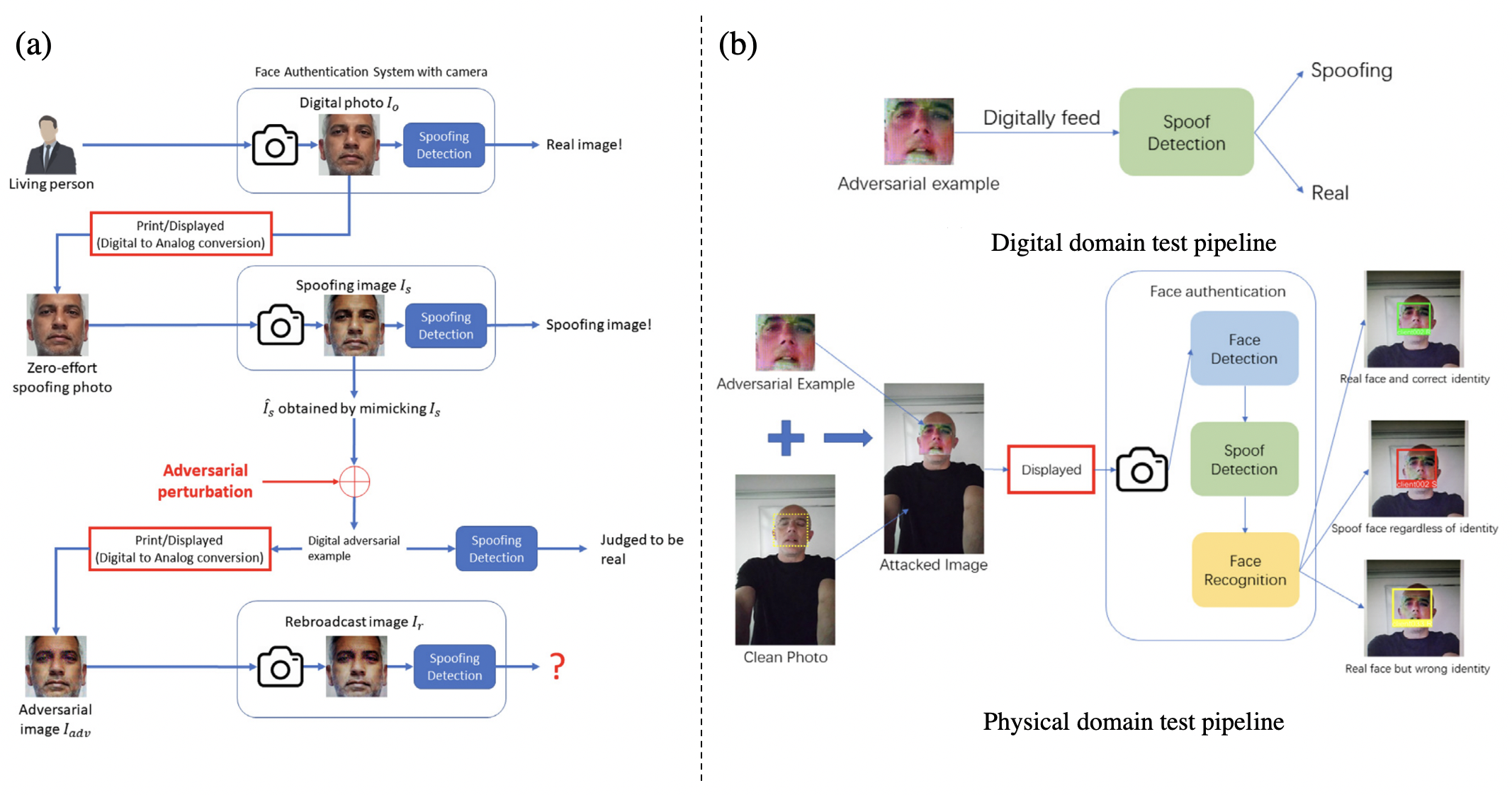}
    \caption{Replay Attack \cite{replayattack}: (a) Adversarial image generation process against an anti-spoofing system, (b) Testing pipeline in both digital and physical domains.}
    \label{fig:replayattack}
\end{figure*}
%AdvPattern \cite{advpattern}

Physical adversarial attacks have extended their reach into the domain of person re-identification tasks. Wang et al. \cite{advpattern} introduced a novel attack algorithm known as \textbf{advPattern}, which generates adversarial patterns on clothing. These patterns have proven to be highly effective in deceiving person re-identification models in real-world scenarios. By leveraging advPattern, the researchers achieved a significant reduction in the top-1 accuracy of person re-identification models, dropping from 87.9\% to 27.1\%. Moreover, in the context of impersonation attacks, advPattern achieved an impressive top-1 accuracy of 47.1\% and a mean Average Precision (mAP) of 67.9\% while successfully impersonating a specific target individual. These results highlight the potential impact and effectiveness of advPattern in undermining the reliability and robustness of person re-identification systems.

%------------------------------------
\subsection{Patch-based Attack}
%------------------------------------
%AdvArcFace \cite{advarcface} adv-patch

\textbf{AdvArcFace}: Pautov et al. \cite{advarcface} conducted an in-depth investigation into the potential of launching physical attacks on the ArcFace algorithm \cite{arcface} using adversarial patches. To this end, they formulated a cosine similarity loss function that minimizes the similarity between the patched photo and the corresponding ground truth. The resulting adversarial patch is designed in grayscale, ensuring its ease of printability. The researchers conducted experiments to assess the effectiveness of the adversarial patch in three distinct scenarios. In the first scenario, they applied an eyeglass patch to the image, while in the other two scenarios, stickers were strategically placed on the nose and forehead. Through comprehensive numerical experiments, the researchers successfully demonstrated the feasibility of launching efficient physical attacks on the ArcFace algorithm in real-world settings.
%------------------------------------
\subsection{Image-based Attack}
%------------------------------------
%Replay Attack \cite{replayattack}
\textbf{Replay Attack}: Zhang et al. \cite{replayattack} conducted a study highlighting the vulnerability of face authentication systems equipped with DNN-based spoof detection modules to adversarial perturbations. Drawing inspiration from the concept of Expectation Over Transformation (EOT) \cite{EOT}, the authors trained the perturbation through a series of transformations, including affine transformations, perspective transformations, adjustments to brightness, Gaussian blurring, and modifications to hue and saturation. To execute physical attacks, the authors displayed the face image containing the adversarial perturbation on a screen and captured it using a cellphone from various viewpoints. By adopting this approach, they were able to evaluate the effectiveness of the adversarial perturbation in real-world scenarios with different viewing conditions. Their research sheds light on the continued vulnerability of face authentication systems to adversarial perturbations, even when equipped with DNN-based spoof detection modules. Figure \ref{fig:replayattack} illustrates the process to generate an adversarial image and the test pipeline.

%------------------------------------
\subsection{Sticker-based Attack}
%------------------------------------
%AdvHAT \cite{advhat}
%TAP \cite{tap} GenAP
%AdvSticker \cite{advsticker} RHDE
%SSOP \cite{PPattack}
%FaceAdv \cite{faceadv}
\begin{figure*}
    \centering
    \includegraphics[width=\textwidth]{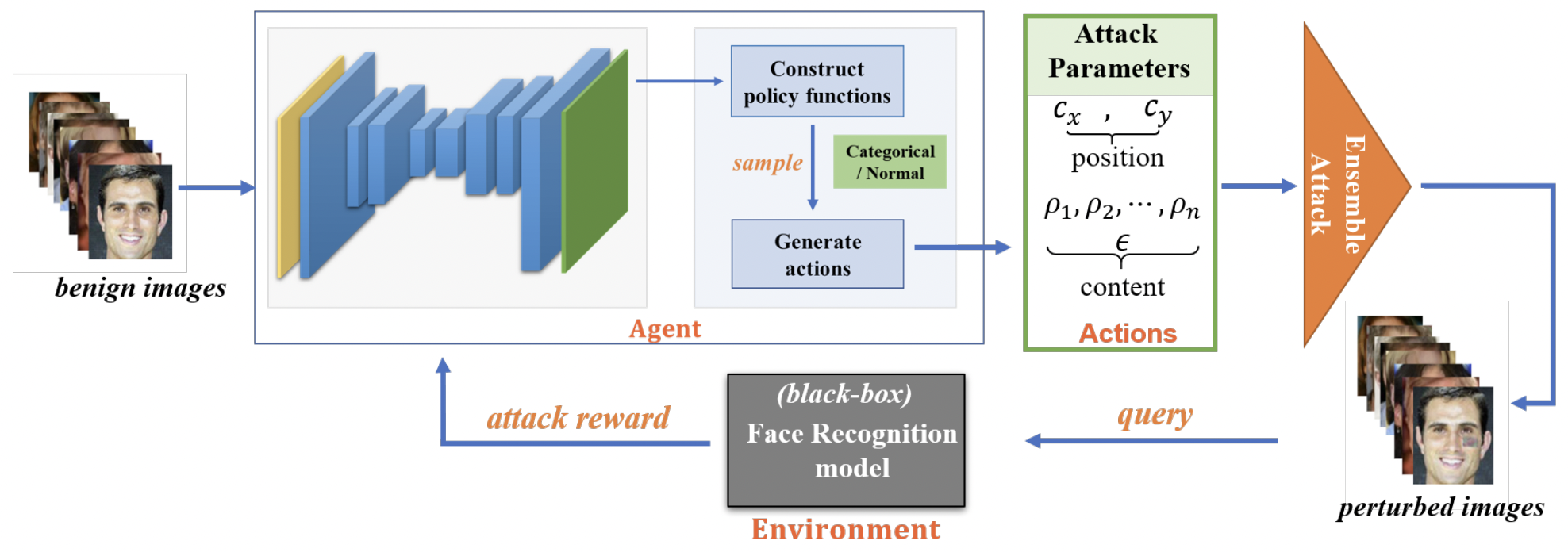}
    \caption{An overview on the simultaneous optimization of positions and perturbations within a reinforcement learning (RL) framework is presented. The process involves inputting a benign image to an agent, which constructs policies using RL principles. Sampling is performed based on these policies to determine specific attack variable solutions (actions). Employing an ensemble attack and querying the target model generates rewards, which are utilized to update the agent's parameters through iterative refinement. This approach facilitates the efficient and effective optimization of both positions and perturbations, contributing to the overall success of the adversarial attack (Figure adapted from \cite{PPattack}).}
    \label{fig:ssop}
\end{figure*}
\textbf{Advhat} \cite{advhat} successfully executed a physically reproducible adversarial attack on the widely recognized public Face ID system \cite{arcface, grother2014face, megaface}. In the digital domain, Advhat employs the Spatial Transformer Layer (STL) \cite{NIPS2015_5854} to project a sticker onto the facial image, creating an adversarial perturbation. In the physical realm, Advhat executes attacks by incorporating a specially designed hat adorned with a distinctive sticker placed on the forehead region. This strategic placement effectively diminishes the similarity between the attacked image and the ground truth class, thereby compromising the accuracy of the Face ID system. These findings underscore the potential vulnerability of the state-of-the-art Face ID system to physical adversarial attacks using the methodology proposed by Advhat.

To enhance the robustness of the attack, Advhat introduces the concept of off-plane sticker transformation. This transformation involves simulating off-plane bending as a parabolic transformation, followed by pitch rotation in three-dimensional space, mimicking the placement of a sticker on the hat. By incorporating these transformations, Advhat improves the attack's resilience to variations in viewing angles and head poses, making it more effective and challenging for the Face ID system to accurately identify individuals.

%\textbf{TAP}: Xiao et al. \cite{tap} proposed an approach to enhance the transferability of adversarial patches for face recognition systems. The authors analyzed the limitations of existing transferability methods, which often suffer from sensitivity to initialization (e.g., MIM \cite{dong2018boosting}) and overfitting issues when the perturbation magnitude is significant. To overcome these challenges, the authors introduced a regularization technique that constrains the adversarial patch to lie within a low-dimensional manifold represented by a generative model. In the optimization process, the adversarial patch is carefully shaped and positioned on the eye area of the face. By employing this specific placement strategy, the authors conducted physical attacks by printing and capturing images of individuals with the adversarial patch applied to their eyes. The experimental results demonstrated the effectiveness of the proposed method in achieving successful adversarial attacks. The TAP approach presented by Xiao et al. \cite{tap} represents a significant advancement in improving the transferability of adversarial patches for face recognition systems. By leveraging a generative model regularization and carefully designing the placement of the patch, the authors successfully demonstrate the potential for physical attacks on face recognition systems, highlighting the importance of addressing the vulnerabilities associated with adversarial patches in real-world scenarios.
The \textbf{TAP} approach, introduced by Xiao et al. \cite{tap}, aims to enhance the transferability of adversarial patches for face recognition systems. The authors identified limitations in existing transferability methods, which are often sensitive to initialization and prone to overfitting when the perturbation magnitude is significant. To overcome these challenges, the authors proposed a regularization technique that constrains the adversarial patch to lie within a low-dimensional manifold represented by a generative model. During the optimization process, the adversarial patch is carefully shaped and positioned on the eye area of the face. This specific placement strategy allows for physical attacks, where images of individuals with the adversarial patch applied to their eyes are printed and captured. Experimental results demonstrated the effectiveness of the TAP approach in achieving successful adversarial attacks on face recognition systems. By leveraging generative model regularization and thoughtful design of patch placement, the authors successfully improve the transferability of adversarial patches, showcasing the potential for physical attacks on face recognition systems. %The TAP approach represents a significant advancement in the realm of adversarial attacks, particularly for face recognition systems. It sheds light on the importance of addressing vulnerabilities associated with adversarial patches in real-world scenarios and calls for further research in developing robust countermeasures to mitigate such threats.

Unlike traditional approaches that focus on optimizing the pattern or color of adversarial perturbations, Wei et al. \cite{advsticker} argue that using a natural image patch alone is sufficient to launch effective attacks on facial recognition models. They propose a general framework for identifying the most sensitive regions of the model that are susceptible to adversarial stickers. To determine the vulnerable areas, the authors employ a traversal method that involves scanning the face image with a patch image, which significantly speeds up the search process. During the optimization stage, the authors incorporate bend and rotation transformations in a three-dimensional space to deform the adversarial patch, mimicking the appearance of a sticker on a real face. In physical attacks, the authors strategically apply the adversarial patch to the attacker's face based on the identified coordinates from the search process. Experimental results demonstrate that a natural image patch placed in the identified vulnerable regions is sufficient to degrade the performance of deep neural network models in both digital and physical conditions.

\textbf{AdvSticker}: The approach presented by Wei et al. \cite{advsticker} represents a novel perspective on adversarial attacks, highlighting the effectiveness of natural image patches in compromising the performance of facial recognition systems. By leveraging the inherent vulnerabilities of the targeted regions, this method provides valuable insights into the design of more practical and impactful attacks, emphasizing the need for robust defense mechanisms against such threats. 
\textbf{SSOP}: In a recent study, Wei et al. \cite{PPattack} highlighted the significance of both the position and pattern of adversarial patches for successful attacks on face recognition systems. Recognizing the importance of optimizing both factors, the authors devised a method to simultaneously optimize the patch position and perturbation against the face recognition model illustrated in Figure \ref{fig:ssop}. To reduce the complexity of the optimization process, the authors employed a UNet architecture to generate the patch position and utilized the attack step from existing gradient-based algorithms, such as I-FGSM \cite{PAE}. This approach effectively reduced the number of optimization variables. Reinforcement learning was then employed to formulate the optimization problem, treating the UNet as the agent, the sampling strategy as the policy, and the specific optimization variables as the action. Rewards were obtained by querying the ensemble model with the generated adversarial examples. In physical attacks, the authors printed the adversarial patch using photo paper to minimize color discrepancy. Experimental results demonstrated the effectiveness of their approach, achieving an average success rate of 66.18\% on captured images. %The study by Wei et al. \cite{PPattack} highlights the importance of jointly optimizing the position and pattern of adversarial patches for face recognition attacks. By leveraging reinforcement learning and employing a UNet-based framework, their method provides insights into improving the success rate of physical attacks. These findings contribute to the ongoing research on adversarial attacks and emphasize the need for robust defenses against such attacks in face recognition systems.

%------------------------------------
\subsection{Light Manipulation-based Attacks}
%------------------------------------
%ALPA \cite{ALPA}

\textbf{ALPA}: The effectiveness of light-based attacks has been extensively explored in classification tasks \cite{OPAD}. Building upon this research, Nguyen et al. \cite{ALPA} investigated the feasibility of light-based adversarial attacks on face recognition systems. Their study focused on two state-of-the-art face recognition models, namely FaceNet \cite{Schroff_2015} and SphereFace \cite{liu2018sphereface}, highlighting the potential vulnerability of these advanced systems to light-based adversarial attacks. To successfully execute the light-based attack, the researchers identified two crucial calibration steps. The first step involved position calibration, ensuring precise projection of the generated adversarial pattern onto the attacker's face. This calibration ensured accurate alignment of the digital adversarial pattern with the physical projection during the attack. The second step, color calibration, aimed to faithfully reproduce the digital adversarial pattern using the projector, ensuring that the projected light pattern accurately matched the intended digital design. To evaluate the vulnerability of face recognition systems to light projection attacks, the researchers targeted both open-source and commercial systems. They conducted experiments involving 50 subjects and assessed the effectiveness of the attacks in both white-box and black-box scenarios, where the attackers had different levels of knowledge about the targeted systems. The results of the study demonstrated the efficacy of light projection attacks in compromising the face recognition systems under investigation. %These findings highlight the susceptibility of face recognition systems to light-based adversarial attacks and underscore the importance of developing robust countermeasures to mitigate such vulnerabilities.

\begin{table*}[!htp]
\centering
      \caption{ Comparison of attack methods against \textbf{Face Recognition} \& \textbf{person re-identification} tasks. IM – Impersonation ; DG – Dodging ; OB - Obfuscation/evade recognition. F – Anywhere in the frame; O – On the target object(s); S – On the sensor (i.e., camera); NO – Creating new object to fool the model; B: Blurring the image.}
  \label{Table:Face_recognition_comparison}
  \begin{tabular}{lcccccc}
    \toprule
       \textbf{Attack}   & \textbf{Attack goal}  & \textbf{Placement} & \textbf{Consider changing view point} & \textbf{Test in physical domain} & \textbf{Test transferability}\\
    \midrule 
          AdvEyeglass \cite{adveyeglass} & IM,DG &  O  &  $\times$   &  \checkmark      & \checkmark   \\   %\checkmark 
          AdvEyeglass2  \cite{adveyeglass2} & IM,DG &  O  &  $\times$   &  \checkmark   &  \checkmark \\  
          AdvHat \cite{advhat} & IM,DG &  O  &  $\times$   &  \checkmark     & \checkmark   \\ 
          AdvArcFace \cite{advarcface} & IM,DG &  O  &  \checkmark   &  \checkmark      &  $\times$ \\ 
          TAP \cite{tap} & IM,DG  & O   &  $\times$   &  \checkmark     & \checkmark   \\ 
          AdversarialMask  \cite{AdversarialMask}  & OB &  O  &  Trained for different angles   &   \checkmark    &  \checkmark  \\ 
          ALPA \cite{ALPA}  & IM, OB & O & \checkmark   &   $\times$      & $\times$\\ 
          CLBAAttack \cite{clbaattack} & IM, DG    &  O  &  $\times$   &  \checkmark      & $\times$  \\ 
          ReplayAttack \cite{replayattack} & IM  &  F  &  $\times$   &  \checkmark     & $\times$ \\ 
          AdvSticker \cite{advsticker} & IM, DG &  O  &  \checkmark   &  \checkmark    & \checkmark\\ 
          SSOP \cite{PPattack} & IM, DG  & O   &  $\times$     &  \checkmark  & \checkmark\\ 
  \bottomrule
\end{tabular} %}
\end{table*}

\begin{table*}[!htp]
\centering
  \caption{Physical adversarial attacks against \textbf{Face recognition} \& \textbf{person re-identification} tasks. Attacker's knowledge, Robustness technique, Stealthiness technique, Physical test type, and Space.}
  \label{Table:Face_recognition_info}
  \begin{tabular}{llllll}
    \toprule
       \textbf{Attack}  & \textbf{Attacker’s}  & \textbf{Robustness} & \textbf{Stealthiness}  & \textbf{Physical}  & \textbf{Space} \\
         & \textbf{Knowledge}  & \textbf{Technique} & \textbf{Technique}  & \textbf{test type}  &  \\
    \midrule 
            AdvEyeglass \cite{adveyeglass}   & White-box  &  TV,NPS & -  & Static &  2D \\
             AdvEyeglass2  \cite{adveyeglass2}  & White-box   & Alignment & - &  Static    &  2D  \\
            AdvHat \cite{advhat} & White-box  &  TV, Bending & Cosine similarity loss  & Static &  2D \\
            AdvArcFace \cite{advarcface}  & White-box  &  TV & -  & Static &  2D \\
            TAP \cite{tap} & White-box  & EOT & GAN  & Static &  2D \\
            AdversarialMask  \cite{AdversarialMask}  & White-box   &  Rendering,TV   &    -   &  Static   &   3D \\
            
            ALPA \cite{ALPA}& White-box   &   Alignment  &    -   &  Static    &  2D  \\
            CLBAAttack \cite{clbaattack}  & White-box, Black-box   &  EOT   &   -    &  Static    &  2D  \\
            ReplayAttack \cite{replayattack}  & White-box   &  EOT   &    PSNR   &  Static    &  2D  \\
            AdvSticker \cite{advsticker}  &  Black-box   & Bending, Rotation & Uses Real stickers &  Static    &  3D  \\
            SSOP \cite{PPattack}  & Black-box   & - & - &  Static    &  2D  \\
  
   \bottomrule
\end{tabular} %}
\end{table*}

\begin{table*}[!htp]
\centering
  \caption{Attacks on \textbf{Face recognition} \& \textbf{person re-identification} task, Datasets, networks, and Code.}
  \label{Table:Face_recognition_dataset}
  \begin{tabular}{llll}
    \toprule
       \textbf{Attack}   & \textbf{Dataset} & \textbf{Network}  & \textbf{Code} \\
    \midrule 
             AdvEyeglass   & PubFig dataset   &  DNNB and DNNC  & https://github.com/mahmoods01/  \\  
                &   &    & accessorize-to-a-crime  \\   
             
             AdvEyeglass2   & PubFig dataset   &   VGG143 and OF143  &  https://github.com/mahmoods01/agns \\  
             %AGN   & Labeled Faces  &  Two of the DNNs were built on the , & https://github.com/mahmoods01/agns  \\  
             %   & in the Wild (LFW) benchmark     &  Visual Geometry Group (VGG) neural network  &   \\ 
             %   &    & Labeled Faces in the Wild (LFW) benchmark      &   \\
             AdvArcFace   &  CASIA-WebFace  &  LResNet100E-IR, multi-task   &  - \\
                &    & cascaded CNN based framework (MTCCN)   &   \\
             ALPA   & the Wild (LFW) benchmark,   &  FaceNet, SphereFace  &  - \\
             AdvHat   &  CASIA-WebFace, MS-Celeb-1M  &  model LResNet100E-IR, ArcFace@ms1m-refine-v2  &https://github.com/papermsucode/advhat   \\
             AdversarialMask & CASIA-WebFace dataset,& ResNet100@ArcFace & https://github.com/AlonZolfi/AdversarialMask\\
             Adv-Makeup & LFW, make-up dataset & LADN with a U-Net structure & https://github.com/TencentYoutuResearch/\\
              &  &  & Adv-Makeup\\
             Face Adv & LFW, VolFace &ArcFace, CosFace and FaceNet & -\\
            TAP  & LFW, CelebA-HQ & FaceNet, CosFace, ArcFace & -\\
            CLBAAttack &   VggFace2   &  ResNet50, MobileFaceNet,SE-IR50, SE-IR100    &    \\
           ReplayAttack  & REPLAY-MOBILE database    &   Fine-tuned VGG-16   & - \\
           AdvSticker  &   LFW, CelebA & FaceNet, CosFace50, ArcFace34 and  & https://github.com/jinyugy21/Adv-Stickers \\
             & & ArcFace50 and one commercial face recognition API& \\
            SOPP & LFW, CelebA & FaceNet, CosFace50, ArcFace34 and ArcFace5, & https://github.com/shighghyujie/newpatch-rl\\
             & & one commercial face recognition & \\
  \bottomrule
\end{tabular} %}
\end{table*}

%% file: semanticSeg.tex
%=================================================
\section{Physical Attacks on Semantic Segmentation}
%=================================================
\label{semantic_segmentation}
Semantic segmentation is a critical task in the field of computer vision, which involves assigning semantic labels to each pixel in an image, thereby enabling a detailed understanding of the image's content. The advent of deep learning, particularly convolutional neural networks (CNNs), has significantly advanced the field of semantic segmentation. These powerful models can learn intricate patterns and contextual information, allowing them to accurately classify pixels and delineate object boundaries. By training on extensive annotated datasets, CNNs acquire the necessary visual representations for precise pixel-level classification. Semantic segmentation finds applications in diverse domains, such as autonomous driving, scene understanding, medical imaging, and augmented reality. It plays a crucial role in higher-level tasks, including object recognition, instance segmentation, and scene understanding, contributing to the development of advanced computer vision systems with enhanced capabilities.

%Physical attacks on semantic segmentation involve manipulating the physical properties of input data to deceive or manipulate the performance of semantic segmentation models. Semantic segmentation focuses on assigning semantic labels to pixels, providing a detailed understanding of the scene at a fine-grained level.

Physical attacks on semantic segmentation pose unique challenges due to the pixel-level nature of the task. Attackers can introduce perturbations or modifications to the physical environment that result in misclassification or mislabeling of pixels by the segmentation model. For instance, by strategically placing specially designed objects or patterns in the scene, an attacker can manipulate the segmentation outputs of the model, leading to incorrect interpretations of the scene.

These attacks have significant implications in real-world scenarios. For example, in autonomous driving systems, attackers could modify the appearance of road markings or signage to deceive the semantic segmentation model, potentially leading to erroneous decisions by the autonomous vehicle. Similarly, in surveillance applications, attackers could manipulate the physical environment to conceal or alter objects of interest, evading accurate semantic segmentation and compromising the overall system's effectiveness.

Table \ref{Table:Semantic_comparison} presents a comprehensive comparison of various adversarial attack methods in the semantic segmentation task. It provides an overview of their attack goals, patch placement strategies, consideration of changing viewpoints, testing in the physical domain, and transferability to other models.
Table \ref{Table:Semantic_info} offers detailed information on adversarial attacks, including the attacker's knowledge level, robustness techniques, stealthiness techniques, physical test types, and space of operation.
Table \ref{Table:Semantic_dataset} provides information on the datasets used, the evaluated networks, and the links to open-source code for the experiments conducted in the semantic segmentation task.

%%%%%%%%%%%%%%%%%%%%%%%%%%%%%%%%%%%%%%%%%%%%%%%%%%%%%%%%%%%%
\begin{figure}
    \centering
    \includegraphics[width=0.5\textwidth]{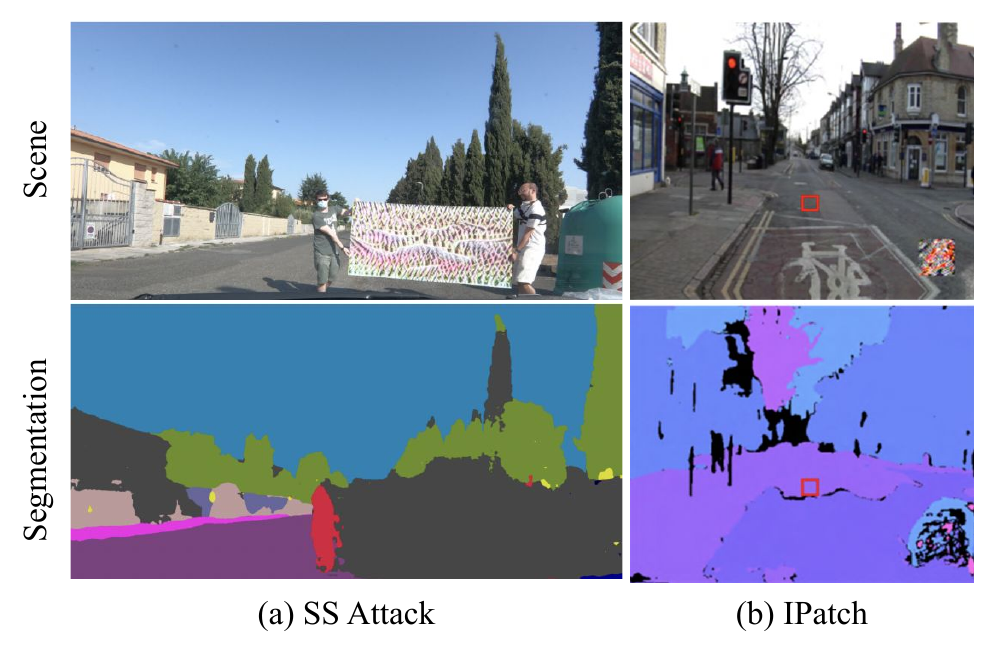}
    \caption{Illustration of different physical attacks on semantic segmentation task: (a) SS Attack \cite{ssattack}, and (b) IPatch \cite{ipatch}.}
    \label{fig:seg}
\end{figure}
%%%%%%%%%%%%%%%%%%%%%%%%%%%%%%%%%%%%%%%%%%%%%%%%%%%%%%%%%%%%
%------------------------------------
\subsection{Patch-based Attacks}
%------------------------------------
%IPatch \cite{ipatch}
%SS Attack \cite{ssattack}

Semantic Segmentation is a task that involves classifying each pixel into predefined categories without distinguishing between object instances \cite{he2018mask}. Nesti et al. \cite{ssattack} developed a novel approach called \textbf{SS Attack}, which focuses on crafting adversarial patches to perturb DNN-based models within the semantic segmentation framework. To optimize the effectiveness of the SS Attack, pixel-wise cross-entropy loss was employed to optimize the patches on a pre-trained ICNet \cite{zhao2018icnet}. The resulting patches were then printed as a $1m \times 2m$ poster in the physical space. Experimental results demonstrate that the SS Attack successfully reduces the accuracy of the baseline semantic segmentation model. However, it should be noted that the attack's performance is significantly degraded when applied in real-world scenarios. Regarding stealthiness, the SS Attack does not specifically consider this aspect in its design. However, for robustness, the SS Attack incorporates the previously mentioned Expectation over Transformation (EOT) technique \cite{EOT}. This helps to improve the robustness of the attack. Additionally, the SS Attack generates diverse and rich scenes using the CARLA Simulator \cite{dosovitskiy2017carla} to create scene-specific patch attacks, enhancing the attack's ability to generalize across different environments. In summary, the SS Attack devised by Nesti et al. presents a method for generating adversarial patches within the semantic segmentation setting. The attack demonstrates effectiveness in reducing model accuracy, although its performance is notably hindered in real-world scenarios. Further considerations for stealthiness and robustness are addressed, with the incorporation of EOT transformations and scene-specific patch attacks using the CARLA Simulator. 
\begin{figure*}
    \centering
    \includegraphics[width=0.8\textwidth]{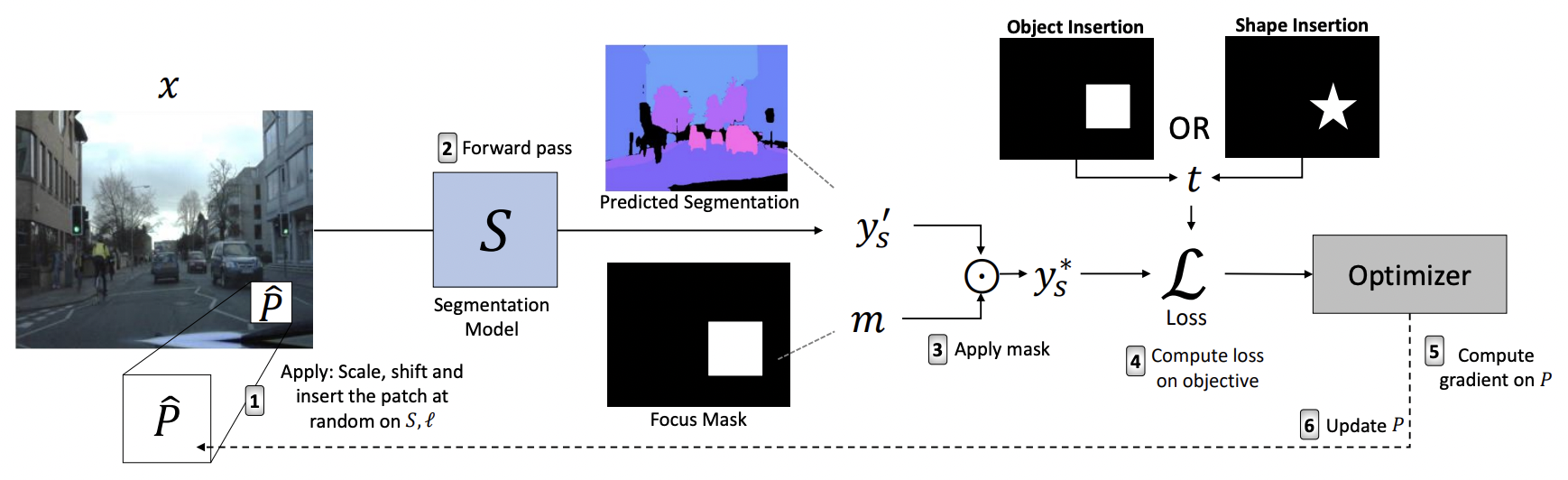}
    \caption{An overview of the IPatch training framework for creating a remote adversarial patch (Figure adapted from \cite{ipatch}).}
    \label{fig:ipatch}
\end{figure*}

\textbf{IPatch} \cite{ipatch} %is a novel method that introduces a remote adversarial patch to deceive object detection systems. Unlike traditional adversarial patches that are physically applied to objects, IPatch operates remotely by leveraging digital techniques. This approach allows for the manipulation of object detections without the need for physical access to the scene. The key idea behind IPatch is to exploit the vulnerabilities of object detection models by strategically placing a patch within the scene. The patch is designed to manipulate the model's decision-making process, causing it to misclassify or ignore certain objects. By carefully optimizing the properties of the patch, such as its size, shape, and color, IPatch can effectively deceive the object detection system while minimizing visual artifacts. The optimization process of IPatch involves iteratively updating the patch to maximize its impact on the model's predictions. This is achieved by leveraging gradient-based optimization techniques and the model's backpropagation mechanism. By backpropagating the gradients through the model, IPatch can identify the most influential regions of the patch and adapt its appearance accordingly. One of the advantages of IPatch is its remote nature, which allows for flexible and dynamic manipulation of object detections. The patch can be digitally transmitted and applied in real-time, enabling attackers to launch adversarial attacks from a distance. This remote capability opens up new possibilities for targeted attacks in surveillance systems, autonomous vehicles, and other real-world applications.

IPatch stands as a different concept in the domain of adversarial attacks, specifically focusing on the development of remote adversarial patches. The essence of IPatch lies in its ability to subtly manipulate the perception of a computer vision system from a distance, targeting a designated region to influence the prediction outcome. As illustrated in Figure \ref{fig:ipatch}, this remote adversarial patch is designed with precision, encapsulating a defined operational area denoted as $m$, and guided by a target pattern $t$. By strategically selecting the region of operation and aligning it with the intended target pattern, IPatch seeks to alter the probabilities associated with specific pixels, thereby influencing the prediction outcome of a desired class within that region. What sets IPatch apart is its emphasis on efficiency and directed optimization. The method zeroes in on the operational area, ensuring that computational efforts are solely dedicated to the relevant region, avoiding energy wastage on extraneous semantic elements. The mechanism for region selection involves the identification of a location $L=(i,j)$, around which a designated square or circular region is marked within $m$. This marked area serves as the locus for subsequent manipulations. Depending on the objective, whether it involves inserting an object or introducing a custom shape, the target pattern $t$ is adjusted accordingly. To generate the patch, they adopt the Expectation over Transformation (EOT) approach, but with a unique twist: they leverage their specialized semantic masks. The objective is to find a patch $P$ that undergoes training to optimize a specific objective function.

\begin{table*}[!htp]
\centering
  \caption{ Comparison of attack methods.  HA – Hiding objects; P – Adding phantom objects; AL - Altering objects. F – Anywhere in the frame; O – On the target object(s); S – On the sensor (i.e., camera); NO – Creating new object to fool the model; B: Blurring the image.}
  \label{Table:Semantic_comparison}
  \begin{tabular}{lcccccc}
    \toprule
       \textbf{Attack}   & \textbf{Attack goal}  & \textbf{Placement} & \textbf{Consider changing view point} & \textbf{Test in physical domain} & \textbf{Test transferability}\\
    \midrule 
          IPatch \cite{ipatch} & HA-AL-P  &  F   &   $\times$     &   $\times$    & \checkmark  \\
          SSAttack \cite{ssattack}  & HA-AL-P  &  NO   &   $\times$     &   \checkmark    & \checkmark  \\
  \bottomrule
\end{tabular} %}
\end{table*}

\begin{table*}[!htp]
\centering
  \caption{Physical adversarial attacks against \textbf{Semantic Segmentation} tasks. Attacker's knowledge, Robustness technique, Stealthiness technique, Physical test type, and Space.}
  \label{Table:Semantic_info}
  \begin{tabular}{llllll}
    \toprule
       \textbf{Attack}  & \textbf{Attacker’s}  & \textbf{Robustness} & \textbf{Stealthiness}  & \textbf{Physical}  & \textbf{Space} \\
         & \textbf{Knowledge}  & \textbf{Technique} & \textbf{Technique}  & \textbf{test type}  &  \\
    \midrule 
           IPatch \cite{ipatch}  &  White-box   &   EOT   &  -  & Static  &  2D  \\
           SSAttack  \cite{ssattack} & White-box  & EOT  & - & Static & 2D  \\
   \bottomrule
\end{tabular} %}
\end{table*}

\begin{table*}[!htp]
\centering
  \caption{Attacks on \textbf{Semantic Segmentation} task, Datasets, networks, and Code.}
  \label{Table:Semantic_dataset}
  \begin{tabular}{llll}
    \toprule
       \textbf{Attack}   & \textbf{Dataset} & \textbf{Network}  & \textbf{Code} \\
    \midrule 
            IPatch & CamVid dataset & Unet++, Linknet, FPN, PSPNet, PAN & https://ymirsky.github.io/\\
            SSAttack & Carla simulator, Cityscapes dataset & SS model& https://github.com/retis-ai/semsegadvpatch\\
            % & & & \\
            % & & & \\
  \bottomrule
\end{tabular} %}
\end{table*}

%% file: depth.tex
%%%%%%%%%%%%%%%%%%%%%%%%%%%%%%%%%%%%%%%%%%%%%%%%%%%%%%%%%%%%
\begin{figure*}
    \centering
    \includegraphics[width=0.8\textwidth]{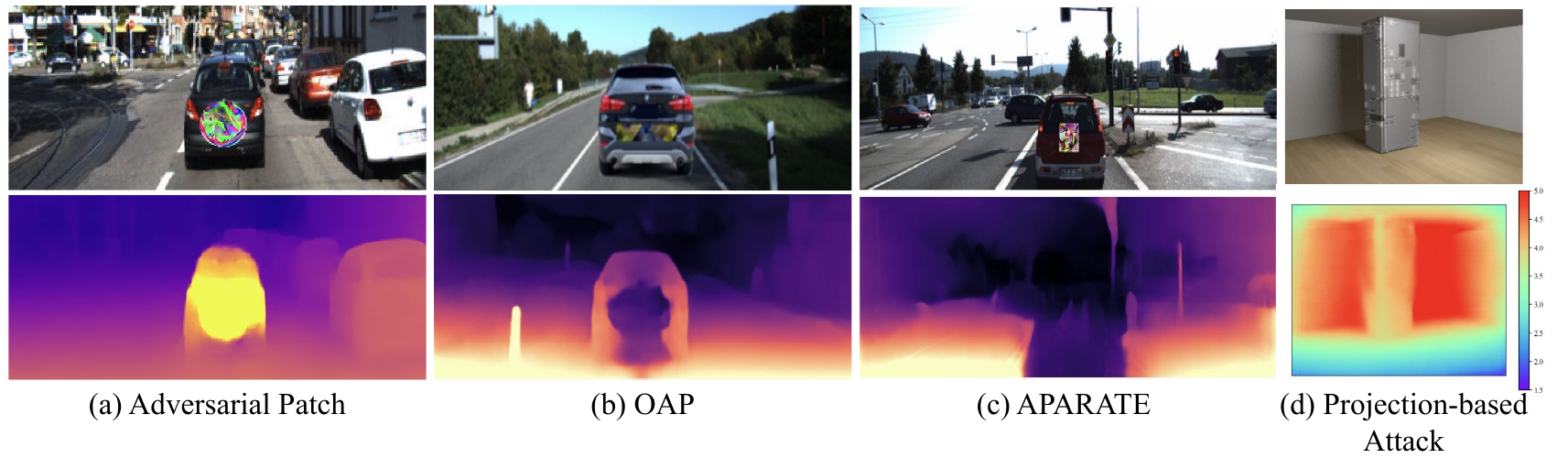}
    \caption{Illustration of different physical attacks on depth estimation task: (a) Adversarial Patch \cite{adversarial-patch}, (b) OAP \cite{OAP}, (c) APARATE \cite{aparate}, and (d) Projection-based Attack \cite{projection}.}
    \label{fig:depth}
\end{figure*}
%%%%%%%%%%%%%%%%%%%%%%%%%%%%%%%%%%%%%%%%%%%%%%%%%%%%%%%%%%%%
%======================================================
\section{Physical Attacks on Monocular Depth Estimation \& Optical Flow Estimation}
%======================================================
\label{depth}
Monocular depth estimation and optical flow estimation are two fundamental tasks in the field of computer vision, which play a crucial role in understanding the spatial layout and motion information within images or video sequences. Monocular depth estimation involves inferring depth or distance information from a single image, enabling the perception of the three-dimensional structure of the scene. Deep learning models, particularly convolutional neural networks (CNNs), have revolutionized monocular depth estimation by learning to extract depth-related cues and capturing complex depth patterns. These models leverage large-scale datasets with corresponding depth annotations to train and accurately predict depth maps.

On the other hand, optical flow estimation focuses on estimating the dense pixel-level motion information between consecutive frames in a video sequence. It enables the understanding of object motion, scene dynamics, and facilitates motion-based tasks such as object tracking and activity recognition. Deep learning models, including CNNs and recurrent neural networks (RNNs), have made significant advancements in optical flow estimation by learning to capture motion patterns and model temporal dependencies. These models are trained on annotated datasets containing pairs of frames and their corresponding optical flow vectors, allowing them to learn the complex relationship between image intensity changes and pixel motion.

Both monocular depth estimation and optical flow estimation are essential for various computer vision applications. Monocular depth estimation provides valuable information about scene geometry, which is crucial for applications such as autonomous navigation, virtual reality, and scene understanding. Optical flow estimation, on the other hand, enables the analysis of object motion, facilitating tasks such as video stabilization, action recognition, and object tracking. The advancements in deep learning models have significantly improved the accuracy and robustness of these tasks, leading to advancements in various real-world applications.

Table \ref{Table:depth_comparison} presents a comprehensive comparison of various adversarial attack methods in the depth estimation task. It provides an overview of the attack goals, patch placement strategies, consideration of changing viewpoints, testing in the physical domain, and transferability to other models.
Table \ref{Table:depth_info} offers detailed information on adversarial attacks, including the attacker's knowledge level, robustness techniques, stealthiness techniques, physical test types, and operational space.
Table \ref{Table:depth_dataset} provides information on the datasets used, the evaluated networks, and the links to open-source code for the experiments conducted in the depth estimation task.

%------------------------------------
\subsection{Depth Estimation}
%------------------------------------

%------------------------------------
\subsubsection{Patch-based Attack}
%------------------------------------
%Adversarial patch \cite{adversarial-patch}
%OAP \cite{OAP}
%APARATE \cite{aparate}

\textbf{Adversarial patch}: The authors in \cite{adversarial-patch} introduced a novel method for performing adversarial patch attacks on CNN-based monocular depth estimation models. Their approach involved training adversarial patches to induce incorrect depth estimations in the vicinity of their respective locations, accounting for various transformations such as perspective transformations, scaling, and translation. The feasibility of their method was demonstrated in real-world scenarios by physically placing printed patterns in the target scenes. Furthermore, the authors conducted an analysis of the behavior of monocular depth estimation models under attack, visualizing the potentially affected regions and activation maps. %This analysis provided insights into the vulnerability and impact of adversarial patches on the performance of monocular depth estimation methods.

Monocular Depth Estimation (MDE) is a critical task in computer vision, used to estimate the distance between a camera and a target object, with significant applications in autonomous driving systems. In recent work, Cheng et al. introduced a novel physical adversarial attack against learning-based MDE models called \textbf{Object-oriented Adversarial Patch (OAP)} \cite{OAP}. This approach generates a physical adversarial patch capable of launching attacks in real-world driving scenarios.
To ensure the effectiveness of the attack, OAP employs a rectangular patch region optimization method to find the optimal patch-pasting region. This optimization strategy results in a mean depth estimation error of over 6 meters and an impressive 93\% Adversarial Success Rate (ASR) in a downstream task, specifically 3D object detection. The researchers also address the challenge of stealthiness by minimizing the patch size while maintaining a successful attack. Additionally, they incorporate a style transfer loss to give the adversarial patch a more unobtrusive appearance, allowing it to blend seamlessly with the surrounding environment.
Robustness is another essential consideration in OAP. The approach applies the Expectation over Transformation (EOT) technique \cite{EOT} and includes various physical transformations, such as size, rotation, brightness, and saturation, during the training stage. This helps the perturbation become resilient to various physical transformations that may occur in real-world scenarios.

\textbf{APARATE} \cite{aparate} (Adaptive Adversarial Patch for CNN-based Monocular Depth Estimation for Autonomous Navigation) is a method designed to manipulate monocular depth estimation systems used in autonomous navigation tasks. By introducing an adaptive adversarial patch, APARATE aims to deceive the depth estimation models and create erroneous depth maps, which can have significant implications for autonomous navigation systems. The key feature of APARATE is its adaptability. The patch is optimized to be adaptive to different scenes, allowing it to effectively deceive the depth estimation model across various environments. This adaptability is achieved through an optimization process that iteratively updates the patch based on the gradients obtained from the depth estimation model's backpropagation. By incorporating the model's feedback, APARATE can dynamically adjust the patch to maximize its impact on the depth estimation process.

\textbf{SAAM} \cite{saam} (Stealthy Adversarial Attack with Monocular Depth Estimation) attack is a patch-based adversarial attack specifically designed to target Monocular Depth Estimation (MDE) models. 
SAAM achieves its stealthy nature by crafting an adversarial patch that seamlessly blends with the scene, making it difficult for human observers to detect any visual anomalies. This patch is meticulously designed to deceive the MDE model, leading to significant depth errors and misinterpretations in the perceived 3D structure of the scene. The attack methodology, presented in Figure \ref{SAAM_overview}, involves iteratively optimizing the adversarial patch in the pixel space while enforcing a semantic constraint to ensure its visual realism. SAAM is equipped to withstand various real-world distortions such as rotation, perspective change, lighting variation, and occlusion, making it effective in different practical scenarios.

%Screen Shot 2023-08-04 at 4.56.51 PM
\begin{figure}[!t]
    \centering
    \includegraphics[width=0.5\textwidth]{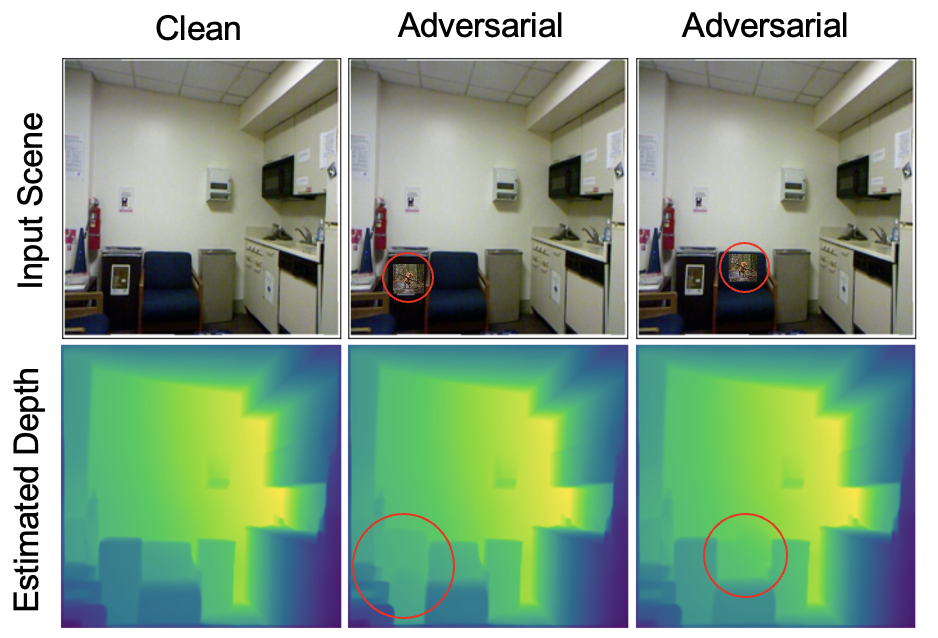}
    \caption{SAAM results: Effectiveness of SAAM in concealing objects: Placing the patch on a target objects changes its predicted depth in a way that the object is blended with the background (Figure adapted from \cite{saam}).}
    \label{fig:conceal}
\end{figure}
%------------------------------------
\subsubsection{Light Manipulation}
%------------------------------------

\textbf{Projection-based attack} \cite{projection} proposes a method to perform physical adversarial attacks on monocular depth estimation networks. The objective is to generate adversarial patches that can be physically placed in the scene to deceive the depth estimation network and produce inaccurate depth predictions. The authors utilize a projection-based approach where they first generate an adversarial depth map using an optimization algorithm. This adversarial depth map is then transformed into a physical adversarial patch using a projector. The patch is designed to be visually inconspicuous and blend seamlessly with the scene. By physically placing the patch in the scene, the authors demonstrate that the depth estimation network is tricked into perceiving the patch as part of the scene, leading to inaccurate depth predictions. To generate the adversarial depth map, the authors formulate an optimization problem that minimizes the discrepancy between the predicted depth from the network and the desired target depth values for the patch region. The optimization process considers both the visual appearance and the geometric consistency of the patch to ensure a convincing attack. Experimental evaluations are conducted on various monocular depth estimation models, demonstrating the effectiveness of the proposed attack method. The results show that the physical adversarial patches can successfully deceive the depth estimation networks and introduce significant errors in the predicted depth maps.

\subsection{Optical Flow Estimation}

%FlowAttack \cite{flowattack}
Optical Flow Estimation (OFE) is a fundamental task in computer vision that aims to estimate the pixel-level 2D motion within an image sequence. It plays a crucial role in various applications such as motion analysis, tracking, and video understanding. Deep learning models, such as FlowNet \cite{ilg2017flownet}, have significantly advanced the accuracy and robustness of OFE by learning to extract motion patterns and capturing spatial and temporal dependencies. Ranjan et al. \cite{flowattack} introduced \textbf{FlowAttack}, a method specifically designed to perturb OFE models. The effectiveness of FlowAttack lies in its utilization of the gradients of pre-trained optical flow networks to update adversarial patches. The approach generates counterpart patches for different networks, allowing for targeted attacks. Experimental results demonstrated that FlowAttack can induce significant errors in encoder-decoder networks while having a lesser impact on spatial pyramid networks. FlowAttack does not explicitly consider stealthiness in its design, focusing primarily on the effectiveness of the attack. However, it should be noted that the adversarial perturbations produced by FlowAttack may not be visually inconspicuous. In terms of robustness, FlowAttack incorporates rotations and scaling transformations to facilitate physical adversarial attacks, enhancing the resilience of the attack strategy against transformations that may be encountered in real-world scenarios.
%%%%%%%%%%%%%%%%%%%%%%%%%%%%%%%%%%%%%%%%%%%%%%%%%%%%%%%%%%%%
\begin{figure}
    \centering
    \includegraphics[width=0.5\textwidth]{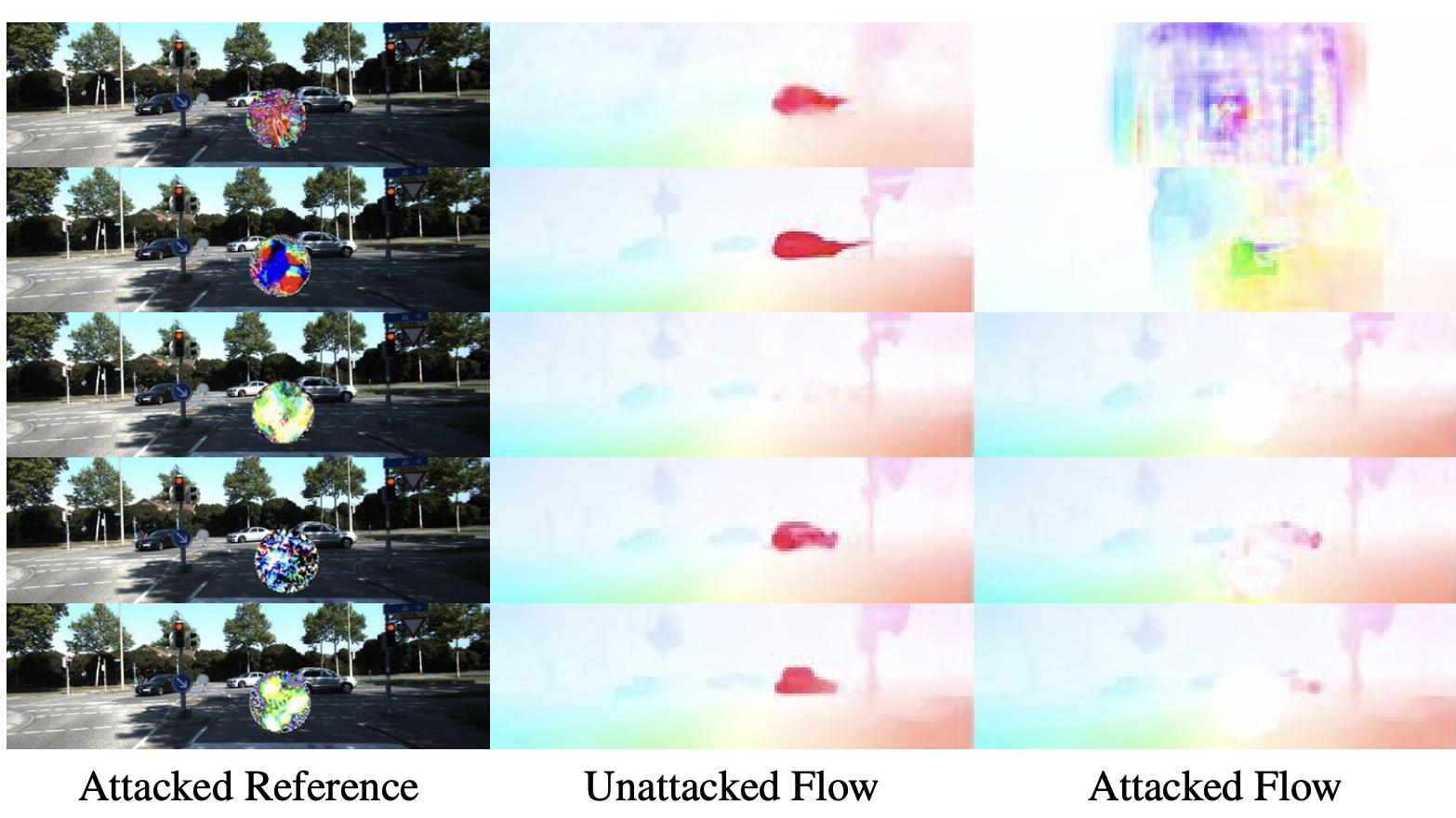}
    \caption{Illustration of different physical attacks on Optical flow: FlowAttack \cite{flowattack}.}
    \label{fig:flow}
\end{figure}
%%%%%%%%%%%%%%%%%%%%%%%%%%%%%%%%%%%%%%%%%%%%%%%%%%%%%%%%%%%%
\begin{table*}[!htp]
\centering
  \caption{ Comparison of attack methods. M - Corrupted estimation; H – Hiding objects. a – Anywhere in the frame; O – On the target object(s).}
  \label{Table:depth_comparison}
  \begin{tabular}{lcccccc}
    \toprule
       \textbf{Attack}   & \textbf{Attack goal}  & \textbf{Placement} & \textbf{Consider changing view point} & \textbf{Test in physical domain} & \textbf{Test transferability}\\
    \midrule 
           Adversarial patch \cite{adversarial-patch}& M  & A   &  $\times$   &   $\checkmark $    & $\times$ \\  %\checkmark 
           OAP \cite{OAP}                            & M &  O  &  $\times$   &   $\checkmark $     & $\times$  \\
           APARATE \cite{aparate}                    & H \& M&  O  &  $\times$   &    $\times$   & $\times$  \\ 
           Projection-based attack \cite{projection} & H \& M& O   &   $\times$  &    $\times$   & $\times$  \\ 
           SAAM \cite{saam}  & H \& M & O   &   $\times$  &    $\times$   & $\times$  \\ 
  \bottomrule
\end{tabular} %}
\end{table*}

\begin{table*}[!htp]
\centering
  \caption{Physical adversarial attacks against \textbf{Depth Estimation} \& \textbf{Optical Flow Estimation} tasks. Attacker's knowledge, Robustness technique, Stealthiness technique, Physical test type, and Space.}
  \label{Table:depth_info}
  \begin{tabular}{llllll}
    \toprule
       \textbf{Attack}  & \textbf{Attacker’s}  & \textbf{Robustness} & \textbf{Stealthiness}  & \textbf{Physical}  & \textbf{Space} \\
         & \textbf{Knowledge}  & \textbf{Technique} & \textbf{Technique}  & \textbf{test type}  &  \\
    \midrule 
            Adversarial patch \cite{adversarial-patch}   & White-box  & EOT, NPS, TV  & -  & Static &  2D \\
            OAP \cite{OAP}   & White-box  & EOT, TV  & Style loss, Content loss, Photorealism regularization loss & Static & 2D  \\
            APARATE \cite{aparate}   & White-box  & EOT, NPS, TV  & - & Static & 2D  \\
            Projection-based attack \cite{projection}   &  Black-box & -  & - & Static & 2D  \\
            SAAM \cite{saam} & White-box  & EOT, TV  & Semantic constraint & Static & 2D  \\
   \bottomrule
\end{tabular} %}
\end{table*}

\begin{table*}[!htp]
\centering
  \caption{Attacks on Monocular Depth Estimation \& Optical Flow Estimation task, Datasets, networks, and Code.}
  \label{Table:depth_dataset}
  \begin{tabular}{llll}
    \toprule
       \textbf{Attack}   & \textbf{Dataset} & \textbf{Network}  & \textbf{Code} \\
    \midrule 
            Adversarial patch & KITTI dataset& Monodepth2, Depthhints, Manydepth & -\\
            OAP               & KITTI dataset & Monodepth2, Depthhints, Manydepth & -\\
            APARATE           & KITTI dataset& Monodepth2, Depthhints, Manydepth & -\\
            Projection-based attack & Indoor scenes & DNN model trained with indoor scenes & -\\
            SAAM & NYUv2& DiverseDept, Monodepth2& -\\
            % & & & \\
  \bottomrule
\end{tabular} %}
\end{table*}

%\begin{table}[!htp]
%\centering
%  \caption{Attacks on Monocular Depth Estimation \& Optical Flow Estimation task, Datasets, networks, and Code.}
%  \label{ssim}
%  \begin{tabular}{llll}
%    \toprule
%       \textbf{Attack}   & \textbf{Dataset} & \textbf{Network}  & \textbf{Code} \\
%    \midrule 
%             & & & \\
%             & & & \\
%             & & & \\
%             & & & \\
%  \bottomrule
%\end{tabular} %}
%\end{table}

%% file: discussion.tex
\section{Open Research Challenges and Future Trends} %\textcolor{red}{\textbf{TODO}}
\label{discussion}

%\subsection{Challenges}

\subsection{No Benchmark for Physical Attacks Available}
One of the challenges in the field of physical adversarial attacks is the absence of a standardized physical adversarial attack benchmark. Unlike digital adversarial attacks, where researchers have established benchmark datasets and evaluation metrics, physical adversarial attacks lack a unified benchmark that could facilitate the comparison and evaluation of different attack methods. Without a standardized benchmark, it becomes difficult to objectively measure the effectiveness, robustness, and stealthiness of physical attacks across different tasks and scenarios.
A physical adversarial attack benchmark would be valuable for researchers and practitioners to test and validate their proposed attack methods in a controlled and consistent manner. It could consist of a diverse set of real-world scenarios, with variations in lighting conditions, camera angles, object poses, and environmental factors. Additionally, the benchmark could include evaluation metrics that capture the impact of physical attacks on the targeted models' performance and robustness.

%Establishing a physical adversarial attack benchmark is a complex task, as it requires carefully curating a large dataset of real-world images and conducting extensive experiments to assess the performance of different attack methods. Moreover, the benchmark should be regularly updated to reflect the evolving challenges and advancements in physical adversarial attacks.
\subsection{Difficult to Reproduce}
Indeed, one of the significant challenges of physical adversarial attacks lies in their reproducibility and manufacturing process. Unlike digital adversarial attacks, which are typically easy to replicate and share due to the nature of digital images, physical attacks involve crafting perturbations in the physical world, which can be more intricate and resource-intensive.

The difficulty in reproducing physical adversarial attacks arises from several factors:
\begin{itemize}
    \item \textbf{Material and Equipment Variability}: Physical attacks often require specific materials, equipment, or settings to create the adversarial perturbations. Variability in the availability and quality of these resources can lead to variations in the attack effectiveness when attempted by different researchers.
    \item \textbf{Environmental Factors}: Physical attacks are sensitive to environmental conditions, such as lighting, camera settings, and background textures. These variations can impact the performance of the attack, making it challenging to reproduce the exact conditions for consistent results.
    \item \textbf{Precision and Calibration}: Crafting physical perturbations with high precision and calibration is crucial for their effectiveness. Small deviations in placement, size, or appearance of the perturbations can significantly affect the success of the attack.
    \item \textbf{Real-World Constraints}: Physical attacks must adhere to real-world constraints, such as size limitations, appearance constraints, and the ability to be applied in diverse scenarios. These constraints can limit the range of possible attack strategies.
    \item \textbf{Cost and Time}: Manufacturing physical adversarial perturbations can be costly and time-consuming, depending on the complexity of the attack. This can discourage researchers from replicating or sharing their attack methods.
\end{itemize}
%To address the issue of reproducibility, researchers can document their attack methodologies, including detailed descriptions of materials, equipment, and settings used during the manufacturing process. Sharing the designs, 3D models, or code for creating physical perturbations can also facilitate reproducibility and foster collaboration in the research community.

%Efforts to establish standardized benchmarks for physical adversarial attacks and the creation of publicly available datasets and evaluation criteria can further contribute to the reproducibility and comparability of different attack methods. Additionally, collaborative research initiatives and open-source contributions can play a vital role in overcoming the challenges associated with manufacturing and reproducing physical adversarial perturbations in various forms.
\subsection{Difficult to Quantitatively Evaluate}
Evaluating the stealthiness of physical adversarial attacks is indeed a challenging task. Unlike other attack metrics, such as accuracy or robustness, which can be quantitatively measured, assessing the stealthiness of an attack requires subjective human perception and judgment. Stealthiness refers to how imperceptible the adversarial perturbation is to human observers, making it difficult to capture with conventional quantitative metrics.

One common approach to evaluating the stealthiness of physical adversarial attacks is through user studies. In these studies, human participants are presented with both the original unaltered image and the image with the adversarial perturbation applied. Participants are then asked to compare the two images and identify any noticeable differences or artifacts introduced by the perturbation. The level of stealthiness is assessed based on how well the adversarial perturbation blends with the original image without being visually detectable by humans.
While user studies provide valuable insights into the stealthiness of physical attacks, they have limitations, such as subjectivity and potential bias. Additionally, conducting user studies can be time-consuming and require a significant number of participants to obtain reliable results.

To address the lack of quantitative metrics for stealthiness, researchers have proposed benchmarking and assessing methods for evaluating the visual naturalness of physical world adversarial attacks. One such effort is the work titled "Towards Benchmarking and Assessing Visual Naturalness of Physical World Adversarial Attacks," \cite{li2023benchmarking} which aims to provide objective metrics for quantifying the visual similarity between adversarial and unaltered images. These metrics focus on assessing the perceptual difference between the original and perturbed images using image quality assessment methods, such as structural similarity index (SSIM), peak signal-to-noise ratio (PSNR), or perceptual similarity index (PSI). By comparing these scores for different attack methods, researchers can gain insights into the stealthiness of the attacks. While such benchmarking efforts are valuable for advancing the assessment of stealthiness, it is essential to recognize that quantifying human perception remains a challenging task. Therefore, user studies are likely to remain a crucial component in evaluating the stealthiness of physical adversarial attacks, as they provide valuable qualitative and perceptual insights that cannot be fully captured by objective metrics alone. Integrating both quantitative metrics and user studies can lead to more comprehensive evaluations of stealthiness and enhance the understanding of the visual impact of physical adversarial attacks on human observers.

By adhering to well-defined reproducibility criteria and utilizing appropriate evaluation metrics, researchers can contribute to the advancement of the field and build on the knowledge of physical adversarial attacks in computer vision tasks. This, in turn, promotes the development of more robust and secure computer vision systems in real-world applications.

%\subsection{Future work}

%\begin{itemize}
%    \item Most of the proposed attacks are under white-box setting. However, despite their success, the white-box setting is not practical.
%    Developing physically deployable universal black-box attacks is more realistic in practice but more difficult.
%    \item Multi-model attacks /Simultaneous attack multitask: in the real world, the attacker cannot get knowledge about the target victim system, such as whether the system is equipped with the image recognition model or object detection model. Therefore, one promising direction to address the above problem is to take multi-task into account during the optimization simultaneously.
%    \item Physical adversarial attacks on videos
%    \item Physical adversarial attacks on new tasks: trajectory prediction, pose estimation, action recognition
%\end{itemize}

\subsection{Investigate Black-box Attacks and Transferability}
%Most of the proposed attacks are under white-box setting. However, despite their success, the white-box setting is not practical. Developing physically deployable universal black-box attacks is more realistic in practice but more difficult.
Future work in the field of physical adversarial attacks should focus on addressing the challenges of black-box settings and developing more practical and deployable attack methods. Universal black-box attacks are more challenging to craft compared to white-box attacks because they need to generalize across different models and environments. Developing effective and efficient methods for generating physically deployable universal black-box attacks is a critical research direction. Investigate the transferability and generalization capabilities of physical adversarial attacks across different scenarios, devices, and datasets. Understanding how well an attack can be transferred from one model or dataset to another is crucial for evaluating its real-world impact and potential threats. By addressing these research directions, we can advance the field of physical adversarial attacks, making it more practical and applicable to real-world scenarios while also better understanding its potential risks and implications. Ultimately, this research will contribute to the development of more secure and robust computer vision systems in various applications.

\subsection{Attacking Multitask-based Systems}
Multi-model attacks, also known as simultaneous attack multitask, are a promising direction to address the challenge of not having prior knowledge about the target victim system in the real world. In scenarios where the attacker cannot determine whether the system employs an image recognition model or an object detection model, considering multiple tasks during the optimization process becomes essential.
The idea behind multi-model attacks is to craft an adversarial perturbation that can effectively fool a wide range of target models, regardless of their specific architectures or functionalities. By optimizing the adversarial perturbation for multiple tasks simultaneously, the attack becomes more versatile and robust, making it more likely to succeed against various computer vision models in real-world scenarios.
To achieve multi-model attacks, researchers explore different optimization techniques and loss functions that can simultaneously optimize the adversarial perturbation for multiple tasks. This approach allows the attacker to generalize the attack to different models without having specific knowledge about the victim system.
By incorporating multi-model attacks into the development of physical adversarial attack methods, researchers can enhance the potential threat of adversarial attacks in real-world settings. However, it is crucial to consider the ethical implications of multi-model attacks, as they can pose significant security risks and challenges in various domains, including autonomous systems, surveillance, and authentication systems. Ethical considerations should guide the responsible use and development of adversarial attack techniques to ensure the safety and security of individuals and society as a whole.

\subsection{Physical Adversarial Attacks on Videos}
Investigating physical adversarial attacks on videos is a compelling area of research with practical implications in various applications, such as surveillance, autonomous driving, and video analysis systems. Unlike single-frame image attacks, video-based attacks require consideration of temporal information and pose unique challenges due to the dynamic nature of videos. One key aspect of physical adversarial attacks on videos is the temporal consistency of the adversarial perturbations. The perturbations should be crafted in such a way that they maintain consistency over consecutive frames, ensuring that the attack remains effective throughout the video sequence. Another critical factor is the consideration of motion and dynamic scene changes. Videos often contain moving objects and changing backgrounds, which can affect the effectiveness of adversarial perturbations. Techniques must be developed to handle motion-induced artifacts and ensure robustness against varying scenes. Furthermore, evaluating the impact of physical adversarial attacks on video-based deep learning models is essential. The evaluation should go beyond single-frame accuracy and encompass temporal robustness, object tracking, and action recognition performance. In the context of video-based attacks, stealthiness is particularly important since any noticeable perturbations could raise suspicions or alert the target system. Ensuring imperceptibility in the temporal domain is a challenging aspect that requires careful consideration during the attack design.

%Addressing the challenges of physical adversarial attacks on videos will contribute to understanding the security vulnerabilities of video-based deep learning models and enable the development of more robust and secure systems. It is crucial to conduct thorough research while considering the ethical implications of these attacks to ensure responsible and ethical use in real-world applications.

\subsection{Attacking New Tasks}
Investigating physical adversarial attacks on new computer vision tasks, such as trajectory prediction, pose estimation, and action recognition, is an emerging and promising research direction with significant implications for real-world applications.
%1. Trajectory Prediction: In trajectory prediction tasks, the goal is to predict the future movement path of objects or agents in a scene. Adversarial attacks in this context could have significant safety implications, especially in autonomous vehicles and robotics. Crafting adversarial perturbations that mislead trajectory prediction models could lead to unexpected and dangerous behaviors, posing a serious threat to safety.
%2. Pose Estimation: Pose estimation aims to predict the 3D pose or joint angles of objects or human bodies from images or videos. Adversarial attacks in this domain could have implications in areas like motion capture, human-robot interaction, and augmented reality. Perturbing pose estimation systems could lead to incorrect pose estimations, resulting in incorrect animations, inaccurate robot control, or compromised AR experiences.
%3. Action Recognition: Action recognition involves identifying and classifying human actions from video sequences. Attacks on action recognition systems could be detrimental in security and surveillance applications. For instance, maliciously perturbing action recognition models could lead to false identifications, impacting decisions in security-sensitive scenarios.
In the context of new tasks, physical adversarial attacks must address specific challenges related to the nature of each task. For example, trajectory prediction tasks require perturbations that are consistent over time to effectively deceive the model. Pose estimation tasks may need adversarial perturbations that preserve the overall structure and alignment of body joints. Action recognition systems may require attacks that maintain the temporal coherence of actions.
Evaluating the robustness and stealthiness of physical adversarial attacks in these tasks is crucial. The evaluation criteria should align with the specific requirements of each task, such as accuracy, temporal consistency, or pose alignment. Researchers need to devise appropriate metrics to assess the performance of physical adversarial attacks in these new domains accurately.
Moreover, exploring transferability across different tasks and evaluating the transferability of physical adversarial attacks from one task to another can provide insights into the generalization and potential risks associated with these attacks. Understanding the transferability of adversarial attacks will shed light on the vulnerability of computer vision models and the potential impact on various real-world applications.
Conducting research responsibly and considering the potential ethical implications of such attacks is paramount. As the development and deployment of physical adversarial attacks continue, researchers must be mindful of the potential harm they may cause in critical applications such as autonomous driving, healthcare, and surveillance. Responsible research and ethical considerations will play a vital role in ensuring the safety, security, and trustworthiness of computer vision systems and their applications in the real world.

%% file: conclusion.tex
\section{Conclusion} %\textcolor{red}{\textbf{TODO}}
\label{conclusion}

%In this survey, we present a comprehensive review of 93 physical adversarial attack approaches based on recent over 190 literature since the year 2016.
In this comprehensive survey paper on physical adversarial attacks in computer vision tasks, we have explored and analyzed the state-of-the-art techniques used to manipulate deep learning models in the real-world. By carefully examining over 190 papers and reviewing 94 adversarial attack methods across various visual tasks, we have provided a comprehensive overview of this rapidly evolving research field.
Furthermore, we delved into the characteristics of physical adversarial attacks, including their robustness techniques, stealthiness strategies, and evaluation metrics. We identified the challenges in quantitatively evaluating stealthiness and discussed efforts to benchmark visual naturalness for physical attacks.

In conclusion, this survey provides researchers and practitioners with a comprehensive understanding of physical adversarial attacks in computer vision. By comprehensively exploring the existing techniques and challenges, we hope to inspire future research to develop robust and secure computer vision systems in the face of adversarial threats in the physical domain. Ultimately, addressing these challenges will contribute to the safe and reliable deployment of deep learning models in real-world applications.